\documentclass[preprint,times]{elsarticle}
\usepackage{csquotes}
\usepackage{amsfonts,amsmath,amssymb,mathrsfs}
\usepackage{bm}
\usepackage{dsfont} % for sets of real, complex, integer etc. numbers
\usepackage{indentfirst}

\usepackage{subcaption}

\usepackage[italicdiff]{physics}

\usepackage[hypertexnames=false]{hyperref}

\usepackage{lineno}

\renewcommand{\le}{\leqslant}
\renewcommand{\ge}{\geqslant}

\newcommand{\kap}{\varkappa}
\newcommand{\eps}{\varepsilon}

\newcommand{\wt}[1]{{\widetilde #1}}
\newcommand{\ol}[1]{{\overline #1}}
\newcommand{\Par}[1]{{(#1)}}

\newcommand{\Ol}[1]{\mkern3.5mu\overline{\mkern-3.5mu {#1} \mkern-0.5mu}\mkern0.5mu}

\newcommand{\Int}[2]{\int\limits_{#1}^{#2}}
\newcommand{\Sum}[2]{\sum\limits_{#1}^{#2}}

\newcommand{\Hilbert}{\mathcal{H}}
\newcommand{\Domain}[1]{{\mathcal{D}(#1)}}
\newcommand{\Range}[1]{{\mathcal{R}(#1)}}
\newcommand{\Reals}{\mathds{R}}

\newcommand{\Space}[2]{{#1\qty(#2)}}
\newcommand{\Ltwo}[1]{{\Space{\mathcal{L}_2}{#1}}}
\newcommand{\Cinf}[1]{{\Space{C_0^\infty}{#1}}}

\newcommand{\Scalar}[2]{{\big(#1, #2\big)}}

\DeclareMathOperator{\sgn}{sgn}
\DeclareMathOperator{\Erf}{Erf}
\renewcommand{\Im}{\mathop{\mathrm{Im}}\nolimits} % Im
\renewcommand{\Re}{\mathop{\mathrm{Re}}\nolimits} % Re
\journal{Annals of Physics}
\begin{document}
%%%%%%%%%%%%%%%%%%%%%%%%%%%%%%%%%%%%%%%%%%%%%%%%%%%%%%%%%%%%%%%%%%%%%%%%%%%%%%%%
\begin{frontmatter}
%%%%%%%%%%%%%%%%%%%%%%%%%%%%%%%%%%%%%%%%%%%%%%%%%%%%%%%%%%%%%%%%%%%%%%%%%%%%%%%%
\title{Topological Pauli Phase and Fractional Quantization of Orbital Angular Momentum in the Problems of Classical and Quantum Physics}

\author[MEPhI]{Konstantin S. Krylov\corref{cor1}}
\cortext[cor1]{Corresponding author}
\ead{krylov@theor.mephi.ru}
\author[MEPhI]{Valeriy M. Kuleshov}
\author[ISRAS,HSE]{Yurii E. Lozovik}
\author[MEPhI]{Vadim D. Mur}

\address[MEPhI]{National Research Nuclear University \enquote{MEPhI}, Kashirskoe Shosse 31, Moscow 115409, Russia}
\address[ISRAS]{Institute for Spectroscopy, Russian Academy of Sciences, 108840 Troitsk, Moscow, Russia}
\address[HSE]{National Research University Higher School of Economics, 109028 Moscow, Russia}
%%%%%%%%%%%%%%%%%%%%%%%%%%%%%%%%%%%%%%%%%%%%%%%%%%%%%%%%%%%%%%%%%%%%%%%%%%%%%%%%
\begin{abstract}
%%%%%%%%%%%%%%%%%%%%%%%%%%%%%%%%%%%%%%%%%%%%%%%%%%%%%%%%%%%%%%%%%%%%%%%%%%%%%%%%
Physical problems for which the existence of non-trivial topological Pauli phase (i.e. fractional quantization of angular orbital angular momenta that is possible in 2D case) is essential are discussed within the framework of two-dimensional Helmholtz, Schroedinger and Dirac equations.

As examples in classical field theory we consider a \enquote{wedge problem}~--- a description of a field generated by a point charge between two conducting half-planes~--- and a Fresnel diffraction from knife-edge.

In few-electron circular quantum dots the choice between integer and half-integer quantization of orbital angular momenta is defined by the Pauli principle. This is in line with precise experimental data for the ground state energy of such quantum dots in a perpendicular magnetic field.

In a gapless graphene, as in the case of gapped one, in the presence of overcharged impurity this problem can be solved experimentally, e.g., using the method of scanning tunnel spectroscopy.
%%%%%%%%%%%%%%%%%%%%%%%%%%%%%%%%%%%%%%%%%%%%%%%%%%%%%%%%%%%%%%%%%%%%%%%%%%%%%%%%
\end{abstract}
%%%%%%%%%%%%%%%%%%%%%%%%%%%%%%%%%%%%%%%%%%%%%%%%%%%%%%%%%%%%%%%%%%%%%%%%%%%%%%%%
\begin{keyword}
%%%%%%%%%%%%%%%%%%%%%%%%%%%%%%%%%%%%%%%%%%%%%%%%%%%%%%%%%%%%%%%%%%%%%%%%%%%%%%%%
topological phase \sep half-integer orbital angular momenta \sep graphene \sep supercharged impurity
%%%%%%%%%%%%%%%%%%%%%%%%%%%%%%%%%%%%%%%%%%%%%%%%%%%%%%%%%%%%%%%%%%%%%%%%%%%%%%%%
\end{keyword}
%%%%%%%%%%%%%%%%%%%%%%%%%%%%%%%%%%%%%%%%%%%%%%%%%%%%%%%%%%%%%%%%%%%%%%%%%%%%%%%%
\end{frontmatter}
%\linenumbers
%%%%%%%%%%%%%%%%%%%%%%%%%%%%%%%%%%%%%%%%%%%%%%%%%%%%%%%%%%%%%%%%%%%%%%%%%%%%%%%%
\section{Introduction}\label{introduction}
%%%%%%%%%%%%%%%%%%%%%%%%%%%%%%%%%%%%%%%%%%%%%%%%%%%%%%%%%%%%%%%%%%%%%%%%%%%%%%%%
It's well known \cite{Dirac1958ClarendonPress, Neumann1955Princeton, Smirnov1964Pergamon, Wightman1967Cargese, Richtmyer1978Springer}, that translation generators, i.e. self-adjoint operators of infinitesimal translations on $a \le q \le b$, form a one-parameter family, $S_\theta = S_\theta^+$,
\begin{equation}\label{eq1}
    S_\theta \Psi(q) = -i\Psi'(q),\quad
    \Domain{S_\theta} = \qty{\Psi \in \Hilbert,\, \Psi' \in \Hilbert;\,
        \Psi(b) = e^{i\theta}\Psi(a)}.
\end{equation}
Here $\Domain{S_\theta}$ is the domain of operator $S_\theta$ in a Hilbert space $\Hilbert = \Ltwo{[a,b]}$ of square-integrable on $[a,b]$ wavefunctions $\Psi(q)$, $\theta$ is topological phase\footnote{On other topological phases in quantum mechanics, see \cite{ShapereWilczek1989WorldScientific, VinitskiiDerbov1990SovPhysUsp}.}, $0 \le \theta \le 2\pi$, and the derivative $\Psi'(q) = d\Psi/dq$ should be read \cite{Smirnov1964Pergamon, Wightman1967Cargese} in the sense of the Schwarz distribution theory \cite{Smirnov1964Pergamon, Richtmyer1978Springer}, see \ref{appendixA}.

In the particular case of two-dimensional rotations when the generalized coordinate $q$ is the angle $\varphi$, $0 \le \varphi \le 2\pi$, and the translation operator $S_\theta$ is the orbital angular momenta operator $L_\theta$ in a plane, topological phase arises as a result of a rotation about angle $\varphi = 2\pi$,
\begin{equation}\label{eq2}
    \Psi(2\pi) = e^{i\theta} \Psi(0),\quad
    \theta = 2\pi\delta,\quad 0 \le \delta < 1.
\end{equation}
This phase determines rotational generator $L_\theta$ and hence the unitary operator $U_\theta$ which describe rotational dynamics of the system \cite{Wightman1967Cargese},
\begin{equation}\label{eq3}
    U_\theta(\alpha) = e^{i\alpha L_\theta},\quad
    U_\theta(\alpha) \Psi(\varphi) = \Psi(\varphi + \alpha).
\end{equation}

In virtue of boundary condition~(\ref{eq2}) the eigenvalues of $L_\theta$, i.e. orbital angular momenta\footnote{in units $\hbar = 1.0546 \cdot 10^{-34}\text{ J\,$\cdot$\,s}$.} $M$, are
\begin{equation}\label{eq4}
    L_\theta \Psi_M(\varphi) = M\Psi_M(\varphi),\quad
    M = \delta + m,\quad 0 \le \delta < 1,\quad m = 0, \pm 1, \pm 2, \ldots
\end{equation}
Its eigenfunctions
\begin{equation}\label{eq5}
    \Psi_M(\varphi) = \frac{1}{\sqrt{2\pi}} e^{iM\varphi}
\end{equation}
form an orthonormal basis in a space of wavefunctions $\Psi(\varphi) \in \Ltwo{[0,2\pi]}$ and implement a multivalued $\delta \ne 0$ irreducible representations of two-dimensional rotational group SO(2) \cite{Hamermesh1962AddisonWesley}.

For T-inversion (more properly, direction of motion inversion) invariant systems \cite{Wigner1959AcadPress}, we have
\begin{equation}\label{eq6}
    \text{1) } \theta = 0,\, \delta = 0;\quad
    \text{2) } \theta = \pi,\, \delta = 1/2,
\end{equation}
see \cite{KowalskiPodlaskiEtAl2002PhysRevA}, and \ref{appendixA}. They correspond to a single- and double-valued representations of SO(2) \cite{Hamermesh1962AddisonWesley, KowalskiPodlaskiEtAl2002PhysRevA} respectively and lead to integer or half-integer quantization of angular orbital angular momentum respectively .

Since single-valuedness of wave function is not a fundamental principle of quantum theory and multi-valued wave functions cannot be excluded a priori\footnote{see also part 6 in \cite{Pauli1980Springer} and Appendix I in \cite{BlattWeisskopf1991Dover}} \cite{Pauli1939HelvPhysActa}, this raises a question, why only integer values of square of angular momenta $l(l+1)$, $l = 0, 1, 2, \ldots$ and its projections on any axis $M = m = 0, \pm1, \ldots, \pm l$ \cite{Dirac1958ClarendonPress} are realized in three-dimensional case.

It seems that this challenge was first issued and solved by Pauli in \cite{Pauli1939HelvPhysActa}, and then studied in details in \cite{Winter1968AnnPhys}. The main reason for integer quantization of angular momenta in three dimensions when its projections operators obey commutation relations
\begin{equation}\label{eq7}
    L_x L_y - L_y L_x = iL_z
\end{equation}
with cyclic commutations of $(x, y, z)$, is the requirement of unitary equivalence for operators of angular momentum projections $L_x$, $L_y$, $L_z$ \cite{Winter1968AnnPhys}.

Heuristically, this can be understood in the following way. The unitary equivalence means that the spectra of these operators must be the same, i.e. satisfy (\ref{eq4}) with the same value of $\delta$. If $\delta \ne 0$, i.e. multi-valued irreducible representation of SO(2) is realized, then according to \cite{Smirnov1964Pergamon}, see (\ref{eqA10}) In \ref{appendixA}, we can choose such a function $\Psi_y(\varphi)$ from the domain of $L_y$, $\Psi_y(\varphi) \in \Domain{L_y}$, that a function $L_y \Psi_y(\varphi)$ from the domain of $L_y$, $L_y \Psi_y(\varphi) \in \Range{L_y} = \Ltwo{[0,2\pi]}$, is not in the domain $L_x$, because $\Domain{L_x} \subset \Ltwo{[0,2\pi]}$. That is why first product of operators in (\ref{eq7}) make no sense.

Thus, the requirement of unitary equivalence of the operators of orbital angular momentum projections $L_x$, $L_y$, $L_z$ in 3D case excludes multi-valued, including double-valued, irreducible representations of SO(3) in coordinate representation (it was proved mathematically in \cite{Winter1968AnnPhys}). However, this exclusion is irrelevant to double-valued representations of this group which leads to half-integer quantization of total angular momenta (including spin).

Fractional values of an orbital angular momenta arise in classical field theory problems with separable variables in cylindrical coordinates. Examples are \enquote{wedge problem} \cite{LandauLifshitz8_1984Pergamon, BatyginToptygin1976AcadPress} and Fresnel diffraction from knife-edge \cite{MorseFeshbach1953McGrawHill}. We briefly discuss these problems in the next section \ref{part2}.

Half-integer quantization of an orbital angular momenta may rise in such quantum mechanical systems as circular quantum dots \cite{Chakraborty1999Elsevier, KouwenhovenAustingTarucha2001RepProgPhys}. It turns out \cite{MurNarozhnyEtAl2008JETPLetters, KuleshovMurEtAl2016FewBodySystems}, that the choice between two possibilities (\ref{eq6}) is specified by the Pauli exclusion principle and this choice is based on precise experimental data \cite{SchmidtTewordtEtAl1995PhysRevB}, derived using one-electron tunnel spectroscopy method. These questions are discussed in \ref{part3}.

Another near-perfect two-dimensional system is graphene \cite{CastroNetoGuineaEtAl2009RevModPhys}. Its electronic properties are described by two-dimensional effective Dirac equation \cite{ZhouGweonEtAl2007NatureMaterials, PereiraKotovCastroNeto2008PhysRevB, Novikov2007PhysRevB}. The spectrum and wavefunctions of gapped graphene as well as resonant scattering of holes in the presence of supercritical Coulomb impurity are discussed in \ref{part4}.

Brief remarks are given in conclusions \ref{conclusions}. In \ref{appendixA} we discuss self-adjoint extensions of an infinitesimal operator on an interval, in \ref{appendixB} the generalization of an expansion of a plane wave accounting for half-integer quantization of an orbital angular momenta is considered, and in \ref{appendixC} boundary conditions for radial Dirac equations are given.
%%%%%%%%%%%%%%%%%%%%%%%%%%%%%%%%%%%%%%%%%%%%%%%%%%%%%%%%%%%%%%%%%%%%%%%%%%%%%%%%
\section{Two-dimensional Helmholtz equation and boundary conditions on half-line}\label{part2}
%%%%%%%%%%%%%%%%%%%%%%%%%%%%%%%%%%%%%%%%%%%%%%%%%%%%%%%%%%%%%%%%%%%%%%%%%%%%%%%%
If variables in classical field theory problem \cite{LandauLifshitz2_1994Butterworth_eng} are separable in cylindrical coordinates and an angle $\varphi$ is in the sector $a \le \varphi \le b$, it is useful to use a complete set of eigenfunctions of the generator of rotations $R_\theta$ acting in the Hilbert space of wave functions $\Psi(\varphi) \in \Ltwo{[a,b]}$,
\[
    R_\theta\Psi(\varphi) = -i\Psi'(\varphi),\quad
    \Domain{R_\theta} = \qty{\Psi \in \Ltwo{[a,b]},\, \Psi' \in \Ltwo{[a,b]};\,
        \Psi(b) = e^{i2\pi\delta}\Psi(a)},
\]
see, e.g., (\ref{eqA6}) in \ref{appendixA}.

The eigenfunctions of this generator in a given sector,
\begin{equation}\label{eq8}
    \Psi_\mu(\varphi) = \frac{1}{\sqrt{b-a}} e^{i\mu\varphi},\quad
    \mu = \frac{2\pi}{(b-a)} (\delta + m),\quad 0 \le \delta < 1,\quad m = 0, \pm 1, \pm 2, \ldots,
\end{equation}
correspond to eigenvalues of an orbital angular momentum $\mu$ and form a complete orthonormal basis in $\Ltwo{[a,b]}$, see \cite{Neumann1955Princeton} and (\ref{eqA8}). If $(b-a) = 2\pi$, then the generator of rotations coincides with the orbital angular momentum operator, $R_\theta = L_\theta$, and its eigenvalues coincide with orbital angular momentum, $\mu = M$, see (\ref{eq4}) in part \ref{introduction}.

However, if $(b-a) \ne 2\pi$, then the angular momentum $\mu$ is fractional and in this case without loss of generality in (\ref{eq8}) it can be set $\delta = 0$, and it can be used the following complete set of real functions (instead of given in (\ref{eq8}))
\begin{equation}\label{eq9}
    \sin\Big[\frac{2\pi n}{(b-a)}\varphi\Big],\quad \cos\Big[\frac{2\pi n}{(b-a)}\varphi\Big],\quad
    n = 0, 1, 2, \ldots,
\end{equation}
leading to the standard Fourier expansion. This set corresponds to the superposition of (\ref{eq8}) with angular momenta
\begin{equation}\label{eq10}
    \mu = \frac{2\pi}{(b-a)}m,\quad m = 0,\pm 1, \pm 2,\ldots,\quad (b-a) \ne 2\pi,
\end{equation}
which are not integer even if $\delta = 0$, instead of (\ref{eq4}).
%%%%%%%%%%%%%%%%%%%%%%%%%%%%%%%%%%%%%%%%%%%%%%%%%%%%%%%%%%%%%%%%%%%%%%%%%%%%%%%%
\subsection{The wedge problem}\label{part21}
\begin{figure}[t]
    \centering
    \includegraphics[width=0.7\textwidth]{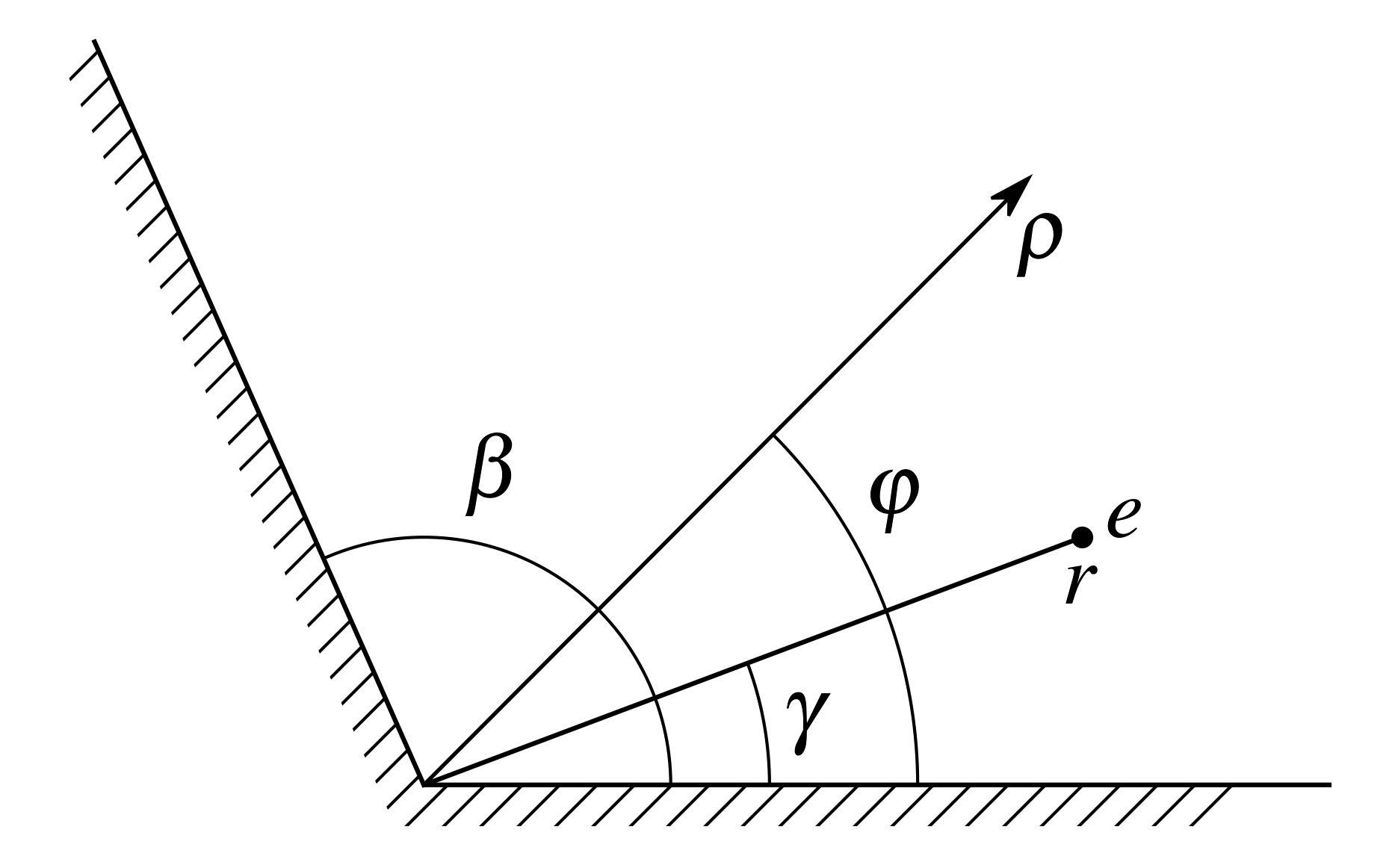}
    \caption{The \enquote{wedge problem}~--- determination of the field created by a point charge $e$ placed between two conducting half-planes.}
    \label{fig1}
\end{figure}

%%%%%%%%%%%%%%%%%%%%%%%%%%%%%%%%%%%%%%%%%%%%%%%%%%%%%%%%%%%%%%%%%%%%%%%%%%%%%%%%
Let us consider a \enquote{wedge problem}\footnote{see \cite{LandauLifshitz8_1984Pergamon} and references therein.} as a significant example. The problem is to find an electric field generated by a point charge $e$ located between two cross conducting half-planes, see Fig.~\ref{fig1}.

The Fourier component of a scalar potential $\Phi(\rho, \varphi, z)$,
\[
    u_\kap(\rho, \varphi) = \frac{1}{\pi} \Int{-\infty}{\infty} \dd{z}
        \Phi(\rho, \varphi, z) \cos(\kap z),
\]
meets the nonhomogeneous 2D Helmholtz equation
\begin{equation}\label{eq11}
    \frac{1}{\rho} \pdv{\rho} \Big(\rho \pdv{\rho} u_\kap\Big) +
        \frac{1}{\rho^2} \pdv[2]{u_\kap}{\varphi} + k^2 u_\kap =
    -\frac{4e}{r} \delta(\rho - r) \delta(\varphi - \gamma),\quad k = i\kap
\end{equation}
with boundary conditions
\begin{align}
    u_\kap(\rho, 0) = u_\kap(\rho, \beta) = 0, \label{eq12}\\
    u_\kap(\rho, \varphi)\Big|_{\rho \to \infty} = 0.
    \tag{\ref*{eq12}${}^\prime$} \label{eq12prime} % see https://tex.stackexchange.com/a/132408
\end{align}

To meet Dirichlet boundary condition (\ref{eq12}), let us set in (\ref{eq9})
\begin{equation}\label{eq13}
    b = 2\beta,\quad a = 2\alpha,\quad \alpha = 0,\quad 0 \le \varphi \le 2\beta.
\end{equation}
Partial solutions of homogeneous equation (\ref{eq11}) that are finite in the origin and vanish at the infinity are:
\[
    R_n(\rho) \sin\Big(\frac{\pi n}{\beta}\varphi\Big),\quad n = 1, 2, 3, \ldots;\quad
    R_n(\rho) =%
    \begin{cases}
        I_\frac{\pi n}{\beta}(\kap \rho), &\rho < r,\\
        K_\frac{\pi n}{\beta}(\kap \rho), &\rho > r,
    \end{cases}
\]
where $I_\nu(x)$ and $K_\nu(x)$ are modified Bessel functions \cite{BatemanErdelyi1_2_1953McGrawHill}

Then it can be shown \cite{BatyginToptygin1976AcadPress} that
\begin{equation}\label{eq14}
    u_\kap(\rho, \varphi) = \frac{8e}{\beta}%
    \begin{cases}
        \Sum{n=1}{\infty} K_\frac{\pi n}{\beta}(\kap r) I_\frac{\pi n}{\beta}(\kap \rho)
            \sin\big(\frac{\pi n}{\beta}\gamma\big) \sin\big(\frac{\pi n}{\beta}\varphi\big),
            &\rho < r,\\
        \Sum{n=1}{\infty} I_\frac{\pi n}{\beta}(\kap r) K_\frac{\pi n}{\beta}(\kap \rho)
            \sin\big(\frac{\pi n}{\beta}\gamma\big) \sin\big(\frac{\pi n}{\beta}\varphi\big),
            &\rho > r,
    \end{cases}
\end{equation}
the field potential is
\begin{equation}\label{eq15}
\begin{gathered}
    \Phi(\rho, \varphi, z) = \frac{e}{\beta\sqrt{2r\rho}} \Int{\eta}{\infty}\qty{
        \frac{1}{\cosh\big(\frac{\pi \zeta}{\beta}\big) -
            \cos\big[\frac{\pi(\varphi - \gamma)}{\beta}\big]} -
        \frac{1}{\cosh\big(\frac{\pi \zeta}{\beta}\big) -
            \cos\big[\frac{\pi(\varphi + \gamma)}{\beta}\big]}} \times\\
    \times \frac{\sinh\big(\frac{\pi\zeta}{\beta}\big)}{\sqrt{\cosh\zeta - \cosh\eta}} \dd{\zeta},\quad
        \cosh\eta = \frac{r^2 + \rho^2 + z^2}{2r\rho},\quad \eta > 0,
\end{gathered}
\end{equation}
and angular momentum in the superposition (\ref{eq14}) $\mu  = \frac{\pi}{\beta}m$, $m = \pm 1, \pm 2, \ldots$ One can see that on the conductor surface, i.e. when $\varphi = 0,\, \beta$, the potential $\Phi = 0$.

If $\beta = \pi$, the wedge becomes a conducting plane and the integral in (\ref{eq15}) can be evaluated easily,
\begin{equation}\label{eq16}
    \Phi(\rho, \varphi, z)\Big|_{\beta = \pi} = \frac{e}{R_+} + \frac{(-e)}{R_-},\quad
    R_\pm = \qty[r^2 + \rho^2 + z^2 - 2r\rho \cos(\varphi \mp \gamma)]^{1/2},
\end{equation}
that is in line with the method of images. Nevertheless, the equations (\ref{eq13}) guarantee the existence of \enquote{fictitious} half-space of the image, $\pi \le \varphi \le 2\pi$, and at the same time, integer values of an orbital angular momentum\footnote{Recall that $\delta = 0$, see (\ref{eq4}).} $\mu = M = m$, $m = \pm 1, \pm 2, \ldots$ in the expansion (\ref{eq14}).

However if $\beta = 2\pi$, the wedge becomes the conducting half-plane. In this case the integral in (\ref{eq15}) can also be evaluated \cite{LandauLifshitz8_1984Pergamon}
\begin{equation}\label{eq17}
    \Phi(\rho, \varphi, z)\Big|_{\beta = 2\pi} = \frac{q_+}{R_+} + \frac{(-q_-)}{R_-},\quad
    q_\pm = \frac{e}{\pi} \arccos\qty[-\frac{\cos[(\frac{\varphi \mp \gamma}{2})]}{\cosh(\frac{\eta}{2})}],
\end{equation}
$R_\pm$ are defined in (\ref{eq16}). Such a value of $\beta$ corresponds to the existence of an image space, $2\pi < \varphi < 4\pi$, the summation of potentials\footnote{Compare with the solution of problem 3.19 in \cite{BatyginToptygin1976AcadPress}.} of denumerable number of images charges and the contribution of half-integer values of an orbital angular momenta $\mu = \frac{1}{2}m$, $m = \pm 1, \pm 2, \ldots$ in (\ref{eq14}).
%%%%%%%%%%%%%%%%%%%%%%%%%%%%%%%%%%%%%%%%%%%%%%%%%%%%%%%%%%%%%%%%%%%%%%%%%%%%%%%%
\subsection{Diffraction from knife-edge}\label{part22}
\begin{figure}[t]
    \centering
    \includegraphics[width=0.7\textwidth]{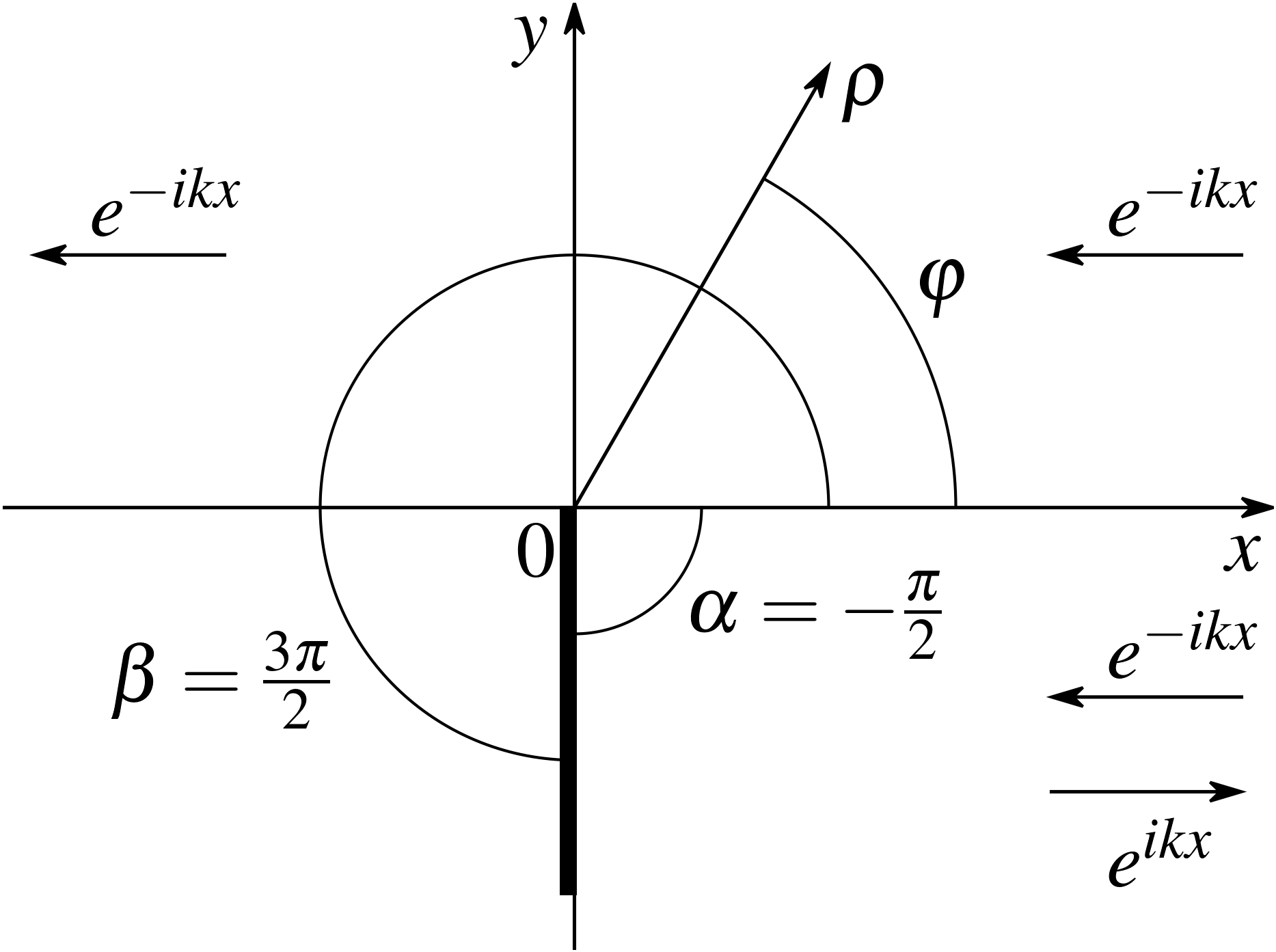}
    \caption{Scattering of a plane monochromatic wave with a frequency $\omega = ck$ on a semi-infinite screen, $x=0$, $y<0$.}
    \label{fig2}
\end{figure}

Our second example is scattering of a plane monochromatic wave with frequency $\omega = ck$ on a semi-infinite screen, $x=0$, $y<0$, which meets the Neumann conditions, see Fig.~\ref{fig2}.

Let us consider a partial solution \cite{MorseFeshbach1953McGrawHill} of a homogeneous Helmholtz equation (\ref{eq11}) which asymptotically $k\rho \gg 1$ turns in a plane wave, see \ref{appendixB},
\begin{equation}\label{eq18}
\begin{aligned}
    u(\rho, \varphi) &= \frac{1}{2} \Sum{n=0}{\infty}
        \eps_n (-i)^\frac{n}{2} J_\frac{n}{2} (k\rho) \cos\qty(\frac{n}{2}\varphi) =\\
    &=\frac{1}{\sqrt{i\pi}} e^{-ik\rho\cos\varphi}\,
        \Phi\qty[\sqrt{2k\rho} \cos\qty(\frac{\varphi}{2})] \underset{k\rho \gg 1}{\simeq}
    e^{-ik\rho\cos\varphi} \Big[1 + O\Big(\frac{1}{(k\rho)}\Big)\Big].
\end{aligned}
\end{equation}
Here $\eps_n$ is a Neumann factor, $\eps_0 = 1$, $\eps_n = 2$ at $n>0$, $J_\nu(x)$ is Bessel function, and the function
\begin{equation}\label{eq19}
    \Phi(z) = \Int{-\infty}{z} e^{it^2} \dd{t} =
        \frac{\sqrt{i\pi}}{3} \Big[1 + \erf\big(e^{-i\frac{\pi}{4}}z\big)\Big]
\end{equation}
is associated with Fresnel integrals and may be expressed through the probability integral \cite{BatemanErdelyi1_2_1953McGrawHill} with complex argument.

Both integer and half-integer values of an orbital angular momentum contribute to the expansion (\ref{eq18}), that's why the period for $\varphi$ of the function $u(\rho, \varphi)$ equals $4\pi$, not $2\pi$. For a screen oriented along $y < 0$, the domain $-\pi/2 < \varphi < 3\pi/2$ is a \enquote{real} space, and a domain $3\pi/2 < \varphi < 7\pi/2$ is a \enquote{fictitious} one. It is necessary to place auxiliary sources\footnote{Compare with (\ref{eq13}) and formulas (\ref{eq16}), (\ref{eq17}) in the preceding section \ref{part21}.} and to meet Neumann boundary conditions,
\begin{equation}\label{eq20}
    \pdv{\varphi}\Psi(\rho, \varphi)\Big|_{\varphi = -\frac{\pi}{2}} =
    \pdv{\varphi}\Psi(\rho, \varphi)\Big|_{\varphi = \frac{3\pi}{2}} = 0.
\end{equation}

To obtain the complete set of functions in first two equations (\ref{eq13}) we set $\alpha = -\pi/2$ and $\beta = 3\pi/2$, that leads to half-integer orbital angular momenta. Noting that
\[
    \pdv{\varphi}\cos(\frac{n}{2}\varphi)\Big|_{-\frac{\pi}{2}} =
        \frac{n}{2}\sin(\frac{n\pi}{4}) =
        -\pdv{\varphi} \cos\Big[\frac{n}{2}(3\pi - \varphi)\Big]\Big|_{-\frac{\pi}{2}}
\]
and
\[
    \pdv{\varphi}\cos\Big[\frac{n}{2}(3\pi - \varphi)\Big]\Big|_{\frac{3\pi}{2}} =
        \frac{n}{2}\sin(\frac{3n\pi}{4}) =
        -\pdv{\varphi} \cos(\frac{n}{2}\varphi)\Big|_{\frac{3\pi}{2}},
\]
we conclude that the Helmholtz equation solution meeting the conditions (\ref{eq20}) is written as \cite{MorseFeshbach1953McGrawHill}
\begin{equation}\label{eq21}
\begin{aligned}
    \Psi(\rho, \varphi) &= u(\rho, \varphi) + u(\rho, 3\pi - \varphi) = \\
    &= \frac{1}{\sqrt{i\pi}} \qty{
        e^{-ik\rho\cos\varphi}\,\Phi\qty[\sqrt{2k\rho} \cos\qty(\frac{\varphi}{2})] +
        e^{ik\rho\cos\varphi}\,\Phi\qty[-\sqrt{2k\rho} \sin\qty(\frac{\varphi}{2})]}.
\end{aligned}
\end{equation}

Accounting for the exact expression for $\Psi(\rho, \varphi)$, one can define the shadow area and reflected wave area for any $\rho$. If $\rho \to \infty$, in different domains of \enquote{real space} for asymptotic behaviour $\Psi(\rho, \varphi)$ we have \cite{MorseFeshbach1953McGrawHill}
\begin{equation}\label{eq22}
    \Psi(\rho, \varphi) \underset{k\rho \gg 1}{\simeq}%
    \begin{cases}
        e^{-ik\rho\cos\varphi} + e^{ik\rho\cos\varphi} + f(\rho, \varphi),
            &-\frac{1}{2}\pi < \varphi < 0,\\
        e^{-ik\rho\cos\varphi} + f(\rho, \varphi), &0 < \varphi < \pi,\\
        f(\rho, \varphi), &\pi < \varphi < \frac{3}{2}\pi,
    \end{cases}
\end{equation}
where
\begin{equation}\tag{\ref*{eq22}${}^\prime$} \label{eq22prime}
    f(\rho, \varphi) = \sqrt{\frac{i}{8\pi k\rho}} e^{ik\rho}
        \bigg[\frac{1}{\sin(\frac{\varphi}{2})} - \frac{1}{\cos(\frac{\varphi}{2})}\bigg],
\end{equation}
compare with asymptotic (\ref{eqB8}) in \ref{appendixB}.

Hence, the function $\Psi(\rho, \varphi)$ is the solution of a given physical problem: a plane wave, $\exp(-ikx)$, comes from the right and meets the screen. In the domain $-\pi/2 < \varphi < 0$ the incident wave reflects from the screen, $\exp(ikx)$, in the domain $0 < \varphi < \pi$ it freely propagates to the left, and in the \enquote{shadow} area, $\pi < \varphi < 3\pi/2$, there is no plane wave, see Fig.~\ref{fig2}.

Nevertheless, in each of these domains there is wave scattered on the screen. Its intensity is asymptotically proportional to the expression
\begin{equation}\label{eq23}
    s(\varphi) = \frac{1}{8\pi k\rho} \bigg[\frac{1}{\sin(\frac{\varphi}{2})} -
        \frac{1}{\cos(\frac{\varphi}{2})}\bigg]^2,\quad
    \varphi \ne 0,\, \pi,
\end{equation}
which is applicable only at large distances, besides the values $\varphi = 0$ and $\pi$ are physically unrealizable because of finite value of the aperture of device. Equation (\ref{eq23}) shows that the edge of the screen seems luminous irrespective of the angle $\varphi$ it is seen at, except $\varphi = \pi/2$.

As it can be seen from (\ref{eq22}), at large distances, $k\rho \gg 1$, and angles $\varphi$ from $\pi - \eps$ to $\pi + \eps$, $\eps \ll 1$, the module $|\Psi(\rho, \varphi)|$ decreases from the value of the order of unity to the value that approximately equals to zero. That's why the line $\varphi = \pi$ is the boundary of the shadow are: below it the intensity is small and above it's large. Near this line one can observe Fresnel diffraction from the screen edge. Mathematically it is associated with so-called Stokes phenomenon \cite{MorseFeshbach1953McGrawHill}.

Both integer and half-integer values of an orbital angular momentum contribute to the solution $\Psi(\rho, \varphi)$ of a problem of scattering of a plane wave on a semi-infinite screen. But the question about superselection \cite{StreaterWightman1964Benjamin} in a superposition of integer and half-integer orbital angular momenta does not rise here because the auxiliary sources for satisfying the Neumann boundary conditions are situated in a \enquote{fictitious} space, $3\pi/2 < \varphi < 7\pi/2$.

Moreover, in problems considered above, see \ref{part21} and \ref{part22}, rotational symmetry is broken. In the next section we discuss quantum mechanical problems with axial symmetry.
%%%%%%%%%%%%%%%%%%%%%%%%%%%%%%%%%%%%%%%%%%%%%%%%%%%%%%%%%%%%%%%%%%%%%%%%%%%%%%%%
%%%%%%%%%%%%%%%%%%%%%%%%%%%%%%%%%%%%%%%%%%%%%%%%%%%%%%%%%%%%%%%%%%%%%%%%%%%%%%%%
\section{Two-dimensional Schroedinger equation with axial symmetry}\label{part3}
%%%%%%%%%%%%%%%%%%%%%%%%%%%%%%%%%%%%%%%%%%%%%%%%%%%%%%%%%%%%%%%%%%%%%%%%%%%%%%%%
Half-integer quantization of an orbital angular momentum can be essential for the energy of few-electron quantum dots and in two-dimensional non-relativistic scattering on an axial-symmetric potential.
%%%%%%%%%%%%%%%%%%%%%%%%%%%%%%%%%%%%%%%%%%%%%%%%%%%%%%%%%%%%%%%%%%%%%%%%%%%%%%%%
\subsection{Circular quantum dots}\label{part31}
%%%%%%%%%%%%%%%%%%%%%%%%%%%%%%%%%%%%%%%%%%%%%%%%%%%%%%%%%%%%%%%%%%%%%%%%%%%%%%%%
In experiments \cite{SchmidtTewordtEtAl1995PhysRevB} the ground states of $N$-electron circular quantum dots, $1 \le N < 30$, in a perpendicular magnetic field, $0 \le B < 16\text{\;T}$ were probed by single-electron tunneling spectroscopy. In such
experiments \cite{Hawrylak1993PhysRevLett} one can reach an accuracy of $\approx 0.015\text{\;meV} = 2.4\cdot10^{-24}\text{\;J}$, that was promoted by maintaining the temperature of the system about $23\text{\;mK} = 1.92\text{\;\text mu eV} = 3.1\cdot10^{-25}\text{\;J}$. That's why in \cite{SchmidtTewordtEtAl1995PhysRevB} precise experimental data for the ground state of two- and three-electron circular quantum dots in perpendicular magnetic field $0 \le B \le 8\text{\;T}$ were obtained.
%%%%%%%%%%%%%%%%%%%%%%%%%%%%%%%%%%%%%%%%%%%%%%%%%%%%%%%%%%%%%%%%%%%%%%%%%%%%%%%%
\subsubsection{$N$-electron circular quantum dots in a perpendicular magnetic field}\label{part311}
%%%%%%%%%%%%%%%%%%%%%%%%%%%%%%%%%%%%%%%%%%%%%%%%%%%%%%%%%%%%%%%%%%%%%%%%%%%%%%%%
In accordance with \cite{Hawrylak1993PhysRevLett, BruceMaksym2000PhysRevB, ReimannManninen2002RevModPhys} the oscillatory model with the parabolic confinement is a good approximation for
low-lying levels in a real circular quantum dot. The Schrodinger equation for reduced energy $\eps(N; B)$ in the symmetric gauge of the vector potential, $\vb{A} = [\vb{B} \cross \vb{r}]/2$, for $N$ particles in the confining potential
\begin{equation}\label{eq24}
    V_\text{cf} = \Sum{a=1}{N} V_a(N) =
        \frac{1}{2}\Sum{a=1}{N} m_* \Omega^2(N) \vb{r}_a^2 + V^\Par{0}(N),
\end{equation}
where $m_*$ is an effective mass of electron, $\Omega(N)$ is the confining frequency, $\vb{r}_a$ is the radius-vector of $a$-th particle in a plane $(x,y)$, and $V^\Par{0}(N)$ is the reference energy level, according to \cite{MurNarozhnyEtAl2008JETPLetters, KuleshovMurEtAl2016FewBodySystems} is written as:
\begin{equation}\label{eq25}
    \qty{-\frac{1}{4Q_L^2(N)} \Sum{a=1}{N}\pdv[2]{\bm{\rho}_a} + \Sum{a=1}{N}\bm{\rho}_a^2 +
        \frac{1}{2}\Sum{a\ne b}{N}\frac{1}{|\bm{\rho}_a - \bm{\rho}_b|}}\Psi_N =
    \eps(N; B)\Psi_N.
\end{equation}
Here the dimensionless parameter $Q_L(N)$ equals
\[
    Q_L(N) = \qty(\frac{m_*a_0^2E_0}{2\hbar^2})^{1/2} =
        \qty(\frac{\mu}{\epsilon^2})^{1/3} \qty(\frac{\text{Ry}}{\hbar\Omega_L(N)})^{1/3},
\]
$\bm{\rho}_a = \vb{r}_a / a_0$ is the reduced radius-vector of $a$-th particle,
\begin{equation}\label{eq26}
    a_0(N) = 2\frac{\epsilon}{\mu} Q_L^2(N) a_B,\quad E_0(N) = \hbar Q_L(N) \Omega_L(N)
\end{equation}
are characteristic size and energy of the dot, $\mu = m_*/m_e$ is the reduced mass, $\epsilon$ is dielectric constant, $a_B = \hbar^2 / m_e e^2 $ is Bohr radius ($m_e$ and $-e$ are electron mass and charge), $\text{Ry} = m_e e^4/2\hbar^2 = 13.6\text{\;eV} = 2.18\cdot10^{-18}\text{\;J}$ is Rydberg constant, and a frequency $\Omega_L(N)$ is a function of the confining frequency:
\[
    \Omega_L(N) = \sqrt{\Omega^2(N) + \omega_L^2},\quad
        \omega_L = \frac{eB}{2m_*c} = \frac{1}{\mu}\frac{B}{B_\text{at}}\omega_\text{at},
\]
where $\omega_L$ is the Larmor frequency,
\[
    B_\text{at} = \frac{cm_e^2e^3}{\hbar^3} = 2.35\cdot10^5\text{\;T},\quad
    \omega_\text{at} = \frac{\text{Ry}}{\hbar} = 2.07\cdot10^{-16}\text{\;s}^{-1}
\]
are atomic magnetic field and frequency, and $\hbar\omega_L = 0.87\cdot B(\text{T})\text{\;meV} = 1.4\cdot10^{-22}\text{\;J}$.

Taking spin into account, we can write for the total energy of the $N$-electron circular quantum dot in perpendicular magnetic field $B$:
\begin{equation}\label{eq27}
    E_{M,\Sigma}(N; B) = \eps_M(N; B) E_0(N) - (M + \mu g\Sigma) \hbar\omega_L +
        V_{M,\Sigma}^\Par{0}(N),\quad N \ge 2,
\end{equation}
where $\hbar M$ and $\hbar\Sigma$ are the conserved orbital and spin momenta projections on the direction of magnetic field, $\eps_M(N; B)$ is the solution of (\ref{eq25}), corresponding to orbital angular momentum projection $\hbar M$, $E_0(N)$ is defined in (\ref{eq26}), and $g$ is Lande factor. For quantum dots in double-barrier structure based on GaAs
\[
    \mu = 0.067,\quad \epsilon = 12.5,\quad g = 0.44.
\]
The energy of such dots was estimated in \cite{KuleshovMurEtAl2016FewBodySystems}.

Unlike the assumption made in \cite{SchmidtTewordtEtAl1995PhysRevB}, phenomenological parameters $\hbar\Omega(N; M,\Sigma)$ and $V_{M,\Sigma}^\Par{0}$ depend not only on the number of electrons in a dot \cite{Hawrylak1993PhysRevLett}, but also on quantum numbers of a system of electrons (in our case~--- total orbital angular momentum projection and spin \cite{MurNarozhnyEtAl2008JETPLetters}).
%%%%%%%%%%%%%%%%%%%%%%%%%%%%%%%%%%%%%%%%%%%%%%%%%%%%%%%%%%%%%%%%%%%%%%%%%%%%%%%%
\subsubsection{Topological phase and half-integer orbital angular momentum in circular quantum dots}\label{part312}
%%%%%%%%%%%%%%%%%%%%%%%%%%%%%%%%%%%%%%%%%%%%%%%%%%%%%%%%%%%%%%%%%%%%%%%%%%%%%%%%
Formally, the equation (\ref{eq25}) is equivalent to the 2D Schrödinger equation for $N$ particles with masses $2Q_L^2$. Let us first
discuss the case of \enquote{heavy particles}, i.e. $Q_L^2 \gg 1$ when we can use quasiclassical $1/Q$-expansion \cite{LozovikMurNarozhny2003}. In the limit $Q_L \to \infty$ ground state is realized by a rigid configuration of electrons which minimizes the potential energy. This configuration is invariant under $2\pi$ rotation around the symmetry axis, that can be used for understanding the quantization of
the angular momentum.

Indeed, the overall phase acquired by the ground-state wave function after the rotation is determined by the total angular momentum $J$ that includes spin. The rotation of a two-dimensional system about $2\pi$ is the identity element of the symmetric group $S_N$ and belongs to the alternating group $A_N$ of even permutations of the set $\{1, 2, \ldots, N\}$. Hence, the rotation of the system about $2\pi$ is equivalent to an even number of pairwise transpositions, so, according to the Pauli principle, the wave function of fermionic system does not change:
\begin{equation}\label{eq28}
    e^{i2\pi J} = 1,\quad J = M + \Sigma = 0, \pm 1, \pm 2,\ldots
\end{equation}
(the total angular momentum equals to the sum of the orbital and spin angular momenta).

For even number of electrons the spin quantum number is integer, so the orbital angular momentum is also integer. For odd number of electrons it is half-integer, thus $M$ is also half-integer. According to (\ref{eq4}) and (\ref{eq28}) it means that $\delta = 1/2$, or that the system is characterized by the topological phase $\theta = \pi$. Therefore, for even number of electrons, $N = 2n$, $n = 1, 2, 3, \ldots$, we have
\begin{equation}\label{eq29}
    \theta = 0,\quad \delta = 0,\quad M = m = 0, \pm 1, \pm 2, \ldots,
\end{equation}
and for odd number of electrons, $N = 2n+1$, $n = 1, 2, 3,\ldots$,
\begin{equation}\label{eq30}
    \theta = \pi,\quad \delta = 1/2,\quad M = m + 1/2, \quad m = 0, \pm 1, \pm 2, \ldots
\end{equation}

Now let us consider the case of three electrons. If total spin $\Sigma = \pm 1/2$, then $M$ an take any half-integer value
as in (\ref{eq30}). But if $\Sigma = \pm 3/2$, an additional symmetry arises in the problem. In this case the symmetry group is $C_{3v}$, see, e.g., \cite{LandauLifshitz3_1981Butterworth}, which is isomorphic to the symmetry group $S_3$. The group $C_{3v}$ consists of rotations around the symmetry axis by $2\pi/3$-fold angles (the $C_3$ group) and reflections at three bisectors of equilateral triangle, which is the equilibrium configuration for three-electron quantum dot. The $C_3$ group is isomorphic to $A_3$, so wave functions of the system at $Q_L \to \infty$ do not change not only under $2\pi$ rotations, but also under $2\pi/3$ and $4\pi/3$ rotations. Thus
\begin{equation}\label{eq31}
    e^{i\frac{2\pi}{3}J} = 1,\quad J = M + \Sigma = 0, \pm 3, \pm 6, \ldots
\end{equation}
This means that the orbital angular momentum projections can take the values
\begin{equation}\label{eq32}
    M = \pm\qty(\frac{3}{2} + 3k),\quad k = 0, 1, 2, \ldots
\end{equation}

To obtain (\ref{eq28}) and (\ref{eq31}) we considered classical limit $Q_L \to \infty$. However, if one varies the
parameter $Q_L$ adiabatically, then the discrete quantum numbers $M$ and $\Sigma$ do not change. Therefore, this result
is valid also at $Q_L \sim 1$ which is typical for real quantum dots.

Thus, for a three-electron quantum dot the Pauli principle leads to half-integer quantization of the orbital angular momentum, i.e. to the value $\delta = 1/2$ corresponding to the topological phase $\theta = \pi$. That's why it is valid to call topological phase $\theta$ the \enquote{Pauli topological phase}. Pauli was first to pay attention to the possibility of fractional quantization of an orbital angular momentum in 2D case. This conclusion is substantiated by experimental data \cite{SchmidtTewordtEtAl1995PhysRevB}.
%%%%%%%%%%%%%%%%%%%%%%%%%%%%%%%%%%%%%%%%%%%%%%%%%%%%%%%%%%%%%%%%%%%%%%%%%%%%%%%%
\subsubsection{Comparison of theoretical calculations with the experimental data}\label{part313}
%%%%%%%%%%%%%%%%%%%%%%%%%%%%%%%%%%%%%%%%%%%%%%%%%%%%%%%%%%%%%%%%%%%%%%%%%%%%%%%%
The variables in the Schroedinger equation (\ref{eq25}) with $N = 2$ can be separated in cylindrical coordinates and then it can be easily solved numerically \cite{LozovikMurNarozhny2003}. Since this approach is not applicable for quantum dots containing more electrons, here we discuss an alternative solution of this problem based on construction of symmetrical coordinates, see \cite{Lubarskii1960Pergamon}. In \cite{KuleshovMurEtAl2016FewBodySystems} it was shown that results of diagonalization are in good agreement with the results of exact (numeric) solution which is possible when variables are separated.

\begin{figure}[t]
    \centering
    \includegraphics[width=0.6\textwidth]{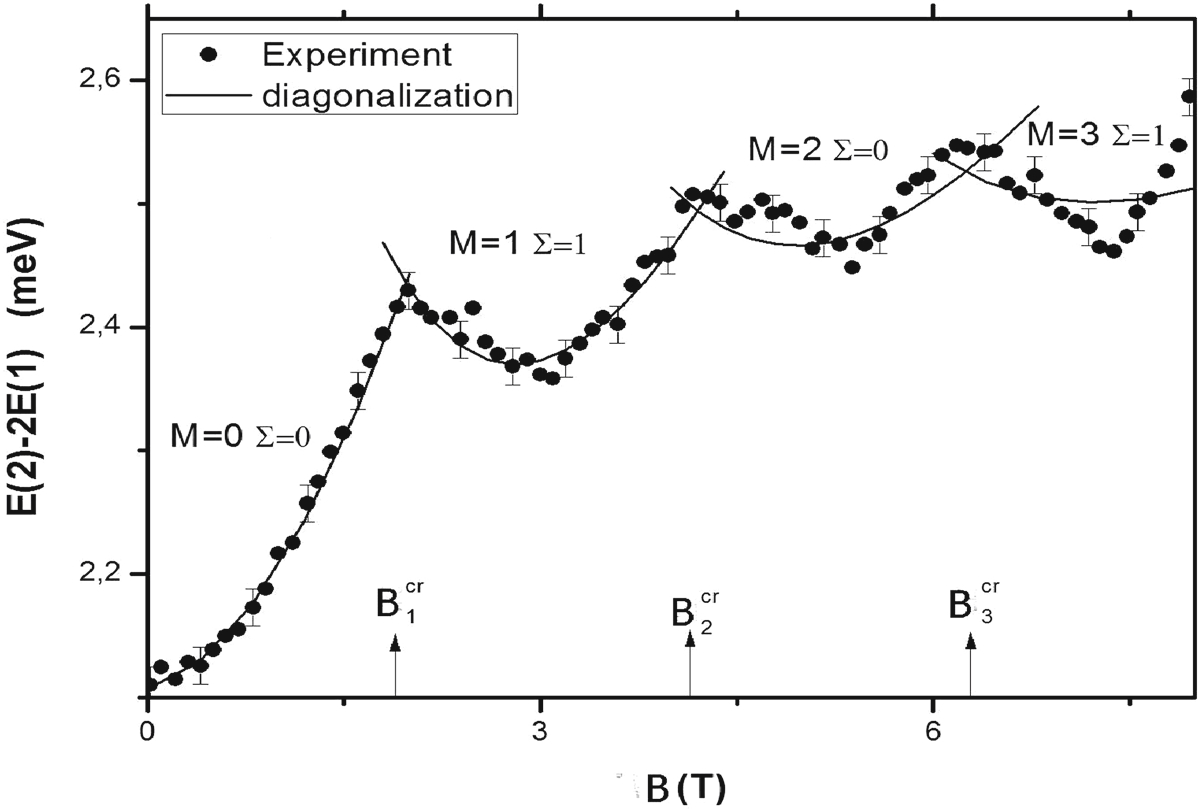}
    \caption{Comparison of the experimental data \cite{SchmidtTewordtEtAl1995PhysRevB} with theoretical calculations. The \emph{solid line} is a result of Hamiltonian diagonalization method which cannot be distinguished from the exact numerical calculation. The experimental points are taken from \cite{SchmidtTewordtEtAl1995PhysRevB}. Above the \emph{curves} there are quantum numbers of the system: total orbital $M$ and spin $\Sigma$ momenta projections on the magnetic field. \emph{Arrows} show the locations of crossings. The experimental accuracy is pointed selectively.}
    \label{fig3}
\end{figure}

On Fig.~\ref{fig3} from \cite{KuleshovMurEtAl2016FewBodySystems} experimental data from \cite{SchmidtTewordtEtAl1995PhysRevB} are compared with theoretical calculations. The confining frequency of one-electron dot $\hbar\Omega(1) = 3.6\text{\;meV} = 5.8\cdot 10^{-22}\text{\;J}$ is taken from \cite{SchmidtTewordtEtAl1995PhysRevB}. The solid line is a result of Hamiltonian diagonalization method which cannot be distinguished from the exact numerical calculation. The experimental points are taken from \cite{SchmidtTewordtEtAl1995PhysRevB}. Above the curves there are quantum numbers of the system: projections of total orbital $M$ and spin $\Sigma$ momenta on the magnetic field. Arrows show the locations of crossings (the values of magnetic field where the ground state symmetry changes).

Values of $M$ and $\Sigma$ are in line with (\ref{eq28}) and (\ref{eq29}) and correspond to total angular momenta $J = 0,\, 2,\, 2,\, 4$.
One can see that when magnetic field is close to zero, the system is in the state with the lowest quantum numbers, $M = 0$ and $\Sigma = 0$. Moreover, the contributions of centrifugal energy and paramagnetic shift are equal to zero and the ground state energy
increases with magnetic field owing to diamagnetic contribution at $0 \le B < 2\text{\;T}$. Next, with the growth of the magnetic field, when the paramagnetic contribution becomes essential, the system goes into state with $M = 1$. According to the Pauli principle, the total spin projection also changes: the value $\Sigma = 0$ becomes $\Sigma = 1$, i.e. at $B = B_1^\text{cr}$ the singlet-triplet crossing occurs. With further increase of magnetic field up to $B = 3\text{\;T}$ the ground state energy decreases due to paramagnetic shift and then increases because of diamagnetic contribution to the energy. At $B \approx 4\text{\;T}$ it becomes advantage for the system to make a transition to the state with $M = 2$, i.e. a transition from $\Sigma = 1$ to $\Sigma = 0$. Hence, the triplet-singlet crossing occurs at $B = B_2^\text{cr}$, and so forth as the magnetic field grows.

\begin{figure}[t]
    \centering
    \includegraphics[width=0.6\textwidth]{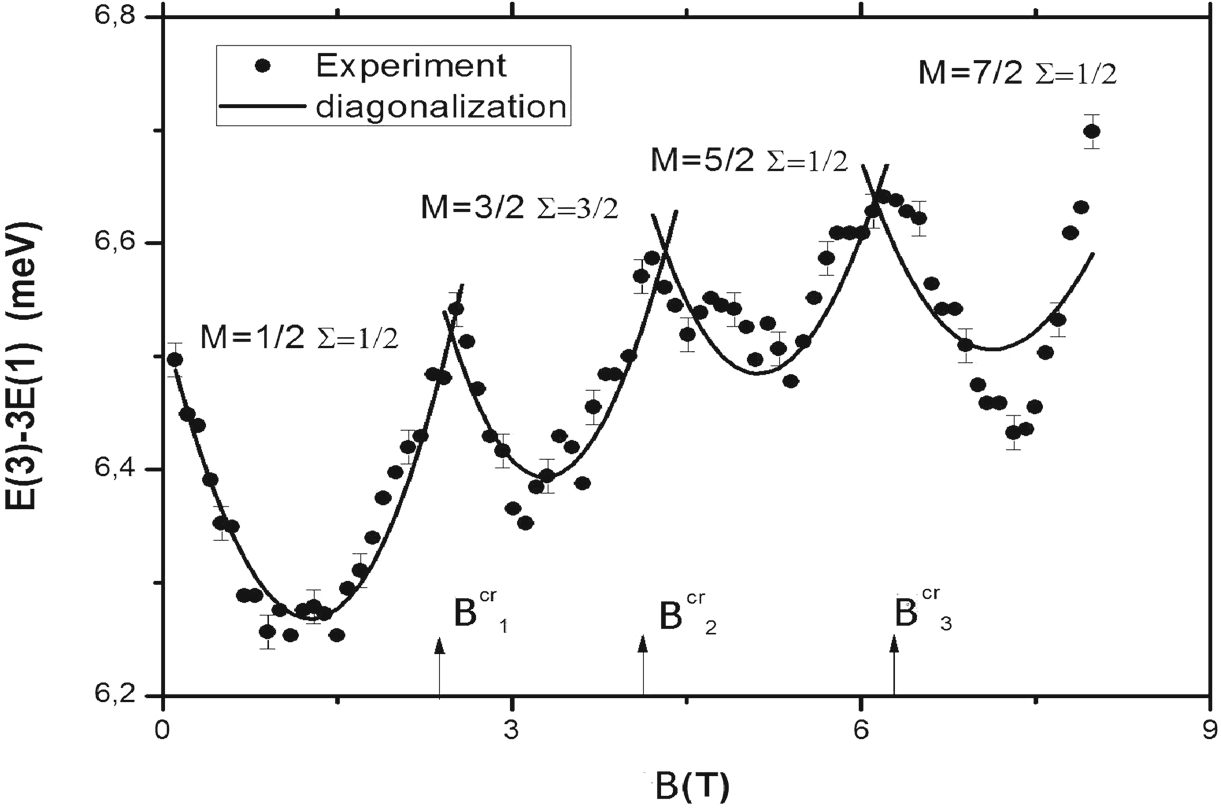}
    \caption{Comparison of calculations of the ground-state effective energy $\Delta E(3; B) = E(3; B) - 3E(1; B)$ of three-electron quantum dot with the experimental data of \cite{SchmidtTewordtEtAl1995PhysRevB}. The legend is the same as in Fig.~\ref{fig3}.}
    \label{fig4}
\end{figure}

On Fig.~\ref{fig4}, which is similar to Fig.~6 in \cite{KuleshovMurEtAl2016FewBodySystems}, the calculation of effective energy $\Delta E(3; B) = E(3; B) - 3E(1; B)$ of the ground state of three-electron circular quantum dots is compared with experimental data \cite{SchmidtTewordtEtAl1995PhysRevB}. The legend is similar to the one on Fig.~\ref{fig3}. The values of $M$ and $\Sigma$ are in agreement with (\ref{eq28}), (\ref{eq30})--(\ref{eq32}) and correspond to total angular orbital angular momenta $J = 1,\, 3,\, 3,\, 4$. One can see from Fig.~\ref{fig4} that three-electron system, as well as two-electron one, makes transitions between states
with different quantum numbers with the growth of magnetic field. The dynamics of these transitions is similar to the two-electron problem discussed above. The difference is that in a weak field, $\omega_L \ll \Omega(3)$, the projections of orbital and spin momenta are not zero, $M = 1/2$ and $\Sigma = 1/2$. Thus, the paramagnetic shift is significant here and it reduces the ground state energy with increase of magnetic field.

Thus, in accordance with the Pauli principle and T-invariance, non-trivial topological phase associated with multiple-valued irreducible representations of the two-dimensional rotation group SO(2) leads to half-integer quantization of orbital angular momentum of the ground state of circular quantum dots with an odd number of electrons.

The value $M = 1/2$ for orbital angular momentum of three-electron circular quantum dot in a weak magnetic field can be detected directly \cite{KuleshovMurEtAl2016FewBodySystems} if one measures the derivative of ground-state energy at $B \to 0$ with more accuracy than it was done in \cite{SchmidtTewordtEtAl1995PhysRevB}. It is determined by the paramagnetic contribution to the energy and does not depend on the shape and parameters of the confining potential.
%%%%%%%%%%%%%%%%%%%%%%%%%%%%%%%%%%%%%%%%%%%%%%%%%%%%%%%%%%%%%%%%%%%%%%%%%%%%%%%%
\subsection{Short-range potential scattering}\label{part32}
%%%%%%%%%%%%%%%%%%%%%%%%%%%%%%%%%%%%%%%%%%%%%%%%%%%%%%%%%%%%%%%%%%%%%%%%%%%%%%%%
At large distances the wave function of 2D finite-range potential scattering problem is given by \cite{LandauLifshitz3_1981Butterworth}
\begin{equation}\label{eq33}
    \Psi_k(\bm{\rho}) = e^{ik\rho\cos\varphi} +
        \frac{f_k(\varphi)}{\sqrt{-i\rho}} e^{ik\rho},\quad k\rho \gg 1,
\end{equation}
where $f_k(\varphi)$ is a dimensionless scattering amplitude. The solution of 2D Schroedinger equation for axially symmetric potential scattering\footnote{here $a_0$ is an effective range.} $U(\rho)$, $\rho = r/a_0$, may be written as a superposition of wave functions of particles with energy $E = E_0 k^2$, $E_0 = \hbar^2 / 2ma_0^2$, and different orbital angular momenta $M$,
\begin{equation}\label{eq34}
    \Psi_k(\bm{\rho}) = \frac{1}{\sqrt{\rho}} \Sum{M}{} A_M F_{kM}(\rho) \Psi_M(\varphi).
\end{equation}
Angle functions $\Psi_M(\varphi)$ are defined in (\ref{eq4})--(\ref{eq6}), where the sum is over integer or half-integer $M$, and radial functions satisfy the following equation
\begin{equation}\label{eq35}
    \dv[2]{\rho} F_{kM}(\rho) + \Big[k^2 - \frac{4M^2 - 1}{4\rho^2} - V(\rho)\Big]F_{kM}(\rho)=  0,\quad
    V(\rho) = \frac{U(\rho)}{E_0}.
\end{equation}

The solution of this equation at large distances
\begin{equation}\label{eq36}
    F_{kM}(\rho) \approx \sqrt{\frac{2}{\pi}} \cos\Big(k\rho - \frac{\pi}{2}M
        - \frac{\pi}{4} + \delta_M\Big),\quad k\rho \gg 1,
\end{equation}
differs from the asymptotic of free equation solution only by a scattering phase $\delta_M(k)$
\begin{equation}\label{eq37}
    F_{kM}^\Par{0} (\rho) = \sqrt{k\rho}\, J_M(k\rho) \approx
        \sqrt{\frac{2}{\pi}} \cos\Big(k\rho - \frac{\pi}{2}M - \frac{\pi}{4}\Big).
\end{equation}
Coefficients $A_M$ should be chosen so that at large distances the expansion (\ref{eq34}) coincides with the asymptotic (\ref{eq33}). Given the expansion (\ref{eqB7}) for a plane wave with integer orbital angular momenta and (\ref{eqB9})~--- the extension of a plane wave to half-integer $M$, this requirement leads to
\begin{equation}\label{eq38}
    A_M = \frac{i^M}{\sqrt{k}} e^{i\delta_M(k)},\quad M = \delta + m,\quad
        \delta = 0,\, 1/2,\quad m = 0, \pm 1, \pm 2, \ldots
\end{equation}
For integer momenta $M$ this result was obtained earlier, see, e.g., (2.21) in \cite{Novikov2007PhysRevB}. Hence, for scattering amplitude we have an expansion
\begin{equation}\label{eq39}
    f_k(\varphi) = \frac{1}{i\sqrt{2\pi k}} \Sum{m=-\infty}{\infty}
        \big(e^{2i\delta_M(k)} - 1\big) e^{iM\varphi},\quad
    M =%
    \begin{cases}
        m = 0, \pm 1, \pm 2, \ldots,\\
        m + 1/2 = \pm 1/2, \pm 3/2, \ldots
    \end{cases}
\end{equation}

Differential and total scattering cross-sections in units of $a_0$ equal
\begin{equation}\label{eq40}
    \dv{\sigma(k, \varphi)}{\varphi} = |f_k(\varphi|^2,\quad
    \sigma(k) = \Int{0}{2\pi} |f_k(\varphi)|^2 \dd{\varphi}.
\end{equation}
Given the orthonormality of $\Psi_M(\varphi)$, for total cross-section we get
\begin{equation}\label{eq41}
    \sigma(k) = \frac{4}{k} \Sum{M}{} \sigma_M(k),\quad \sigma_M(k) = \sin^2\delta_M(k),\quad
    M =%
    \begin{cases}
        m = 0, \pm 1, \pm 2, \ldots,\\
        m + 1/2 = \pm 1/2, \pm 3/2, \ldots
    \end{cases}
\end{equation}
where $\sigma_M(k)$ is a partial cross-section and summation is over integer and half-integer orbital angular momenta $M$. If the value is integer, then $M = m$, see \cite{SternHoward1967PhysRev, Barton1983AmJPhys}. In two dimensions the optical theorem is written as:
\begin{equation}\label{eq42}
    \sigma(k) = \sqrt{\frac{8\pi}{k}} \Im f_k(0).
\end{equation}

Note that the approach to 2D scattering problem with the possible half-integer quantization of orbital angular momentum is analogous to the Coulomb scattering problem with its integer quantization.
%%%%%%%%%%%%%%%%%%%%%%%%%%%%%%%%%%%%%%%%%%%%%%%%%%%%%%%%%%%%%%%%%%%%%%%%%%%%%%%%
\subsubsection{Two-dimensional Coulomb scattering problem}\label{part321}
%%%%%%%%%%%%%%%%%%%%%%%%%%%%%%%%%%%%%%%%%%%%%%%%%%%%%%%%%%%%%%%%%%%%%%%%%%%%%%%%
The exact solution of 2D Schroedinger equation with Coulomb attractive potential for scattering problem is \cite{SternHoward1967PhysRev, Barton1983AmJPhys}:
\begin{equation}\label{eq43}
    \Psi(x, y;\, E) = \frac{1}{\sqrt{\pi}} e^{\frac{\pi}{2k}}\, \Gamma\Big(\frac{1}{2} - \frac{i}{k}\Big)
        e^{ikx}\, \Phi\Big[\frac{i}{k},\, \frac{1}{2};\, 2ik\rho\sin^2\qty(\frac{\varphi}{2})\Big].
\end{equation}
Here the Coulomb units are used, $k = \sqrt{2E}$, $\Phi(a,\, c;\, z)$ us a degenerate hypergeometric Humbert function \cite{BatemanErdelyi1_2_1953McGrawHill}, and a normalization constant is chosen so that the amplitude of a plane wave in
\begin{equation}\label{eq44}
\begin{aligned}
    \Psi(x, y;\, E) &= \bigg[1 - \frac{k + 2i}{4k^3\rho\sin^2\qty(\frac{\varphi}{2})}\bigg]
        e^{ikx - \frac{i}{k}\ln\qty[2k\rho\sin^2\qty(\frac{\varphi}{2})]} + \\ &+
    \bigg[1 + \frac{ik^2 + 3k -2i}{4k^3\rho\sin^2\qty(\frac{\varphi}{2})}\bigg]
        \frac{f_C(\varphi)}{\sqrt{-i\rho}} e^{ik\rho + \frac{i}{k}\ln(2k\rho)},\quad
    k\rho \gg 1
\end{aligned}
\end{equation}
is equal to unity. For the Coulomb scattering amplitude we have
\begin{equation}\label{eq45}
    f_C(\varphi) = -i\frac{\Gamma\big(\frac{1}{2} -
        \frac{i}{a_Z k}\big)}{\Gamma\big(\frac{i}{a_Z k}\big)}
        \frac{e^{\frac{i}{a_Z k} \ln\qty[\sin^2\qty(\frac{\varphi}{2})]}}
            {\sqrt{2k\sin^2\qty(\frac{\varphi}{2})}},\quad
    a_Z = \frac{\hbar^2}{Zme^2},\quad k = \sqrt{\frac{2mE}{\hbar^2}}.
\end{equation}

If we compare the definition of a Coulomb amplitude in (\ref{eq44}) and definitions of short-range potentials scattering amplitude in (\ref{eq33}) and the asymptotic (\ref{eqB8}) of the generalization of a plane wave for half-integer momenta, we can see that in both cases only the principal term in the asymptotic expansion of exact solutions of equations is used.

Nevertheless, scattering amplitude and differential cross-section can be obtained according to (\ref{eq39}), at least for integer quantization of orbital angular momentum \cite{Novikov2007PhysRevB}. Particularly, the amplitude (\ref{eq45}) corresponds to Coulomb partial scattering matrix
\begin{equation}\label{eq46}
    S_m = e^{2i\delta_m} = \frac{\Gamma\Big(|m| + \frac{1}{2} - \frac{i}{a_Z k}\Big)}
        {\Gamma\Big(|m| + \frac{1}{2} + \frac{i}{a_Z k}\Big)},\quad
    m = 0, \pm 1, \pm 2, \ldots,
\end{equation}
and slow particles' scattering phases are
\begin{equation}\label{eq47}
    \delta_m = -\frac{1}{a_Z k} \ln\Big(\frac{1}{e a_Z k}\Big) - \frac{\pi}{2} |m| +
        \Big(\frac{m^2}{2} - \frac{1}{4}\Big) a_Z k,\quad a_Z k \ll 1.
\end{equation}
The anomalous growth of scattering phases at $k \to 0$ owes to slow decreasing of the Coulomb potential at large distances. That's why special techniques \cite{LandauLifshitz3_1981Butterworth} are required when calculating zero-angle scattering amplitudes.
%%%%%%%%%%%%%%%%%%%%%%%%%%%%%%%%%%%%%%%%%%%%%%%%%%%%%%%%%%%%%%%%%%%%%%%%%%%%%%%%
\subsubsection{The scattering with generalization of a plane wave to integer and half-integer orbital angular momenta}\label{part322}
%%%%%%%%%%%%%%%%%%%%%%%%%%%%%%%%%%%%%%%%%%%%%%%%%%%%%%%%%%%%%%%%%%%%%%%%%%%%%%%%
Following the analogy given above, let's take the following function as a generalization of a plane wave
\begin{equation}\label{eq48}
    v(\rho, \varphi) = \frac{1}{2} \Sum{m = -\infty}{\infty} \qty{
        i^m J_m(k\rho) e^{im\varphi} + i^{|M|} J_{|M|}(k\rho) e^{iM\varphi}},\quad
    M = m + 1/2,
\end{equation}
where summation is both over integer $M = m$ and over half-integer orbital angular momenta, $M = m + 1/2$. Both values gives asymptotically similar contribution, see \ref{appendixB}, particularly (\ref{eqB1}) and (\ref{eqB11}). Acting like in the beginning of \ref{part32}, when deriving (\ref{eq38}), instead of (\ref{eq39}) we get
\begin{equation}\label{eq49}
    \wt{f}_k (\varphi) = \frac{1}{i\sqrt{8\pi k}} \Sum{M}{} \big(e^{2i\delta_M(k)} - 1\big)
        e^{iM\varphi}.
\end{equation}
Here summation is over all integer $M = m$ and half-integer $M = m + 1/2$, $m = 0, \pm 1, \pm 2, \ldots$, and the scattering amplitude $\wt{f}_k(\varphi)$ corresponds to the asymptotic (\ref{eq33}). Since functions $\exp(iM\varphi)$ both at integer and half-integer values of $M$ are orthogonal on $0 \le \varphi \le 4\pi$, see \ref{appendixA}, we get instead of (\ref{eq41})
\begin{equation}\label{eq50}
    \wt{\sigma}(k) = \Int{0}{4\pi} \big|\wt{f}_k(\varphi)\big|^2 \dd{\varphi} =
        \frac{1}{4k} \Sum{M}{} \sin^2 \delta_M(k).
\end{equation}
Nevertheless, the optical theorem is still has the form (\ref{eq42}),
\begin{equation}\label{eq51}
    \wt{\sigma}(k) = \sqrt{\frac{8\pi}{k}} \Im \wt{f}_k(0).
\end{equation}
so the solution
\begin{equation}\label{eq52}
    \wt{\Psi}_k(\bm{\rho}) = \frac{1}{\sqrt{k\rho}} \Sum{M}{} i^M e^{i\delta_M(k)}
        F_{kM}(\rho) \Psi_M(\varphi),
\end{equation}
where summation is over both integer and half-integer orbital angular momenta, is not contrary to the unitarity.

This solution, along with one of the problem of plane wave scattering on half-infinite screen (\ref{eq21}), contains contributions from both integer and half-integer orbital angular momenta. So, it arises the question about superselection rule, i.e. the realization\footnote{In the case of circular quantum dots, see section~\ref{part31}, the choice between integer and half-integer values of orbital angular momenta is determined by the Pauli exclusion principle, which is not applicable in one-particle problem.} of single- or double-valued representations of two-dimensional rotation group $SO(2)$. Let's illustrate physical consequences of it.
%%%%%%%%%%%%%%%%%%%%%%%%%%%%%%%%%%%%%%%%%%%%%%%%%%%%%%%%%%%%%%%%%%%%%%%%%%%%%%%%
\subsubsection{Impenetrable disk scattering}\label{part323}
%%%%%%%%%%%%%%%%%%%%%%%%%%%%%%%%%%%%%%%%%%%%%%%%%%%%%%%%%%%%%%%%%%%%%%%%%%%%%%%%
The scattering phases for impenetrable disk of radius $R$ are
\begin{equation}\label{eq53}
    \exp[2i\delta_M(k)] = -\frac{H_{|M|}^\Par{2}(kR)}{H_{|M|}^\Par{1}(kR)},\quad
    M = \delta + m,\quad \delta = 0,\, 1/2,\quad m = 0, \pm 1, \pm 2, \ldots,
\end{equation}
where $H_\nu^\Par{1,2}(x)$ are Hankel functions \cite{BatemanErdelyi1_2_1953McGrawHill}.

In the case of slow particles scattering we get
\begin{equation}\label{eq54}
    \delta_M(k) = -\frac{(kR)^{2|M|}}{(2|M| - 2)!! (2|M|)!!},\quad M \ne 0,\quad kR \ll 1.
\end{equation}
Here $(2n-1)!! = 2^n \Gamma(n+1/2) / \sqrt{\pi}$, where $(-1)!! = 1$. Thus, for half-integer orbital angular momenta
\begin{equation}\label{eq55}
    \delta_M(k) = -\frac{\pi}{\Gamma(m + 1/2)\Gamma(m + 3/2)} \qty(\frac{kR}{2})^{2m+1},\quad
    M = m + 1/2,\quad m \ge 0,\quad kR \ll 1.
\end{equation}
At the same time for integer $M \ne 0$ we have
\begin{equation}\label{eq56}
    \delta_M(k) = -\frac{\pi m}{(m!)^2} \qty(\frac{kR}{2})^{2m},\quad
    M = m,\quad m \ge 1,\quad kR \ll 1,
\end{equation}
and for $m = 0$ it occurs a large logarithm,
\begin{equation}\label{eq57}
    \delta_0(k) = -\frac{\pi}{2}\, \frac{1}{\ln(1/kR)},\quad kR \ll 1.
\end{equation}
Let's note that results (\ref{eq56}) and (\ref{eq57}) coincide with the expressions for phases of long-wave scattering by a cylinder of radius $R$ with Dirichlet boundary conditions, see, e.g., \cite{MorseFeshbach1953McGrawHill}. They correspond to scattering phases of plane electromagnetic wave with E-vector in the direction of axis of cylinder. For half-integer values of $M$ we obtain for differential and total cross-section:
\begin{equation}\label{eq58}
    \dv{\sigma(k, \varphi)}{\varphi} = \frac{8}{\pi} kR^2 \cos^2\qty(\frac{\varphi}{2}),\quad
    \sigma(k) = 8kR^2,\quad kR \ll 1.
\end{equation}
In the case of integer values of $M$ the situation is different,
\begin{equation}\label{eq59}
    \dv{\sigma(k, \varphi)}{\varphi} = \frac{\pi R}{2(kR) \ln^2(1/kR)},\quad
    \sigma(k) = \frac{\pi^2 R}{(kR) \ln^2(1/kR)},\quad kR \ll 1,
\end{equation}
i.e. the long-wave scattering is isotropic and like in the problem of scattering by a cylinder of radius $R$, it grows with increase of the wavelength \cite{MorseFeshbach1953McGrawHill}.

In the case of short waves, according to the Hankel functions asymptotic at large arguments \cite{BatemanErdelyi1_2_1953McGrawHill} we have
\begin{equation}\label{eq60}
    \delta_M(k) \simeq%
    \begin{cases}
        kR - \frac{\pi}{2}|M| - \frac{\pi}{4}, &|M| \lesssim kR,\quad kR \gg 1,\\
        0, &|M| \gtrsim kR \gg 1.
    \end{cases}
\end{equation}
For the total scattering cross-section we obtain
\begin{equation}\label{eq61}
    \sigma(k) \simeq \frac{4}{k} \Sum{M = -kR}{kR} \sin^2\Big(kR - \frac{\pi}{2}|M| -
        \frac{\pi}{4}\Big) \simeq 4R,\quad kR \gg 1,
\end{equation}
and this result is valid both for half-integer and integer orbital angular momenta, see \cite{MorseFeshbach1953McGrawHill}.

Thus, the short-wave cross-section is independent of the type of orbital angular momentum quantization, but for long-wave cross-section the type of quantization is essential. The way to determine which type of orbital angular momenta quantization is realized in 2D scattering is based on this fact.

The near perfect two-dimensional system is graphene and in the next section we consider the spectrum and wave functions of graphene with axially symmetric impurity.
%%%%%%%%%%%%%%%%%%%%%%%%%%%%%%%%%%%%%%%%%%%%%%%%%%%%%%%%%%%%%%%%%%%%%%%%%%%%%%%%
\section{Two-dimensional Dirac equation in a strong axially symmetric electric field}\label{part4}
%%%%%%%%%%%%%%%%%%%%%%%%%%%%%%%%%%%%%%%%%%%%%%%%%%%%%%%%%%%%%%%%%%%%%%%%%%%%%%%%
The electron characteristics of gapped graphene can be described by the effective two-dimensional Dirac equation with external axially symmetric potential which vanishes at infinity,
\begin{equation}\label{eq62}
    \Big[-i\bm{\sigma}\pdv{\bm{\rho}} + V(\rho) + \sigma_3\Big] \Psi_\eps(\bm{\rho}) =
        \eps \Psi_\eps(\bm{\rho}).
\end{equation}
Here $\bm{\sigma} = (\sigma_1, \sigma_2)$ and $\sigma_3$ are the Pauli matrices, $\bm{\rho} = (\rho\cos\varphi, \rho\sin\varphi)$, $\Psi_\eps(\bm{\rho})$ is the two-component wave function for energy $\eps$. We use \enquote{relativistic} system of units $\hbar = v_F = m_* = 1$, where $v_F$ is the velocity at the Fermi surface and $m_*$ is the effective mass of electron, so the energy unit is $mv_F^2$, and the length unit $l_F = \hbar/m_*v_F$ is \enquote{Compton} length. Detail are described in \cite{ZhouGweonEtAl2007NatureMaterials, PereiraKotovCastroNeto2008PhysRevB, Novikov2007PhysRevB}, \cite{CastroNetoGuineaEtAl2009RevModPhys} and references given there.

Owing to the axial symmetry, the conserving quantum number is the total angular momentum $J = M + 1/2$, i.e., the eigenvalue of the generator for the two-dimensional rotations, $-i\pdv{\varphi} + \frac{1}{2}\sigma_3$ \cite{DiVincenzoMele1984PhysRevB}. The eigenvalues $M$ of orbital angular momentum and its eigenfunctions are defined in (\ref{eq4}--\ref{eq6}). For a wave function with total angular momentum $J$ we have
\begin{equation}\label{eq63}
    \Psi_{\eps, J} (\bm{\rho}) = \frac{1}{\sqrt{2\pi\rho}} e^{iJ\varphi}
        \mqty(e^{-i\varphi/2}F(\rho) \\ ie^{i\varphi/2}G(\rho)),\quad
    J = M + 1/2.
\end{equation}
Functions $F(\rho)$ and $G(\rho)$ satisfy the two-dimensional Dirac equation,
\begin{equation}\label{eq64}
    H_D \Psi_{\eps, J}(\rho) = \eps \Psi_{\eps, J}(\rho),\quad
    \Psi_{\eps, J}(\rho) = \mqty(F(\rho) \\ G(\rho)),\quad
    H_D = \mqty(V(\rho) + 1                & \frac{J}{\rho} + \dv{\rho} \\
                \frac{J}{\rho} - \dv{\rho} & V(\rho) - 1),
\end{equation}
which up to the notation\footnote{The equation (\ref{eq64}) coincides with the set of equations (12.7) from \cite{AkhiezerBerestetskii1965QED} with the substitution $F \to rg(r)$, $G \to -rf(r)$, $J \to -\kap$, where $\kap = \pm 1, \pm 2, \ldots$ is the Dirac quantum number expressed as \enquote{$j$} in \S71 of \cite{Dirac1958ClarendonPress}.} coincides with the set of equations for the radial functions for the three-dimensional problem when $J = -\kap = \pm 1, \pm 2, \ldots$ In the two-dimensional case, however, the values $J = 0$ and half-integer values $J = m + 1/2$ where $M = m = 0, \pm 1, \pm 2, \ldots$ are possible.

For potentials which vanish $V(\rho) \to 0$ at large distances $\rho \to \infty$, the values $\eps \ge 1$ and $\eps \le 1$ correspond to the upper and lower (the \enquote{Dirac sea}, deformed by an external field) continua of the solutions of the Dirac equation, respectively, whereas the range $-1 < \eps < 1$ corresponds to the discrete spectrum. Since the eigenvalues of the differential operator $H_D$ can be arbitrarily large, the self-adjoint operator $\wt{H}$, associated with $H_D$, is unbounded one. With the explicit form of the Hamiltonian $H_D$ it can be shown that the boundary conditions of wave functions at $\rho = 0$ are determined by the behaviour of the potential $V(\rho)$ at small distances. These boundary conditions provide a self-adjointness of $\wt{H}$, see \ref{appendixC}.
%%%%%%%%%%%%%%%%%%%%%%%%%%%%%%%%%%%%%%%%%%%%%%%%%%%%%%%%%%%%%%%%%%%%%%%%%%%%%%%%
\subsection{Scattering by a short-range potential}\label{part41}
%%%%%%%%%%%%%%%%%%%%%%%%%%%%%%%%%%%%%%%%%%%%%%%%%%%%%%%%%%%%%%%%%%%%%%%%%%%%%%%%
In a scattering problem with short range potential the asymptotic of two-component wave function of the 2D Dirac equation has the form:
\begin{equation}\label{eq65}
    \Psi_\eps(\bm{\rho}) \simeq \frac{1}{\sqrt{2|\eps|}}
        \mqty(\sqrt{|\eps + 1|} \\ \pm \sqrt{|\eps - 1|}) e^{ik\rho\cos\varphi} +
    \frac{f_k(\varphi)}{\sqrt{-i2|\eps|\rho}}
        \mqty(\sqrt{|\eps + 1|} \\ \pm i\sqrt{|\eps - 1|} e^{i\varphi}) e^{ik\rho},\quad
    k\rho \gg 1,
\end{equation}
see, e.g., \cite{Novikov2007PhysRevB}. Here $\pm \equiv \sgn\eps$ is the energy sign, $k = \sqrt{\eps^2 - 1} > 0$ is the wave vector, $f_k(\varphi)$ is the 2D scattering amplitude which defines differential, total and transport cross-section (in the units of $l_F$),
\begin{equation}\label{eq66}
    \dv{\sigma(k; \varphi)}{\varphi} = |f_k(\varphi)|^2,\quad
    \sigma(k) = \Int{0}{2\pi} |f_k(\varphi)|^2 \dd{\varphi},\quad
    \sigma_\text{tr}(k) = \Int{0}{2\pi} (1 - \cos\varphi) |f_k(\varphi)|^2 \dd{\varphi}.
\end{equation}
With the scattering amplitude specified in (\ref{eq65}) (with the term $\exp(i\pi/4)$) the optical theorem takes the form given in \cite{Novikov2007PhysRevB}
\begin{equation}\label{eq67}
    \sigma(k) = \sqrt{\frac{8\pi}{k}} \Im f_k(0),
\end{equation}
and coincides with (\ref{eq42}) in non-relativistic case.

The exact solution of 2D Dirac equation that coincides with the asymptotic (\ref{eq65}) at $k\rho \gg 1$ may be found as an expansion in the eigenfunctions (\ref{eq5}) of the orbital angular momentum operator,
\begin{equation}\label{eq68}
    \Psi_\eps(\bm{\rho}) = \Sum{m = -\infty}{\infty} A_M \frac{1}{\sqrt{2|\eps|}}
        \mqty(\sqrt{|\eps + 1|} F_M(k\rho) e^{iM\varphi} \\
              \pm i\sqrt{|\eps - 1|} G_M(k\rho) e^{i(M+1)\varphi}),
\end{equation}
where $M = \delta + m$, $\delta = 0, 1/2$, $m = 0, \pm 1, \pm 2, \ldots$ Acting like in the section \ref{part32}, we get for the expansion coefficients
\begin{equation}\tag{\ref*{eq68}${}^\prime$} \label{eq68prime}
    A_M = i^{J - 1/2} e^{i\delta_J},\quad J = M + 1/2,
\end{equation}
and for scattering amplitude we have:
\begin{equation}\label{eq69}
    f_k(\varphi) = \frac{1}{i\sqrt{2\pi k}} \Sum{J}{} \qty(e^{2i\delta_J(k)} - 1)
        e^{i(J - 1/2)\varphi},\quad
    \begin{cases}
        J = \pm 1/2, \pm 3/2, \ldots,\\
        J = 0, \pm 1, \pm 2, \ldots,
    \end{cases}
\end{equation}
where summation is over either half-integer \cite{Novikov2007PhysRevB} or integer angular momenta $J$, including zero. Hence it follows the full cross-section expressed in terms of scattering phases $\delta_J(k)$,
\begin{equation}\label{eq70}
    \sigma(k) = \frac{4}{k} \Sum{J}{} \sigma_J(k),\quad \sigma_J(k) = \sin^2\delta_J(k),
\end{equation}
where $\sigma_J(k)$ is a partial cross-section. The transport cross-section is:
\begin{equation}\label{eq71}
    \sigma_\text{tr}(k) = \frac{2}{k} \Sum{J}{} \sin^2[\delta_{J+1}(k) - \delta_1(k)].
\end{equation}
In (\ref{eq70}) and (\ref{eq71}) the summation, as in (\ref{eq69}), is over total angular momentum, i.e. over half-integer values of $J$, as in \cite{Novikov2007PhysRevB}, or over integer values corresponding to half-integer values of orbital angular momentum $M$.
%%%%%%%%%%%%%%%%%%%%%%%%%%%%%%%%%%%%%%%%%%%%%%%%%%%%%%%%%%%%%%%%%%%%%%%%%%%%%%%%
\subsubsection{Electron scattering by a neutral impurity}\label{part411}
%%%%%%%%%%%%%%%%%%%%%%%%%%%%%%%%%%%%%%%%%%%%%%%%%%%%%%%%%%%%%%%%%%%%%%%%%%%%%%%%
As an example let us consider quasiparticles scattering on a neutral impenetrable impurity of radius $Rl_F$, which is much larger than a graphene lattice constant (heavy atom). For electron scattering phases, $\eps > 1$, we have
\begin{equation}\label{eq72}
    e^{2i\delta_J(k)} = -\frac{H_{|J-1/2|}^\Par{2}(kR)}{H_{|J-1/2|}^\Par{1}(kR)},\quad
    J =%
    \begin{cases}
        \pm 1/2, \pm 3/2, \ldots,\\
        0, \pm 1, \pm 2, \ldots,
    \end{cases}
\end{equation}
in full accordance with non-relativistic case, see (\ref{eq53}). This implies the symmetry of scattering phases,
\begin{equation}\label{eq73}
    \delta_J(k) = \delta_{1-J}(k).
\end{equation}

In the case of slow particles, i.e. near the upper continuum boundary, as in section \ref{part323} we get
\begin{equation}\label{eq74}
    \delta_J(k) = -\frac{(kR)^{2|J - 1/2|}}{(2|J - 1/2| - 2)!!\, (2|J - 1/2|)!!},\quad
    J \ne 1/2,\quad kR \ll 1.
\end{equation}
Hence, if $J$ is integer:
\begin{equation}\label{eq75}
    \delta_1(k) = \delta_0(k) = -kR,\quad
    \delta_2(k) = \delta_{-1}(k) = -\frac{1}{3} (kR)^3,
\end{equation}
and for half-integer angular momenta it follows
\begin{equation}\label{eq76}
    \delta_{3/2}(k) = \delta_{-1/2}(k) = -\frac{\pi}{4} (kR)^2,\quad
    \delta_{5/2}(k) = \delta_{-3/2}(k) = -\frac{\pi}{32} (kR)^4,
\end{equation}
and if $J = 1/2$ the scattering phase at $k \to 0$ descents logarithmically,
\begin{equation}\label{eq77}
    \delta_{1/2}(k) = -\frac{\pi}{2}\, \frac{1}{\ln(1/kR)},\quad kR \ll 1.
\end{equation}

Thus, if total angular momentum $J$ is integer, the differential cross-section is
\begin{equation}\label{eq78}
    \dv{\sigma(k; \varphi)}{\varphi} \equiv |f_k(\varphi)|^2 =
        \frac{8}{\pi} kR^2 \cos^2(\varphi/2),\quad kR \ll 1,
\end{equation}
so for total and transport cross-sections we have the following equations
\begin{equation}\label{eq79}
    \sigma(k) = 2\sigma_\text{tr}(k) = 8kR^2,\quad kR \ll 1.
\end{equation}
If $J$ takes half-integer values, the situation is different (as in a non-relativistic case),
\begin{equation}\label{eq80}
    \dv{\sigma(k; \varphi)}{\varphi} = \frac{\pi R}{2(kR) \ln^2(1/kR)},\quad
    \sigma(k) = \sigma_\text{tr}(k) = \frac{\pi^2R}{(kR) \ln^2(1/kR)},
\end{equation}
i.e. the scattering is isotropic and it grows with the wavelength.

In the opposite case of small wavelengths at large arguments according to the Hankel function asymptotic we have
\begin{equation}\label{eq81}
    \delta_J(k) \simeq%
    \begin{cases}
        kR - \frac{\pi}{2} |J - \frac{1}{2}| - \frac{\pi}{4},
            &|J - \frac{1}{2}| \lesssim kR,\quad kR \gg 1,\\
        0, & |J - \frac{1}{2}| \gtrsim kR \gg 1,
    \end{cases}
\end{equation}
so for the total cross-section we obtain
\begin{equation}\label{eq82}
    \sigma(k) \simeq \frac{4}{k} \Sum{-kR}{kR} \sin^2 \Big(kR -
        \frac{\pi}{2}\, \Big|J - \frac{1}{2}\Big| - \frac{\pi}{4}\Big) \simeq 4R,\quad kR \gg 1,
\end{equation}
where the symmetry relation (\ref{eq73}) is used.

Hence, the short-wave scattering properties do not depend on the way of orbital angular momentum quantization, while for long-wave case these properties differ for integer and half-integer orbital angular momenta.
%%%%%%%%%%%%%%%%%%%%%%%%%%%%%%%%%%%%%%%%%%%%%%%%%%%%%%%%%%%%%%%%%%%%%%%%%%%%%%%%
\subsubsection{Hole scattering by a neutral impurity}\label{part412}
%%%%%%%%%%%%%%%%%%%%%%%%%%%%%%%%%%%%%%%%%%%%%%%%%%%%%%%%%%%%%%%%%%%%%%%%%%%%%%%%
For scattering of quasiparticles with energy $\eps < -1$, i.e. holes with $\ol{\eps} = -\eps > 0$ and a wave vector $k = \sqrt{\ol{\eps}^2 - 1}$, instead of (\ref{eq72}) for scattering phases we have
\begin{equation}\label{eq83}
    e^{2i\ol{\delta}_J(k)} = -\frac{H_{|J+1/2|}^\Par{2}(kR)}{H_{|J+1/2|}^\Par{1}(kR)},\quad
    J =%
    \begin{cases}
        \pm 1/2, \pm 3/2, \ldots,\\
        0, \pm 1, \pm 2, \ldots,
    \end{cases}
\end{equation}
that leads to the symmetry relation
\begin{equation}\label{eq84}
    \ol{\delta}_J(k) = \ol{\delta}_{-(J+1)}(k).
\end{equation}

In the case of slow antiparticles scattering, i.e. near the lower continuum boundary, we have instead of (\ref{eq74})
\begin{equation}\label{eq85}
    \ol{\delta}_J(k) = -\frac{(kR)^{2|J + 1/2|}}{(2|J + 1/2| - 2)!!\, (2|J + 1/2|)!!},\quad
    J \ne -1/2,\quad kR \ll 1.
\end{equation}
The substitution of $J - 1/2$ for $J + 1/2$ is made due to the fact that near the upper continuum boundary the orbital angular momentum of upper component, $M = J - 1/2$, of the Dirac spinor (\ref{eq68}) is principal, while near the lower continuum boundary the principal component is the lower one, $M + 1 = J + 1/2$.

For integer values of total angular momentum $J$ we get
\begin{equation}\label{eq86}
    \ol{\delta}_{-1}(k) = \ol{\delta}_0(k) = -kR,\quad
    \ol{\delta}_{-2}(k) = \ol{\delta}_1(k) = -\frac{1}{3} (kR)^3,
\end{equation}
for half-integer $J$, we have respectively
\begin{equation}\label{eq87}
    \ol{\delta}_{1/2}(k) = \ol{\delta}_{-3/2}(k) = -\frac{\pi}{4} (kR)^2,\quad
    \ol{\delta}_{3/2}(k) = \ol{\delta}_{-5/2}(k) = -\frac{\pi}{32} (kR)^4,
\end{equation}
and for $J = -1/2$ the phase becomes logarithmically small,
\begin{equation}\label{eq88}
    \ol{\delta}_{-1/2}(k) = -\frac{\pi}{2}\, \frac{1}{\ln(1/kR)},\quad kR \ll 1.
\end{equation}
The expressions for differential, total and transport cross-sections of antiparticle scattering are fully identical with given above.

In the problem considered there are only scattering states and there is no discrete spectrum. Lt us consider the opposite case of a strong short-range attraction potential.
%%%%%%%%%%%%%%%%%%%%%%%%%%%%%%%%%%%%%%%%%%%%%%%%%%%%%%%%%%%%%%%%%%%%%%%%%%%%%%%%
\subsection{Solutions of the Dirac equation for a deep rectangular well}\label{part42}
%%%%%%%%%%%%%%%%%%%%%%%%%%%%%%%%%%%%%%%%%%%%%%%%%%%%%%%%%%%%%%%%%%%%%%%%%%%%%%%%
For half-integer values of the orbital angular momentum $M = m + 1/2$, $m = 0, \pm 1, \pm 2, \ldots$, i.e. for integer values of the total angular momentum $J = M + 1/2$, the two-dimensional radial Dirac equation (\ref{eq64}) coincides, as already noted at the beginning of the section \ref{part4}, with the radial equation of the relativistic three-dimensional problem, if we set
\begin{equation}\label{eq89}
    J = -\kap =%
    \begin{cases}
        \phantom{-}l + 1,\; \wt{l}, &j = |\kap| - 1/2,\\
        -l,\; -(\wt{l} + 1), &j = \kap - 1/2.
    \end{cases}
\end{equation}
Here $\kap = \pm 1, \pm 2, \ldots$ is the Dirac quantum number, $j$ is the three-dimensional total angular momentum, $l$ and $\wt{l}$ are the orbital angular momenta of the upper and lower components of the Dirac spinor $\Psi_{\eps, J}(\rho) \equiv \Psi_{\eps, \kap}(\rho)$.

Therefore, this case has an additional physical interest related, in particular, to the question of the validity of the single-particle Dirac equation in a strong electrostatic field decreasing at infinity, which was apparently first posed in the work of Schiff, Snyder and Weinberg \cite{SchiffSnyderWeinberg1940PhysRev}. Using the example of a spherically symmetric rectangular potential well, they showed that as the well deepens, the electronic levels move from the bottom of the upper continuum of solutions of the Dirac equation to the upper boundary of the lower continuum. At the same time, the authors of \cite{SchiffSnyderWeinberg1940PhysRev} believed that the difficulty of interpretation arises if the depth of the well exceeds the so-called “critical” value at which the given level reaches the boundary of the lower continuum.

This conclusion is shared by the authors of the monograph \cite{AkhiezerBerestetskii1965QED}, who believe that this difficulty is associated, as in the case of the relativistic Coulomb problem with the nuclear charge $Z > Z_\text{cr}$ \cite{PomeranchukSmorodinsky1945JPhysUSSR, ZeldovichPopov1972SovPhysUsp}, with the spontaneous production of electron-positron pairs by a strong field, so that the problem “cannot be solved within the framework of the quantum mechanics of a single particle” \cite[p.~120]{AkhiezerBerestetskii1965QED}.

However, papers \cite{PopovMur1974SovJNuclPhys, MurPopov1976TheorMathPhys_FermionCase, KuleshovMurEtAl2015PhysicsUspekhi, KuleshovMurEtAl2017JETP} show arguments based on the general principles of quantum theory that the one-particle Dirac equation remains valid also in the region of supercritical fields. Here we illustrate this statement with a model of a narrow rectangular potential well.
%%%%%%%%%%%%%%%%%%%%%%%%%%%%%%%%%%%%%%%%%%%%%%%%%%%%%%%%%%%%%%%%%%%%%%%%%%%%%%%%
\subsubsection{Scattering phase and poles of the scattering matrix of $s$-states}\label{part421}
%%%%%%%%%%%%%%%%%%%%%%%%%%%%%%%%%%%%%%%%%%%%%%%%%%%%%%%%%%%%%%%%%%%%%%%%%%%%%%%%
For potential
\begin{equation}\label{eq90}
    V(\rho) = -\theta(R - \rho) V,
\end{equation}
where $\theta(x)$ is the Heaviside step function, equation (\ref{eq64}) is solved in terms of Bessel functions. So, for example, for the partial phase of elastic scattering $\delta_\kap$ in states with $\kap = -1$, i.e. for $s$-states, we have \cite{KrylovMurFedotov2020EurPhysJC}
\begin{equation}\label{eq91}
    \delta_{-1}(k) = \delta^\Par{s}(k) - kR,\quad
    \cot\delta^\Par{s}(k) = \frac{1}{kR}\left\{
        1 - \frac{(\eps + 1)}{(V + \eps + 1)}\big[
            1 - KR\cot(KR)\big]\right\},
\end{equation}
where $K = \sqrt{(V + \eps)^2 - 1}$ and $k = \sqrt{\eps^2 - 1}$ are the wavevectors of the particle inside and outside the well.

The poles of the partial matrix of elastic scattering $S_\kap(k) = \exp[2i \delta_\kap(k)]$ with $k = i\lambda$, i.e. the equation $\cot\delta_\kap(i\lambda) = i$, determine the discrete spectrum of the problem. For $\kap = -1$ this gives the equation for the spectrum of $ns_{1/2}$-states,
\begin{equation}\label{eq92}
    KR\cot(KR) = -\lambda R - \frac{V}{(1 + \eps)}(1 + \lambda R),
    \quad \lambda = \sqrt{1 - \eps^2},\quad -1 \le \eps \le 1.
\end{equation}
In the case of a narrow well, $R \ll 1$, this equation can be represented as:
\begin{equation}\label{eq93}
    V = \frac{n\pi}{R} - (2\eps + 1) + \Big[(1 + \eps) \sqrt{1 - \eps^2} +
        \frac{1 - 2\eps(1 + \eps)}{2n\pi}\Big]R + O\qty(R^2),
\end{equation}
$n = 1, 2, \ldots$ is the radial quantum number. By taking $\eps = -1$ and $n = 1$ here, we get the critical value of the well depth for the ground state,
\begin{equation}\label{eq94}
    V_\text{cr}^\Par{s} = \frac{\pi}{R} + 1 + \frac{1}{2\pi}R + O\qty(R^2),
\end{equation}
which is consistent with the result of \cite{PopovMur1974SovJNuclPhys}. When $V > V_\text{cr}^\Par{s}$, the discrete level $k_d^\Par{s}$ “dives” into the lower continuum, turning into a quasistationary state with complex energy. To determine it, one can use the equation (\ref{eq92}), which near the lower continuum boundary is convenient to represented as \cite{PopovMur1974SovJNuclPhys}
\begin{equation}\label{eq95}
    V_\text{cr}^\Par{s} - V = a_2^\Par{s}\lambda^2 + a_3^\Par{s}\lambda^3,
\end{equation}
where $a_2^\Par{s} = 1 - R/2\pi$, $a_3^\Par{s} = -R/2$ for $n = 1$ and $R \ll 1$.

\begin{figure}[t]
    \centering
    \includegraphics[width=0.6\textwidth]{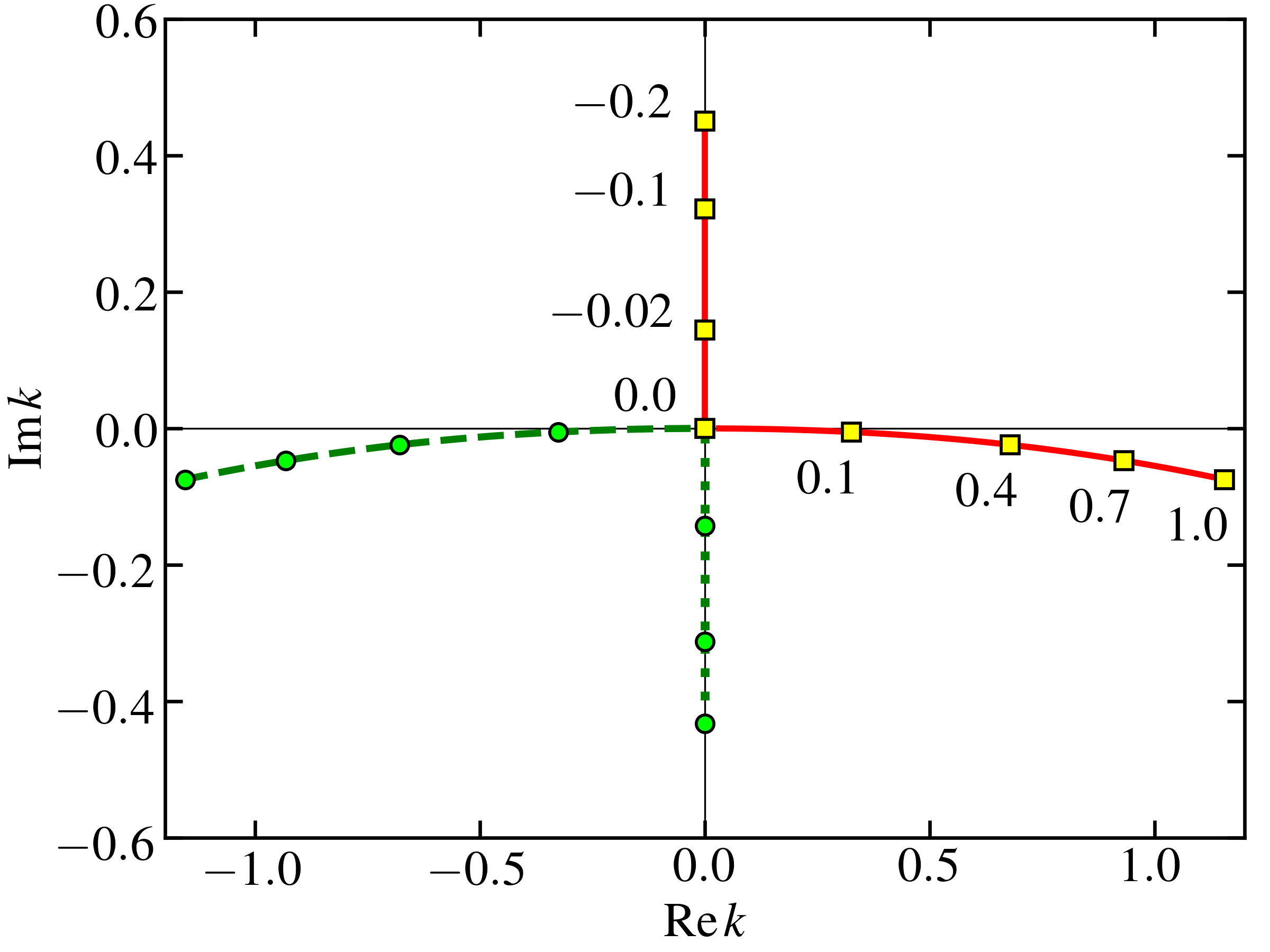}
    \caption{The trajectories of the poles of the $S$-matrix in the complex $k$-plane near the boundaries of the lower continuum for $\kap = -1$ and $R = 1/5$. The solid line with $\Re k = 0$ corresponds to a discrete level, the dotted line~--- to a virtual one. The solid line with $\Re k > 0$ corresponds to the Breit--Wigner pole, the dashed line with $\Re k < 0$~--- to the second pole, located far from the physical domain. The tick marks indicate the values of $\big(V - V_c^{(s)}\big)$.}
    \label{fig5}
\end{figure}

\begin{figure}[t]
    \centering
    \includegraphics[width=0.6\textwidth]{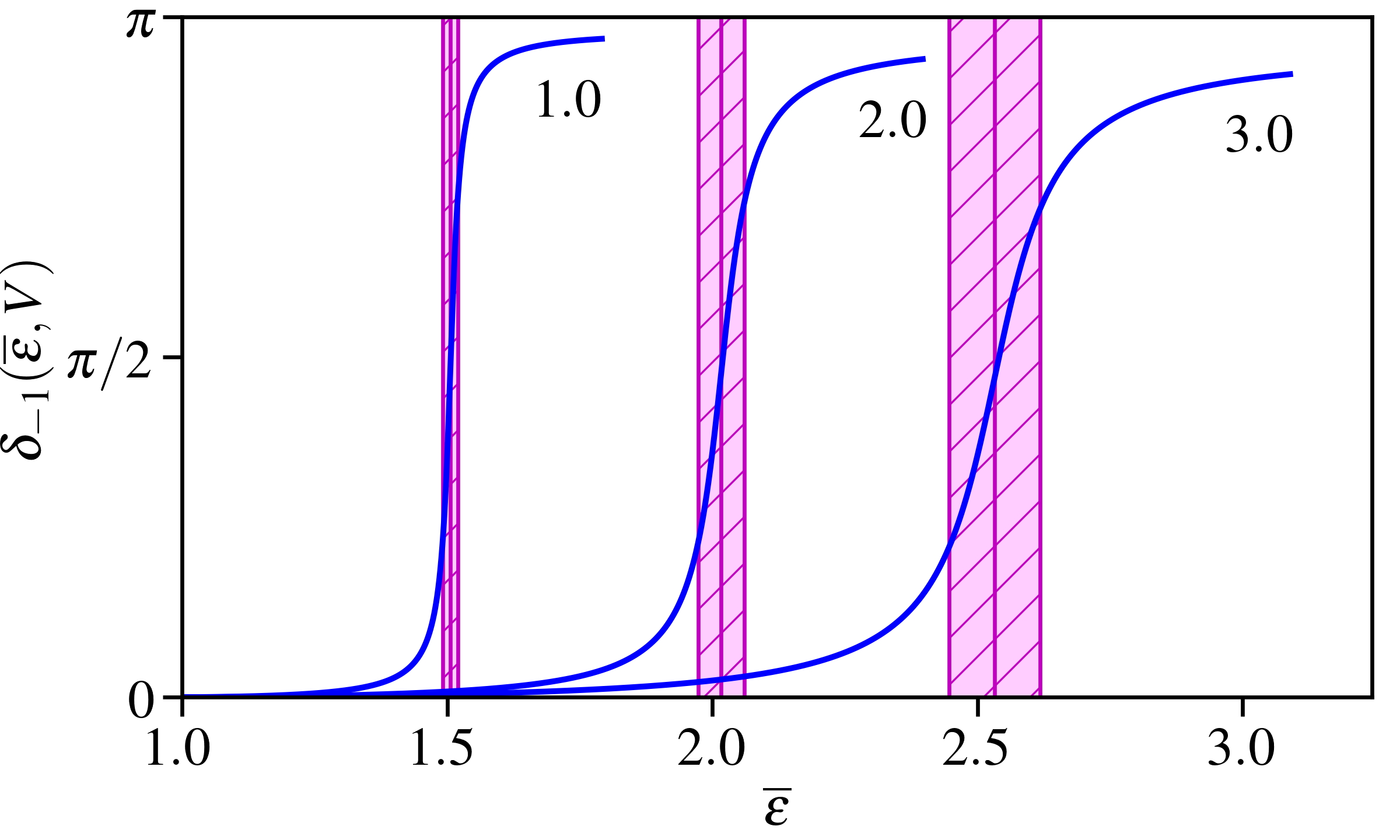}
    \caption{Scattering phases near the boundary of the lower continuum for $\kap = -1$ and $R = 1/20$. The shaded areas show the positions and widths of the Breit--Wigner resonances. The numbers indicate the values of $\big(V - V_c^{(s)}\big)$.}
    \label{fig6}
\end{figure}

Analytic continuation to the domain $V > V_\text{cr}^\Par{s}$, $\lambda = -ik$ gives
\begin{equation}\label{eq96}
    k_\pm = k'_\pm - ik'',\quad
    k'_\pm = \pm\qty(1 + R/4\pi) \qty(V - V_\text{cr}^\Par{s})^{1/2},\quad
    k'' = \tfrac{1}{4}R \big(V - V_\text{cr}^\Par{s}\big).
\end{equation}
The motion of the poles of the matrix $S_{-1}(k)$ in the complex $k$-plane as a function of supercriticality $(V - V_\text{cr}^\Par{s})$ is shown in Fig.~\ref{fig5} (Fig.~2\textit{a} in \cite{KrylovMurFedotov2020EurPhysJC}). The pole closest to the physical region will be called the Breit--Wigner pole, $k_+ \equiv k_\text{BW}$. For the energy of such a $s_{1/2}$-level in the lower continuum, $\eps = -\sqrt{k^2 + 1}$, we get
\begin{equation}\label{eq97}
    \wt{\eps}^\Par{s}_\text{BW} = -\wt{\eps}_0^\Par{s} +
        \tfrac{i}{2}\wt{\gamma}^\Par{s},\quad
    \wt{\eps}_0^\Par{s} = 1 + \tfrac{1}{2}\qty(1 + \tfrac{1}{2\pi}R)\qty(V -
        V_\text{cr}^\Par{s}),\quad
    \wt{\gamma}^\Par{s} = \tfrac{1}{2}R \qty(V - V_\text{cr}^\Par{s})^{3/2}.
\end{equation}
The unusual sign of the imaginary part, $\wt{\gamma}^\Par{s} > 0$, is coming from the use of a nonsecond-quantized approach. The Breit--Wigner level $\wt{\eps}^\Par{s}_\text{BW}$ in the Dirac sea corresponds to a quasistationary state of a positron with energy
\begin{equation}\label{eq98}
    \ol{\eps}^\Par{s}_\text{qs} = -\wt{\eps}^\Par{s}_\text{BW} =
        \wt{\eps}_0^\Par{s} - \tfrac{i}{2}\wt{\gamma}^\Par{s},\quad
    \wt{\eps}_0^\Par{s} > 1,\quad \wt{\gamma}^\Par{s} > 0.
\end{equation}
At a small supercriticality, this quasidiscrete level can manifest itself as a Breit--Wigner resonance in the elastic scattering of a positron with the width $\wt{\gamma}^\Par{s}$. The threshold behavior of the width is determined by the permeability of the centrifugal barrier for slow particles,
\begin{equation}\label{eq99}
    D \sim k^{2L+1},\quad k \to 0.
\end{equation}
Here one should specify an orbital angular momentum $L$ to the orbital angular momentum $\wt{l}$ of the lower component of the solution (\ref{eq64}) \cite{PopovMur1974SovJNuclPhys}. In the case under consideration, we have $L = \wt{l} = 1$, in agreement with the result ($S_{-1}(k)$) for the width \ref{eq95}. Fig.~\ref{fig6} (Fig.~4 in \cite{KrylovMurFedotov2020EurPhysJC}) shows the phase of elastic scattering of a positron as a function of its energy $\ol{\eps} = -\eps$. It can be seen that when energy $\ol{\eps}$ is close to the resonance position $\wt{\eps}_0^\Par{s}$, the phase changes abruptly across the width $\wt{\gamma}^\Par{s}$, which ensures the appearance of a resonance in the scattering in the same way as in the nonrelativistic theory of scattering \cite{Taylor1972}.

Similarly, one can consider the motion of the poles of the matrix $S_{-1}(k)$ near the boundary of the upper continuum. However, here, contrary to (\ref{eq95}), the expansion begins with a linear term,
\begin{equation}\label{eq100}
    V - V_b^\Par{s} = c_1^\Par{s}\lambda + c_2^\Par{s}\lambda^2,\quad
    c_1^\Par{s} = 2R,\quad c_2^\Par{s} = 1 + \frac{3}{2\pi}R,\quad
    V_b^\Par{s} = \frac{\pi}{R} - 3 - \frac{3}{2\pi}R,
\end{equation}
and the first bound $s_{1/2}$-level appears when $V = V_b^\Par{s}$. By taking $\lambda = -ik$ here, we get
\begin{equation}\label{eq101}
    k_\pm^\Par{s} = \pm iR\sqrt{\big(V - V_b^\Par{s}\big)/R^2 + 1} - iR.
\end{equation}

\begin{figure}[t]
    \centering
    \includegraphics[width=0.6\textwidth]{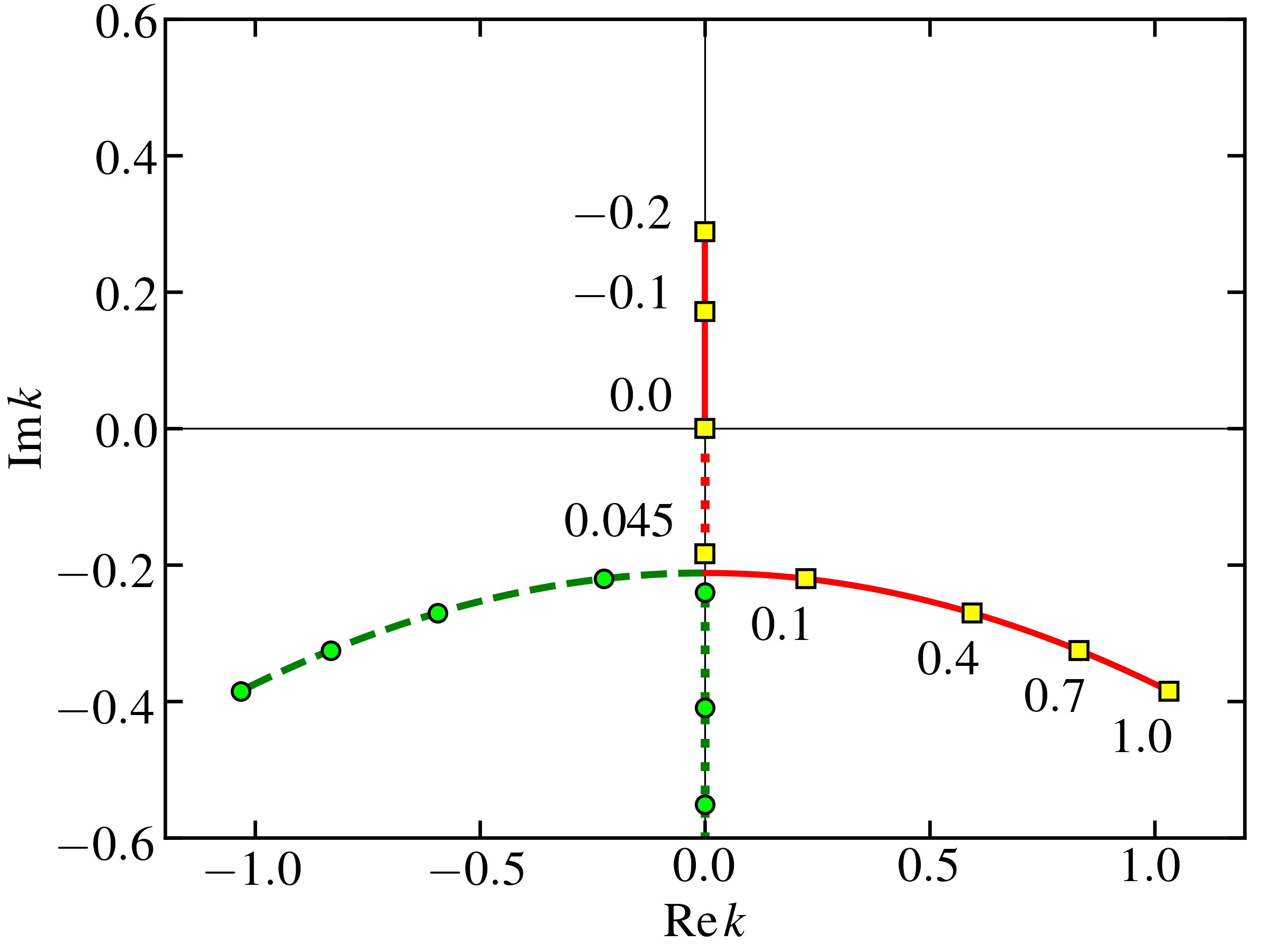}
    \caption{The trajectories of the poles of the $S$-matrix in the complex $k$-plane near the boundaries of the upper continuum for $\kap = -1$ and $R = 1/5$. The tick marks indicate the values of $\big(V_b^{(s)} - V\big)$. The legend is the same as in Fig.~\ref{fig5}.}
    \label{fig7}
\end{figure}

As the depth of the well decreases, the discrete level $k_d^\Par{s} = k_+^\Par{s}$ moves down along the positive imaginary axis of the $k$-plane. When $V = V_b^\Par{s}$, it reaches the boundary of the upper continuum, and with further decrease of $V$ it is “pushed” into it, turning into a virtual level $k_v^\Par{s}$ rather than into a quasistationary one, since $l = 0$ and there is no centrifugal barrier. At the same time, the second virtual level $k_{v'}^\Par{s}$ moves towards it, for which $k_{v'}^\Par{s} = -2iR$ at $V = V_b^\Par{s}$. When $V = V_b^\Par{s} - R^2$ they collide, then go into the complex plane, see Fig.~\ref{fig7} (Fig.~2\textit{b} in \cite{KrylovMurFedotov2020EurPhysJC}).
%%%%%%%%%%%%%%%%%%%%%%%%%%%%%%%%%%%%%%%%%%%%%%%%%%%%%%%%%%%%%%%%%%%%%%%%%%%%%%%%
\subsubsection{Scattering phase and poles of the scattering matrix of $p$-states}\label{part422}
%%%%%%%%%%%%%%%%%%%%%%%%%%%%%%%%%%%%%%%%%%%%%%%%%%%%%%%%%%%%%%%%%%%%%%%%%%%%%%%%
For $np_{1/2}$-states, i.e. at $\kap = 1$, the discrete spectrum is determined by the equation
\begin{equation}\label{eq102}
    KR\cot(KR) = -\lambda R + \frac{V}{(1 - \eps)}(1 + \lambda R),\quad
    \lambda = \sqrt{1 - \eps^2},\quad -1 \le \eps \le 1,
\end{equation}
which for $R \ll 1$ is equivalent to equality
\begin{equation}\label{eq103}
    V = \frac{n\pi}{R} - (2\eps - 1) - \Big[(1 - \eps) \sqrt{1 - \eps^2} -
        \frac{1 + 2\eps(1 - \eps)}{2n\pi}\Big]R + O\qty(R^2).
\end{equation}
The first bound $1p_{1/2}$-level appears when $V = V_b^\Par{p}$,
\begin{equation}\label{eq104}
    V_b^\Par{p} = \frac{\pi}{R} - 1 + \frac{1}{2\pi}R + O\qty(R^2),
\end{equation}
and the equation (\ref{eq102}) at a small underboundness, $(V - V_b^\Par{p}) \ll 1$, can be represented as
\begin{equation}\label{eq105}
    V - V_b^\Par{p} = a_2^\Par{p}\lambda^2 + a_3^\Par{p}\lambda^3,\quad
    a_2^\Par{p} = 1 + \tfrac{1}{2\pi}R,\quad a_3^\Par{p} = -\tfrac{1}{2}R.
\end{equation}
Comparison with the expansion (\ref{eq95}) shows that the motion of the poles of the $S$-matrix when $\kap = 1$ near the boundary of the upper continuum with decreasing depth of the well is similar to the motion of the poles of the $S$-matrix with $\kap = -1$ in the vicinity of $\eps = -1$ with increasing $V$, cf. Fig. \ref{fig5}.

For the energy of the quasistationary Breit--Wigner state of the electron, we obtain
\begin{equation}\label{eq106}
    \eps_\text{qs}^\Par{p} = \eps_\text{BW}^\Par{p} =
        \eps_0^\Par{p} - \tfrac{i}{2}\gamma^\Par{p},\quad
    \eps_0^\Par{p} = 1 + \tfrac{1}{2}\qty(1 - \tfrac{1}{2\pi}R)
        \big(V_b^\Par{p} - V\big),\quad
    \gamma^\Par{p} = \tfrac{1}{2} R\big(V_b^\Par{p} - V\big)^{3/2} > 0,
\end{equation}
i.e. the usual sign in front of the width $\gamma^\Par{p}$, as in the nonrelativistic scattering theory \cite{Taylor1972}. The index of the degree of underboundness $(V_b^\Par{p} - V)$ is determined here by the orbital angular momentum of the upper component of the spinor, $l = 1$, $l + 1/2 = 3/2$.

Conversely, the motion of poles with $\kap = 1$ near the boundary of the lower continuum is similar to the motion of poles with $\kap = -1$ for $\eps \sim 1$. In this case, the Eq. (\ref{eq103}) gives
\begin{equation}\label{eq107}
    V_\text{cr}^\Par{p} - V = c_1^\Par{p}\lambda + c_2^\Par{p}\lambda^2,\quad
    c_1^\Par{p} = 2R,\quad c_2^\Par{p} = 1 - \tfrac{3}{2\pi}R,\quad
    V_\text{cr}^\Par{p} = \frac{\pi}{R} + 3 - \frac{3}{2\pi}R,
\end{equation}
and the signs of the coefficients $c_1^\Par{p}$ and $c_2^\Par{p}$ are just such that at $V > V_\text{cr}^\Par{p}$ the level goes to the second, unphysical sheet \cite{MurPopov1976TheorMathPhys_FermionCase}.

Indeed, setting $\lambda = -ik$, at $R \ll 1$ we get
\begin{equation}\label{eq108}
    k_\pm^\Par{p} = \pm i\sqrt{\big(V_\text{cr}^\Par{p} - V\big) + R^2} - iR.
\end{equation}
For $V < V_\text{cr}^\Par{p} + R^2$, the real level with an increase in $V$ moves down the imaginary axis of the $k$-plane, turning into a virtual level at $V_\text{cr}^\Par{p} < V < V_\text{cr}^\Par{p} + R^2$. The second virtual level moves towards it. When $V = V_\text{cr}^\Par{p} + R^2$ they collide, then go into the complex plane. And the quasidiscrete level with energy $\wt{\eps}_-^\Par{p}$,
\begin{equation}\label{eq109}
    \wt{\eps}_-^\Par{p} = -\wt{\eps}_0^\Par{p} + \tfrac{i}{2}\wt{\gamma}^\Par{p},\quad
    \wt{\eps}_0^\Par{p} = 1 + \tfrac{1}{2}\qty(V - V_\text{cr}^\Par{p}),\quad
    \wt{\gamma}^\Par{p} = R\big(V - V_\text{cr}^\Par{p}\big)^{1/2},
\end{equation}
moves to the right, being closer to the physical region. For the energy of the quasidiscrete level of the positron in the $p_{1/2}$-state, we obtain
\begin{equation}\label{eq110}
    \ol{\eps}_\text{qs}^\Par{p} = -\wt{\eps}_-^\Par{p} =
        \wt{\eps}_0^\Par{p} - \tfrac{i}{2}\wt{\gamma}^\Par{p},\quad
    \wt{\eps}_0^\Par{p} > 1,\quad \wt{\gamma}^\Par{p} > 0.
\end{equation}
The dependence of the width $\wt{\gamma}^\Par{p}$ on supercriticality is determined in this case by the orbital angular momentum $\wt{l} = 0$, which means the absence of a centrifugal barrier.
%%%%%%%%%%%%%%%%%%%%%%%%%%%%%%%%%%%%%%%%%%%%%%%%%%%%%%%%%%%%%%%%%%%%%%%%%%%%%%%%
\subsubsection{Generalization to the case of arbitrary values $\kap \ne \pm 1$}\label{part423}
%%%%%%%%%%%%%%%%%%%%%%%%%%%%%%%%%%%%%%%%%%%%%%%%%%%%%%%%%%%%%%%%%%%%%%%%%%%%%%%%
The values $\kap = \pm 1$ considered in sections \ref{part421} and \ref{part422} are in some sense distinguished: the expansions (\ref{eq100}) and (\ref{eq107}), respectively for $\kap = -1$ and $\kap = 1$, begin with a linear term. But if $|\kap| \ge 2$, then all similar expansions begin with a quadratic term, since the orbital angular momenta $l$ and $\wt{l}$ for such values of $\kap$ are always nonzero.

Therefore, for the energies of Breit--Wigner resonances near the boundary of the upper continuum, i.e. for quasistationary states of an electron, in complete analogy with equality (\ref{eq106}) we have
\begin{equation}\label{eq111}
    \eps_\text{qs}^\Par{\kap} = \eps_\text{BW}^\Par{\kap} = \eps_0^\Par{\kap} -
        \tfrac{i}{2}\gamma^\Par{\kap},\quad
    \eps_0^\Par{\kap} > 1,\quad \gamma^\Par{\kap} > 0.
\end{equation}
And the threshold dependence of the width $\gamma^\Par{\kap}$ is determined, as before, by the centrifugal barrier:
\begin{equation}\label{eq112}
    \gamma^\Par{\kap} \sim \big(V_b^\Par{\kap} - V\big)^{l + 1/2},\quad
    l = j + \tfrac{1}{2}\sgn\kap > 0.
\end{equation}

At the same time, near the boundary of the lower continuum, for Breit--Wigner energy, similarly to equality (\ref{eq97}), we get
\begin{equation}\label{eq113}
    \wt{\eps}^\Par{\kap}_\text{BW} = -\wt{\eps}_0^\Par{\kap} +
        \tfrac{i}{2}\wt{\gamma}^\Par{\kap},\quad
    \wt{\eps}_0^\Par{\kap} > 1,\quad \wt{\gamma}^\Par{\kap} > 0
\end{equation}
with the threshold dependence for the width
\begin{equation}\label{eq114}
    \wt{\gamma}^\Par{\kap} \sim \big(V - V_\text{cr}^\Par{\kap}\big)^{\wt{l} + 1/2},\quad
    \wt{l} = j - \tfrac{1}{2}\sgn\kap > 0.
\end{equation}

It should be noted that the inequality $\wt{\gamma}^\Par{\kap} > 0$ holds for a short-range potential of arbitrary shape. The positivity\footnote{In the review \cite{ZeldovichPopov1972SovPhysUsp} $\wt{\gamma}^\Par{\kap}$ was interpreted as the probability of spontaneous electron-positron pair production, see also the monograph \cite{GreinerMullerRafelski1985}.} of $\wt{\gamma}^\Par{\kap}$ was substantiated in the paper \cite{MurPopov1976TheorMathPhys_FermionCase} in the framework of the effective radius approximation developed there for the Dirac equation, see also \cite{PopovEletskiiMur1976SovPhysJETP} By means of this approximation, the expansions, which generalize (\ref{eq95}) and (\ref{eq105}), were also established.

An invalid sign in front of $\wt{\gamma}^\Par{\kap}$ in (\ref{eq113}) indicates that it is necessary to move to the second quantized theory. The Dirac radial Hamiltonian (\ref{eq64}) with the potential (\ref{eq90}) is a self-adjoint operator, see, for example, \ref{appendixC}. Its eigensolutions form a complete system of functions, which according to Furry \cite{Furry1951PhysRev} can be used to quantize a single-particle system. In the Furry picture, solutions of the Dirac equation in the lower continuum, in complete analogy with the solutions of the free equation, correspond to states of a positron with energy $\ol{\eps} = -\eps > 1$. In the non-second quantized theory, i.e. in the one-particle approach, they correspond to states in the “Dirac sea” distorted by the external field.

Thus, the Breit--Wigner poles with energy (\ref{eq113}) correspond to quasistationary states of a positron with energy
\begin{equation}\label{eq115}
    \ol{\eps}_\text{qs}^\Par{\kap} = -\wt{\eps}_\text{BW}^\Par{\kap} =
        \wt{\eps}_0^\Par{\kap} - \tfrac{i}{2}\wt{\gamma}^\Par{\kap},\quad
    \wt{\eps}_0^\Par{\kap} > 1,\quad \wt{\gamma}^\Par{\kap} > 0
\end{equation}
with a negative, as it should be, imaginary part. In the case of small supercriticality, such quasidiscrete levels can manifest themselves as resonances in the scattering of positrons by a supercritical well. If the energy of the positron $\ol{\eps} > 1$ lies in the region of abrupt change in the partial scattering phase $\delta_\kap(k)$, see Fig.~\ref{fig6} with $\kap = -1$, then a resonance in its scattering occurs, and the partial cross section corresponds to the Breit--Wigner formula
\begin{equation}\label{eq116}
    \sigma_\kap(\ol{\eps}) = \sin^2 \delta_\kap(k) =
    \frac{\big(\wt{\gamma}^\Par{\kap} / 2\big)^2}
         {\big(\ol{\eps} - \wt{\eps}_0^\Par{\kap}\big)^2 +
            \big(\wt{\gamma}^\Par{\kap} / 2\big)^2}.
\end{equation}

Due to the fact that the partial scattering phases $\delta_\kap(k)$ are real, see the equation (\ref{eq91}) and Fig.~\ref{fig6} in the case $\kap = -1$, the elastic scattering matrix of positrons $S_\kap = \exp[2i\delta_\kap(k)]$ is unitary. Therefore, in accordance with the quantum scattering theory, there are no inelastic processes in the channel with a given $\kap$, including the spontaneous production of electron-positron pairs.

A remarkable property of the Dirac equation should be emphasized. As in the nonrelativistic case, pushing an electron level into the upper continuum upon a decrease of the well depth leads to emergence of a quasistationary state in a scattering of an electron with energy $\eps > 1$. However, diving of the electron level into the lower continuum upon deepening of the well results in emergence of a quasistationary state in a scattering of a positron with energy $\ol{\eps} = -\eps > 1$.

Therefore, the resonant scattering of positrons at small supercriticality, similar to the resonant scattering of electrons at small underboundness, cannot serve as evidence in favor of the spontaneous production of electron-positron pairs in the supercritical region of depth of the well. It should be noted that the authors of \cite{GodunovMachetVysotsky2017EurPhysJ} do not agree with this conclusion. They believe that the resonant scattering of positrons in the relativistic Coulomb problem with the nuclear charge $Z > Z_\text{cr}$ are precisely a signature of the spontaneous $e^+e^-$-pair production process.

Let us discuss a similar problem in gapped graphene with a Coulomb impurity and the possibility of experimental verification of the statements made above.
%%%%%%%%%%%%%%%%%%%%%%%%%%%%%%%%%%%%%%%%%%%%%%%%%%%%%%%%%%%%%%%%%%%%%%%%%%%%%%%%
\subsection{The Coulomb problem in gapped graphene}\label{part43}
%%%%%%%%%%%%%%%%%%%%%%%%%%%%%%%%%%%%%%%%%%%%%%%%%%%%%%%%%%%%%%%%%%%%%%%%%%%%%%%%
The radial Dirac equation (\ref{eq64}) with the Coulomb attraction potential, $V_C = -q/\rho$, $q > 0$, is equivalent to the relativistic Coulomb problem with $J = -\kap \ne 0$, $q = Z\alpha$, where $Z$ is a nucleus charge, and $\alpha = e^2/\hbar c$ is the Sommerfeld fine structure constant. It is known for a long time \cite{Sommerfeld1916AnnDerPhys} that in this problem the point-like charge model is an idealization because it becomes meaningless if the nucleus charge $Z > Z_s = \alpha^{-1} \simeq 137$, see \ref{appendixC}.

Pomeranchuk and Smorodinsky \cite{PomeranchukSmorodinsky1945JPhysUSSR} showed that this difficulty can be obviated by accounting for the finite size of a nucleus, i.e. the modification of the Coulomb potential at short distances\footnote{In this case the Hamiltonian $H$, associated with the Dirac Hamiltonian $H_D$, see (\ref{eqC3}), is self-adjoint operator \cite{Dirac1958ClarendonPress}.},
\begin{equation}\label{eq117}
    V_R(\rho) = -\frac{q}{R}%
    \begin{cases}
        R/\rho, &\rho \ge R,\\
        f\qty(\rho/R), &\rho \le R,
    \end{cases}
\end{equation}
where $f(0) = \text{const}$, $f(1) = 1$. In this work the \enquote{rectangular cut-off} of the Coulomb potential, $f(\rho/R) = 1$, was used. In this case the Dirac equation can be solved analytically. In this case the discrete level dives monotonically with the growth of $Z$ and at \enquote{critical} value of $Z$, \cite{PomeranchukSmorodinsky1945JPhysUSSR}, $Z = Z_\text{cr}$, it reaches the lower continuum boundary of the Dirac equation.

In graphene we have the parameter $\alpha_F = e^2/\hbar v_F \sim 1$ which is analogous to the fine structure constant $\alpha$. So, in graphene, even at effective charges $Z \gtrsim 1$ the Coulomb potential regularization at short distances is needed. The \enquote{cut-off} radius of the Coulomb potential should be much larger than the unit cell size, so that the transition to a continuous description using the effective two-dimensional Dirac equation is possible,
\begin{equation}\label{eq118}
    1 \gg R \gg \frac{a_\text{CC}}{l_F} \simeq 5.2 \cdot 10^{-3}.
\end{equation}

The last estimation is given for the gapped graphene deposited onto a SiC substrate. According to \cite{ZhouGweonEtAl2007NatureMaterials, PereiraKotovCastroNeto2008PhysRevB}, the gap size $\Delta = 2m_*v_F^2 = 0.26\text{\;eV} = 4.17 \cdot 10^{-20} \text{\;J}$, the effective coupling constant $\alpha_F = 0.4$, the distance between carbon nuclei $a_\text{CC} = 1.42\text{\;Å} = 1.42 \cdot 10^{-10}\text{\;m}$,
\begin{equation}\label{eq119}
    v_F = 5.59 \cdot 10^8\text{\;cm/s} = 5.59 \cdot 10^6\text{\;m/s},\quad
    m_* = 7.65 \cdot 10^{-4} m_e,\quad l_F = 271\text{\;Å},
\end{equation}
so the transition to a continuous limit is valid.

The \enquote{rectangular cut-off} of the Coulomb potential at small distances \cite{PomeranchukSmorodinsky1945JPhysUSSR} not only provides self-adjointness of the Hamiltonian (\ref{eqC3}), but also allows us to single out a specific self-adjoint Hamiltonian from one-parametrical families (\ref{eqC14}) and (\ref{eqC18}), i.e. to fix parameters $\theta_\sigma(J)$ and $\theta_\tau(J)$ analytically.

According to \cite{KuleshovMurEtAl2017JETP} we have
\begin{equation}\label{eq120}
    \tan\theta_\sigma(J; R) = \frac{(\sigma + J)}{(\sigma - J)}
        \frac{\qty[qJ_{\pm(J-1/2)}(q) \pm (\sigma - J) J_{\pm(J+1/2)}(q)] R^{-\sigma}}
             {\qty[qJ_{\pm(J-1/2)}(q) \mp (\sigma + J) J_{\pm(J+1/2)}(q)] R^{\sigma\phantom{-}}},
\end{equation}
where $\sigma = \sqrt{J^2 - q^2} > 0$, $J_\nu(x)$ is the Bessel function, and upper (lower) signs correspond to $J > 0$ ($J < 0$). If $q > |J|$, then, according to (\ref{eqC17}) we get
\begin{equation}\label{eq121}
    \exp[2i\theta_\tau(J; R)] = \frac{(i\tau + J)}{(i\tau - J)}
        \frac{\qty[qJ_{\pm(J-1/2)}(q) \pm (i\tau - J) J_{\pm(J+1/2)}(q)] R^{-i\tau}}
             {\qty[qJ_{\pm(J-1/2)}(q) \mp (i\tau + J) J_{\pm(J+1/2)}(q)] R^{i\tau\phantom{-}}},
\end{equation}
where $\tau = \sqrt{q^2 - J^2} > 0$.

\begin{figure}[t]
    \centering
    \includegraphics[width=0.6\textwidth]{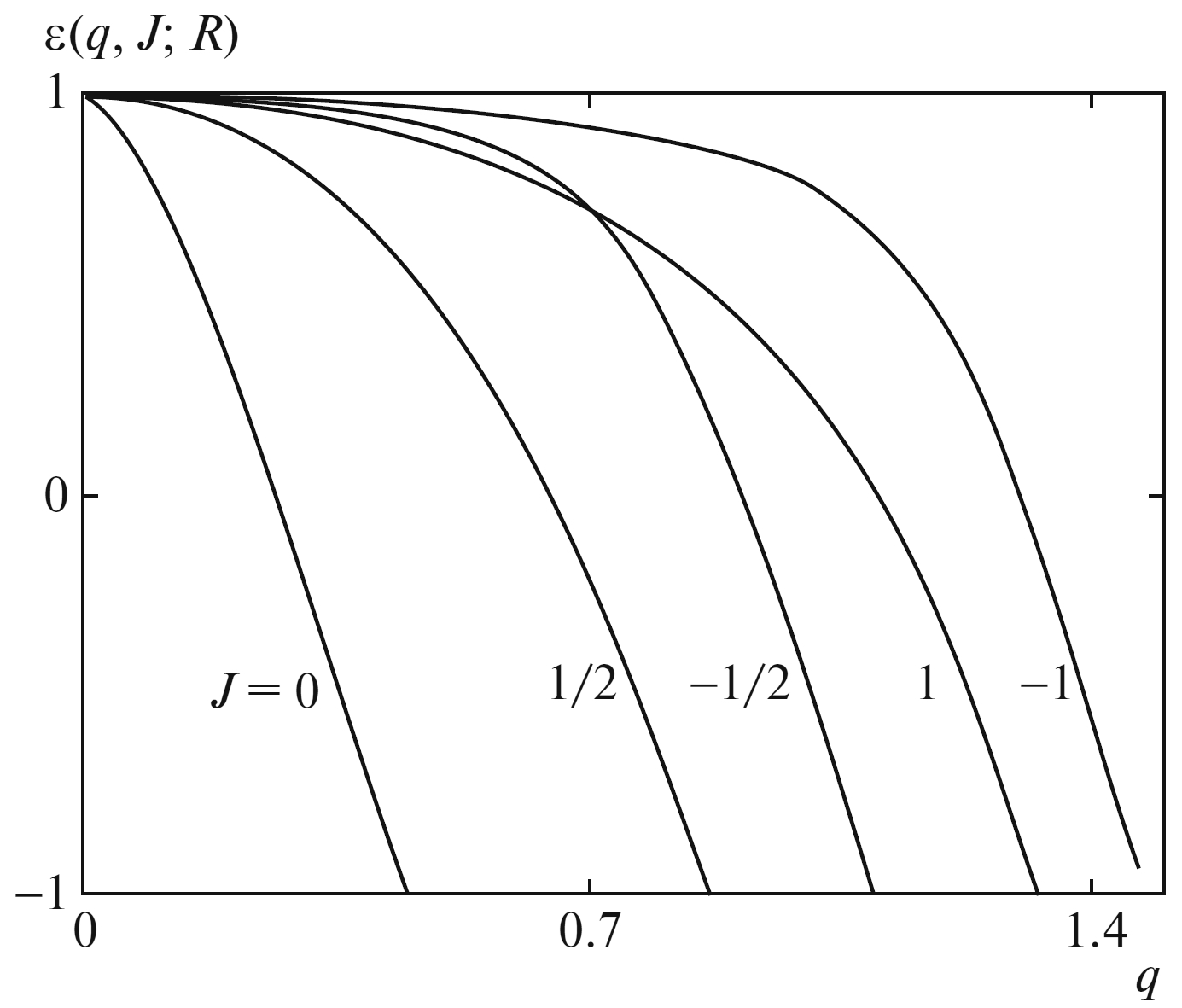}
    \caption{Energy $\eps(q, J; R)$ of the lowest level with a given value of the orbital angular momentum $M = J - 1/2$ as a function of charge $q = Z\alpha_F$ for the radius of rectangular cutoff of the Coulomb potential $R = 1/20$. The numbers at the curves are the values of the quantum number $J$.}
    \label{fig8}
\end{figure}

If $J = 0$, then instead of (\ref{eq121}) we have the equation
\begin{equation}\label{eq122}
    \exp[2i\theta_\tau(J=0; R)] = \exp\!\big[2iq\ln\big(\tfrac{1}{R_0}\big)\big],\quad
    R_0 = R\exp(-f_0),\quad f_0 = \Int{0}{1} f(\xi) \dd{\xi}.
\end{equation}
Thus, if the total angular momentum is $J = 0$, then the analytical solution of the set (\ref{eq64}) with potential (\ref{eq117}) is possible with any modification of the Coulomb potential at small distances. For polynomial cut-off in three-dimensional problem we have \cite{KuleshovMurEtAl2017JETP}
\begin{equation}\label{eq123}
    f_0(n) = \frac{n + 2}{n + 1},\quad f_0(\infty) = 1,\quad
    f_0(2) = \frac{4}{3},\quad f_0(1) = \frac{3}{2},\quad f_0(0) = 2,
\end{equation}
where $n = \infty$ corresponds to uniform impurity charge distribution over a sphere $R$ (rectangular cut-off), and $n = 2$~--- over a ball of the same radius.
%%%%%%%%%%%%%%%%%%%%%%%%%%%%%%%%%%%%%%%%%%%%%%%%%%%%%%%%%%%%%%%%%%%%%%%%%%%%%%%%
\subsubsection{Discrete spectrum}\label{part431}
%%%%%%%%%%%%%%%%%%%%%%%%%%%%%%%%%%%%%%%%%%%%%%%%%%%%%%%%%%%%%%%%%%%%%%%%%%%%%%%%
Setting $r = 2\lambda \rho$, $\lambda = \sqrt{1 - \eps^2}$ and following Gordon's approach \cite{Gordon1928ZPhys}, see also the monograph \cite{AkhiezerBerestetskii1965QED}, we get the solution of (\ref{eqC11}) decreasing at the infinity,
\begin{equation}\label{eq124}
    \mqty(F \\ G) = C \sqrt{1 \pm \eps}\; e^{-r/2}\, r^\sigma \qty{\Psi(a, c; r) \pm
        \qty(\frac{q}{\lambda} - J) \Psi(a+1, c; r)}.
\end{equation}
Here $C$ is a normalization factor, $\sigma = \sqrt{J^2 - q^2} > 0$, $\Psi(a, c; r)$ is the Tricomi function \cite{BatemanErdelyi1_2_1953McGrawHill}, $a = \sigma - \eps q/\lambda$ and $c = 1 + 2\sigma$ are its parameters, and upper (lower) signs correspond to function $F$ ($G$). At small distances and\footnote{The special case $q = q_s = |J|$ was discussed in \cite{KuleshovMurEtAl2017JETP}.} $\sigma \ne 0$ this solution has the same structure as (\ref{eqC12}), and
\begin{equation}\label{eq125}
    u_{\pm \sigma} = \frac{\Gamma(\mp 2\sigma) (2\lambda)^{\pm\sigma}}
                          {\Gamma\qty(1 \mp \sigma - \frac{\eps}{\lambda}q)}
        \big[q\sqrt{1 - \eps} - (J \pm \sigma \sqrt{1 + \eps})\big],
\end{equation}
where $\Gamma(z)$ is the Euler gamma-function.

\begin{figure}[t]
    \centering
    \includegraphics[width=0.6\textwidth]{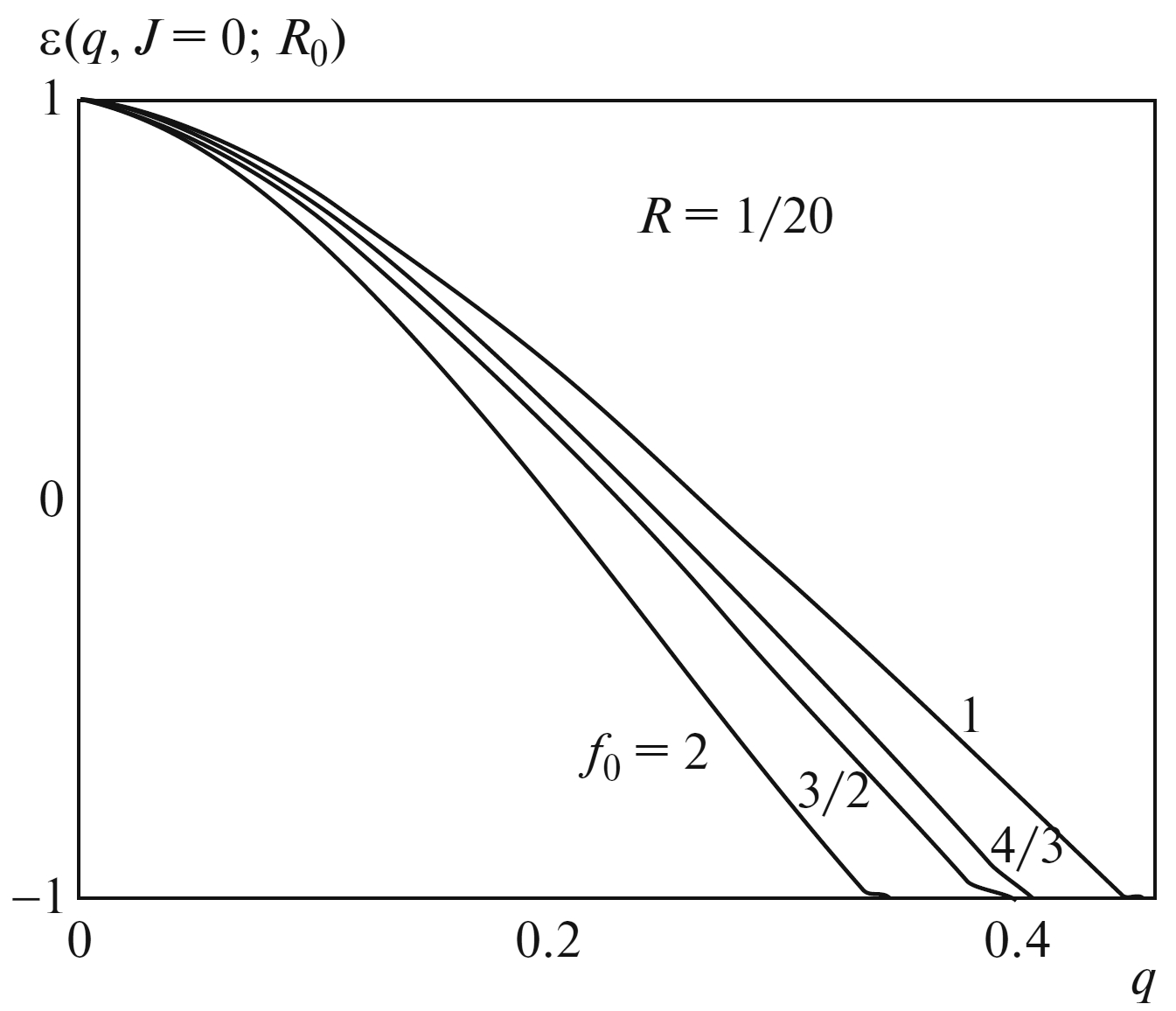}
    \caption{Energy $\eps(q, J = 0; R_0)$ of the ground state level as a function of charge $q = Z\alpha_F$ for a Coulomb potential cutoff radius of $R = 1/20$ for various shapes of the cutoff function. The numbers at the curves are the values of $f_0$; the values of the effective cutoff radius are given in Table \ref{tab2}.}
    \label{fig9}
\end{figure}

The boundary condition (\ref{eqC13}) and the explicit expression (\ref{eq120}) for the phase $\theta_\sigma$ determine the equation for discrete spectrum at $0 < q < |J|$ in analytical form. If $q > |J|$ the expression (\ref{eq125}) with the substitution $\sigma = i\tau$, $\tau = \sqrt{q^2 - J^2} > 0$ is valid. Taking into account the boundary condition (\ref{eqC17}) and the expression (\ref{eq121}) for the phase $\theta_\tau$, we get the algebraic equation for the spectrum for given charges. In particular, for the critical impurity charge,
\[
    q_\text{cr}^\Par{n}(J) = Z_\text{cr}^\Par{n}(J) \alpha_F,
\]
when the $n$-th level with quantum number $J$ reaches the lower continuum boundary, $\eps = -1$, we have
\begin{equation}\label{eq126}
    \arg\Gamma\qty(2i\sqrt{\big[q_\text{cr}^\Par{n}(J)\big]^2 - J^2}) =
    \sqrt{\big[q_\text{cr}^\Par{n}(J)\big]^2 - J^2}\, \ln\!\big[2q_\text{cr}^\Par{n}(J)\big] -
        \theta_\tau(J) + \pi n,\quad n = 0, 1, 2, \ldots
\end{equation}
And if $J = 0$, then for the phase $\theta_\tau(J = 0; R)$ one should use the expression (\ref{eq122}), which is determined for any form of cut-off function, i.e. for the short-range Coulomb problem (\ref{eq117}).

The ground state energy $\eps(q, J; R)$ dependence on the impurity charge $q = Z\alpha_F$ for the cut-off radius $R = 1/20$ for several values of $J$ are given on Fig.~\ref{fig8} (Fig.~1 in \cite{KuleshovMurEtAl2017JETP}). The radius $Rl_F = 13.5\text{\;Å}$ 	by about an order of magnitude larger then a distance $a_\text{CC} = 1.42\text{\;Å}$ between carbon nuclei, that is necessary to describe the electronic properties of graphene deposited on SiC substrate within the effective 2D Dirac equation. The ground state energy $\eps(q, J=0; R)$ for different radii $R_0 = R\exp(-f_0)$, $R = 1/20$ and several values of $f_0$, i.e. for different cut-offs, is given on Fig.~\ref{fig9} (Fig.~2 in \cite{KuleshovMurEtAl2017JETP}).

\begin{table}[t]
    \caption{The values of critical charge $q_\text{cr}^{(n)}(J) = Z_\text{cr}^{(n)}(J) \alpha_F$ for which the lowest ($n = 0$) and the first excited ($n = 1$) levels with a given value of the orbital angular momentum $M = J - 1/2$ reach the boundaries of the lower continuum for a cutoff radius of $R = 1/20$.}
    \centering
\begin{tabular}{|c|c|c|c|c|}
    \hline
    & \multicolumn{4}{|c|}{$R = 1/20$}\\ \cline{2-5}
    & $J = 1/2$ & $J = -1/2$ & $J = 1$ & $J = -1$\\ \hline
    $q_\text{cr}^\Par{0}(J)$ & 0.90 & 1.14 & 1.36 & 1.51\\ \hline
    $q_\text{cr}^\Par{1}(J)$ & 1.61 & 1.91 & 1.97 & 2.19\\
    \hline
\end{tabular}
\label{tab1}
\end{table}

\begin{table}[t]
\caption{The values of critical charge of the ground, $q_\text{cr}^{(0)}$, and the first excited, $q_\text{cr}^{(1)}$, levels with orbital angular momentum $M = -1/2$ for various forms of the cutoff function and the corresponding effective cutoff radii $R 0 = R\exp(-f_0)$ for $R = 1/20$.}
\centering
\begin{tabular}{|c|c|c|c|c|}
    \hline
    & \multicolumn{4}{|c|}{$R = 1/20$, $J = 0$}\\ \cline{2-5}
    & $f_0 = 1$, & $f_0 = 4/3$, & $f_0 = 3/2$, & $f_0 = 2$,\\
    & $R_0 = 0.018$ & $R_0 = 0.013$ & $R_0 = 0.011$ & $R_0 = 0.007$\\ \hline
    $q_\text{cr}^\Par{0}(0)$ & 0.46 & 0.41 & 0.40 & 0.35\\ \hline
    $q_\text{cr}^\Par{1}(0)$ & 1.33 & 1.23 & 1.18 & 1.06\\
    \hline
\end{tabular}
\label{tab2}
\end{table}

The values of critical charge $q_\text{cr}^\Par{n}(J) = Z_\text{cr}^\Par{n}(J)\alpha_F$, when the ground ($n = 0$) and first excited level ($n = 1$) with the given angular momentum, $J = 1/2,\, -1/2,\, 1,\, -1$, reaches the lower continuum boundary are given in Table~\ref{tab1} (for $R = 1/20$). The values of critical charge for the ground, $q_\text{cr}^\Par{0}(0)$, and first excited, $q_\text{cr}^\Par{1}(0)$, level with $J = 0$ for different cut-off functions are given in Table~\ref{tab2}.

\begin{figure}[t]
    \centering
    \includegraphics[width=0.6\textwidth]{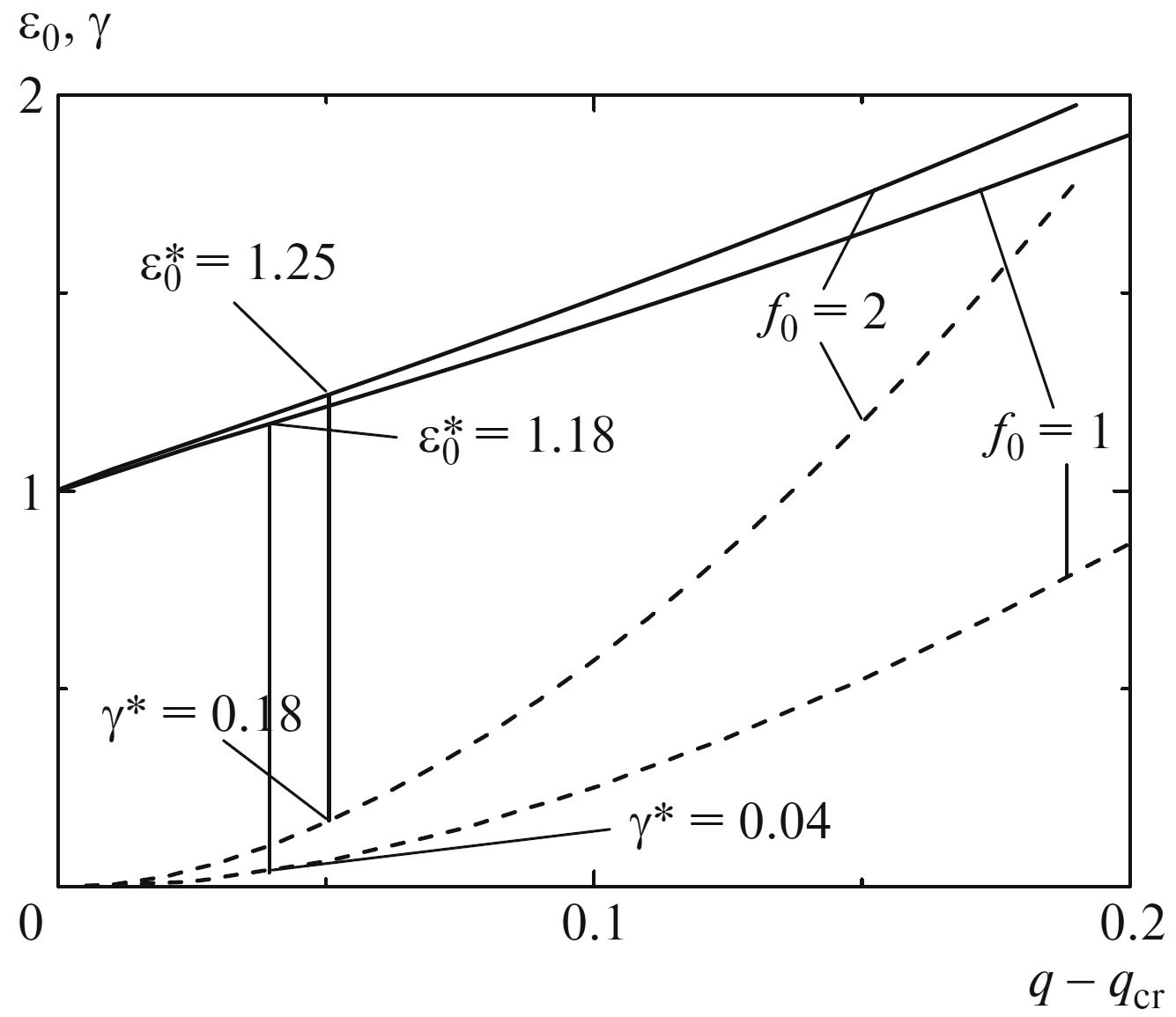}
    \caption{Energy $\eps_0(q, J = 0; R_0)$ of quasidiscrete levels (solid curves) and the widths $\gamma(q, J = 0; R_0)$ of levels (dashed curves) as a function of supercriticality, i.e., as a function of the difference $q - q_\text{cr}$ for $R = 1/20$ for two shapes of the cutoff function of the Coulomb potential. The numbers at the curves are the values of $f_0$ ($f_0 = 1$ corresponds to the uniform distribution of charge over a sphere of radius $R$, and $f_0 = 2$, to the distribution over a ball of the same radius) and the positions $\eps_0^*$ and widths $\gamma^*$ of resonances corresponding to the poles of the scattering matrix (see Fig.~\ref{fig14} below).}
    \label{fig10}
\end{figure}
%%%%%%%%%%%%%%%%%%%%%%%%%%%%%%%%%%%%%%%%%%%%%%%%%%%%%%%%%%%%%%%%%%%%%%%%%%%%%%%%
\subsubsection{Quasidiscrete levels in the lower continuum}\label{part432}
%%%%%%%%%%%%%%%%%%%%%%%%%%%%%%%%%%%%%%%%%%%%%%%%%%%%%%%%%%%%%%%%%%%%%%%%%%%%%%%%
The parameters $\theta_\sigma(J)$ and $\theta_\tau(J)$ completely determine the wave functions and the energy spectrum of the problem (\ref{eq64}) with the Coulomb potential, $V(\rho) = V_C(\rho) = -q/\rho$, for any charge $q = Z\alpha_F$. Let's note, without giving the explicit form of wave functions in the lower continuum $q > q_\text{cr}$, that the boundary conditions (\ref{eqC17}) and expressions (\ref{eq121}), (\ref{eq122}) for the parameter $\theta_\tau(J)$ determine the partial phases $\delta_J(k)$ and the partial matrix $S_J = \exp(2i\delta_J)$ of elastic scattering in closed form \cite{KuleshovMurEtAl2017JETP},
\begin{equation}\label{eq127}
    \exp[2i\delta_J(k)] = \frac{\alpha^* - \beta}{\beta^* - \alpha},\quad
    \alpha = \exp\!\qty(\frac{\pi}{2}\tau - i\eta_\tau)a,\quad
    \beta = \exp\!\qty(-\frac{\pi}{2}\tau - i\eta_\tau)b.
\end{equation}
Here $k = \sqrt{\eps^2 - 1}$,\; $\eps < -1$,\; $\tau = \sqrt{q^2 - J^2} > 0$,\; $q > q_\text{cr}$,
\begin{equation}\label{eq128}
    a = \frac{q\sqrt{-\eps + 1} + (iJ - \tau) \sqrt{-\eps - 1}}
             {\Gamma\qty(1 - i\tau - i\tfrac{\eps}{k}q)},\quad
    b = \frac{q\sqrt{-\eps + 1} - (iJ - \tau) \sqrt{-\eps - 1}}
             {\Gamma\qty(1 - i\tau + i\tfrac{\eps}{k}q)}
\end{equation}
and 
\begin{equation}\label{eq129}
    e^{2i\eta_\tau(J)} = \frac{(2k)^{-i\tau} \Gamma(2i\tau)}{(2k)^{i\tau} \Gamma(-2i\tau)}
        e^{2i\theta_\tau(J)}.
\end{equation}

\begin{figure}[t]
    \begin{subfigure}{0.47\textwidth}
        \centering
        \includegraphics[width=1.0\textwidth]{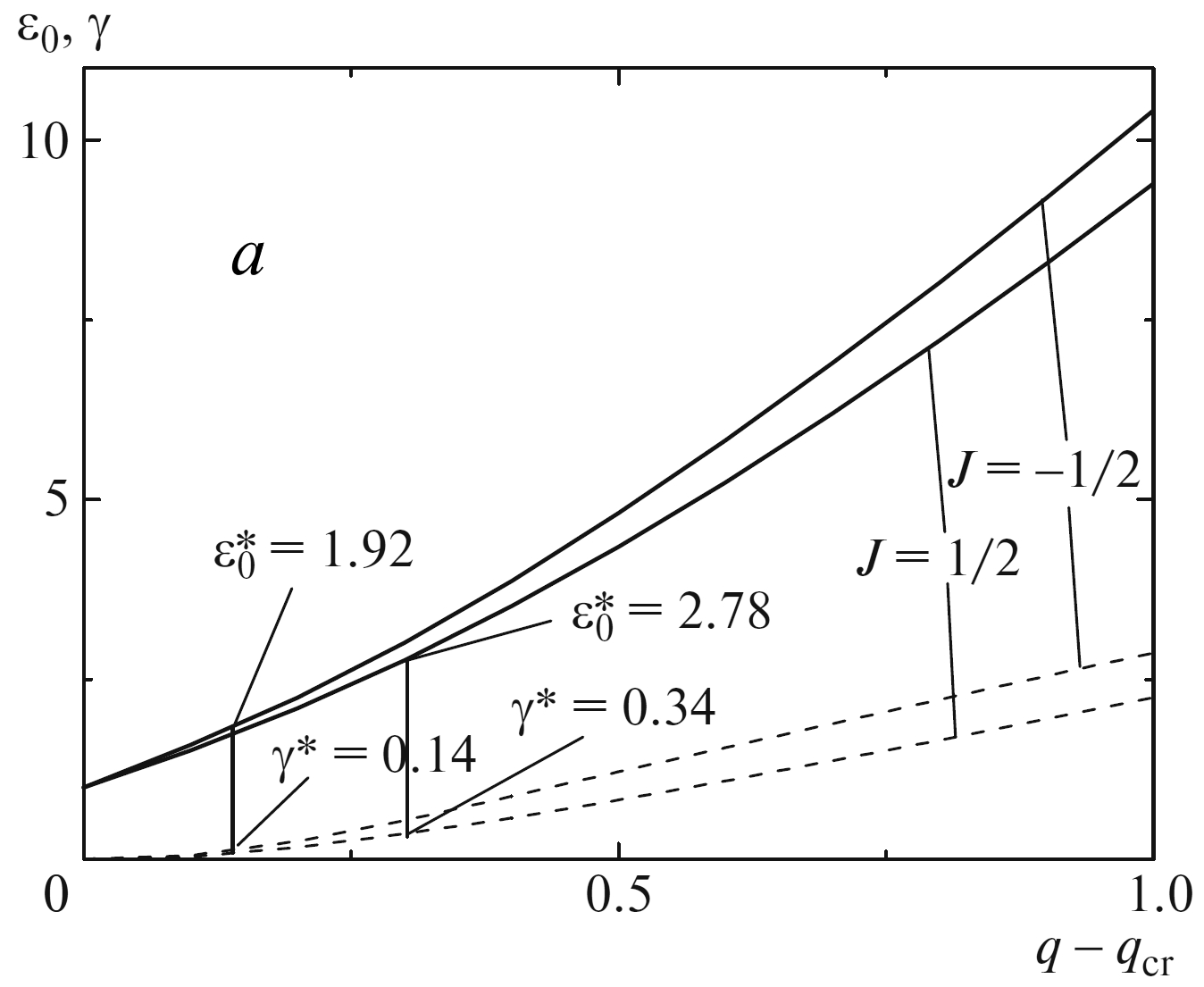}
    \end{subfigure}
    \hfill
    \begin{subfigure}{0.47\textwidth}
        \centering
        \includegraphics[width=1.0\textwidth]{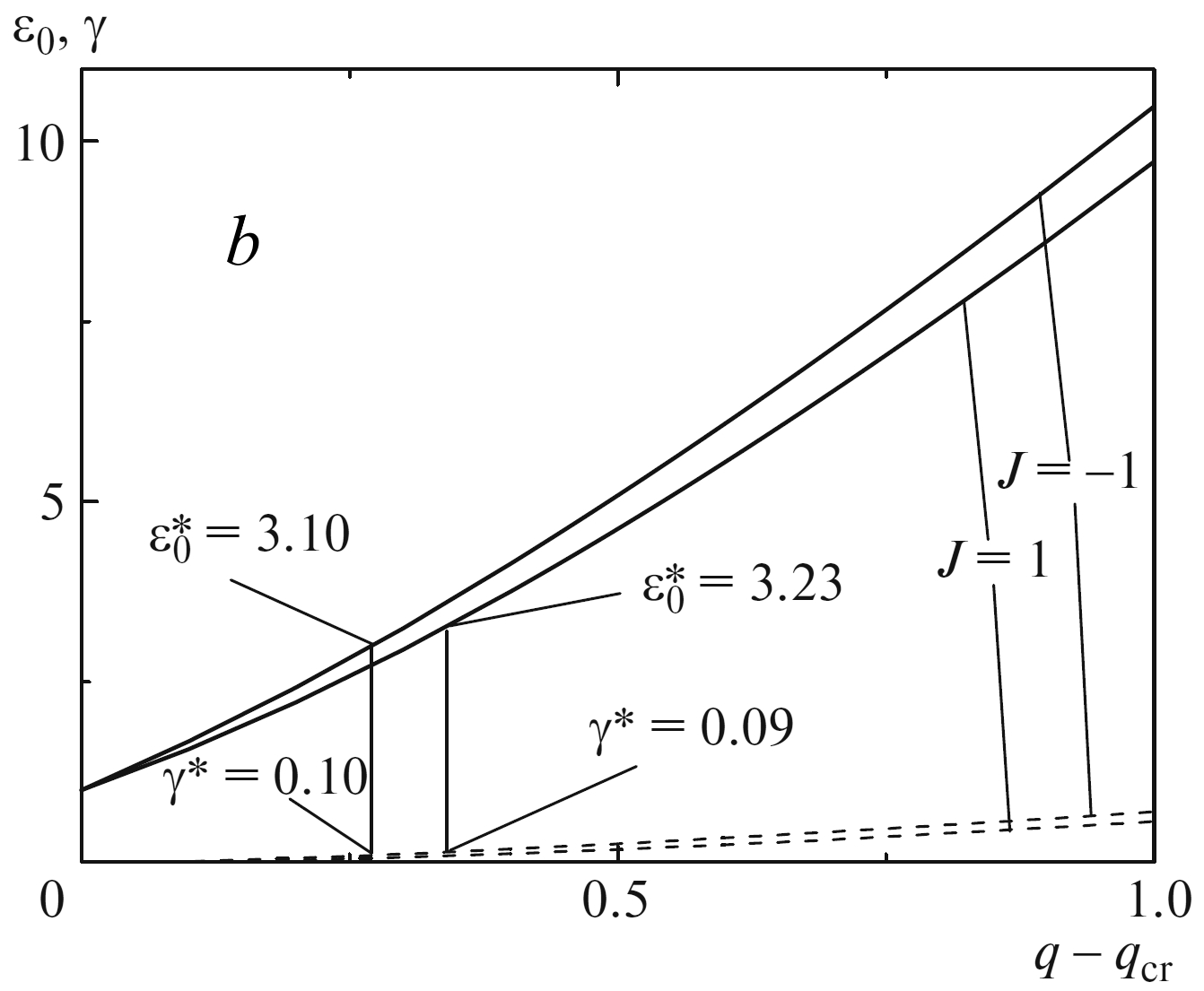}
    \end{subfigure}
    \caption{Energy $\eps_0(q, J; R)$ of quasidiscrete levels (solid curves) and the widths $\gamma(q, J; R)$ of levels (dashed curves) as a function of the difference $q - q_\text{cr}$ for $R = 1/20$. The numbers at the curves indicate the values of the total momentum $J$, as well as the positions $\eps_0^*$ and widths $\gamma^*$ of resonances (see Fig.~\ref{fig15}); (\textit{a}) for $J = 1/2$ at $q - q_\text{cr} = 0.3$ and $J = -1/2$ at $q - q_\text{cr} = 0.16$; (\textit{b}) for $J = 1$ at $q - q_\text{cr} = 0.34$ and $J = -1$ at $q - q_\text{cr} = 0.29$.}
    \label{fig11}
\end{figure}

The scattering matrix poles, i.e. the equation $\exp[-i\delta_J(k)] = 0$, determine the complex energies of the Gamov quasistationary states:
\begin{equation}\label{eq130}
    \frac{\qty[q\sqrt{-\eps + 1} + (iJ - \tau)\sqrt{-\eps - 1}\,]}
         {\qty[q\sqrt{-\eps + 1} + (iJ + \tau)\sqrt{-\eps - 1}\,]}
    \frac{\Gamma\qty(1 + i\tau - i\frac{\eps}{k}q)}{\Gamma\qty(1 - i\tau - i\frac{\eps}{k}q)} =
    e^{-\pi\tau} \exp[2i\eta_\tau(J; R)].
\end{equation}
The solution of this equation determine both the position $\eps_0$ and width $\gamma$ of quasidiscrete level:
\begin{equation}\label{eq131}
    \eps = -\eps_0 + \tfrac{i}{2}\gamma,\quad \eps_0 > 1,\quad \gamma > 0,
\end{equation}
compare to (\ref{eq113}) in \ref{part423}. Fig.~\ref{fig10} (Fig.~3 in \cite{KuleshovMurEtAl2017JETP}) illustrates the dependence of $\eps_0$ and $\gamma$ on the value $(q - q_\text{cr})$ (supercriticality), and also on the cut-off function form at $J = 0$. Fig.~\ref{fig11}\textit{a} (Fig.~4\textit{a} in \cite{KuleshovMurEtAl2017JETP}) gives the dependence of the position $\eps_0$ and width $\gamma$ of quasidiscrete level on supercriticality for $J = 1/2$ and $J = -1/2$ and the cut-off radius $R = 1/20$. Fig.~\ref{fig11}\textit{b} (Fig.~4\textit{b} in \cite{KuleshovMurEtAl2017JETP}) gives the same dependence for $J = 1$ and $J = -1$. From these figures one can see that near the lower continuum boundary, $q \to q_\text{cr}$, the width of quasistationary state is small, $\gamma \ll \eps_0 \sim 1$. This is because the system (\ref{eq64}) with the potential (\ref{eq117}) is equivalent to the problem with low-permeable barrier.

\begin{figure}[t]
    \begin{subfigure}{0.47\textwidth}
        \centering
        \includegraphics[width=1.0\textwidth]{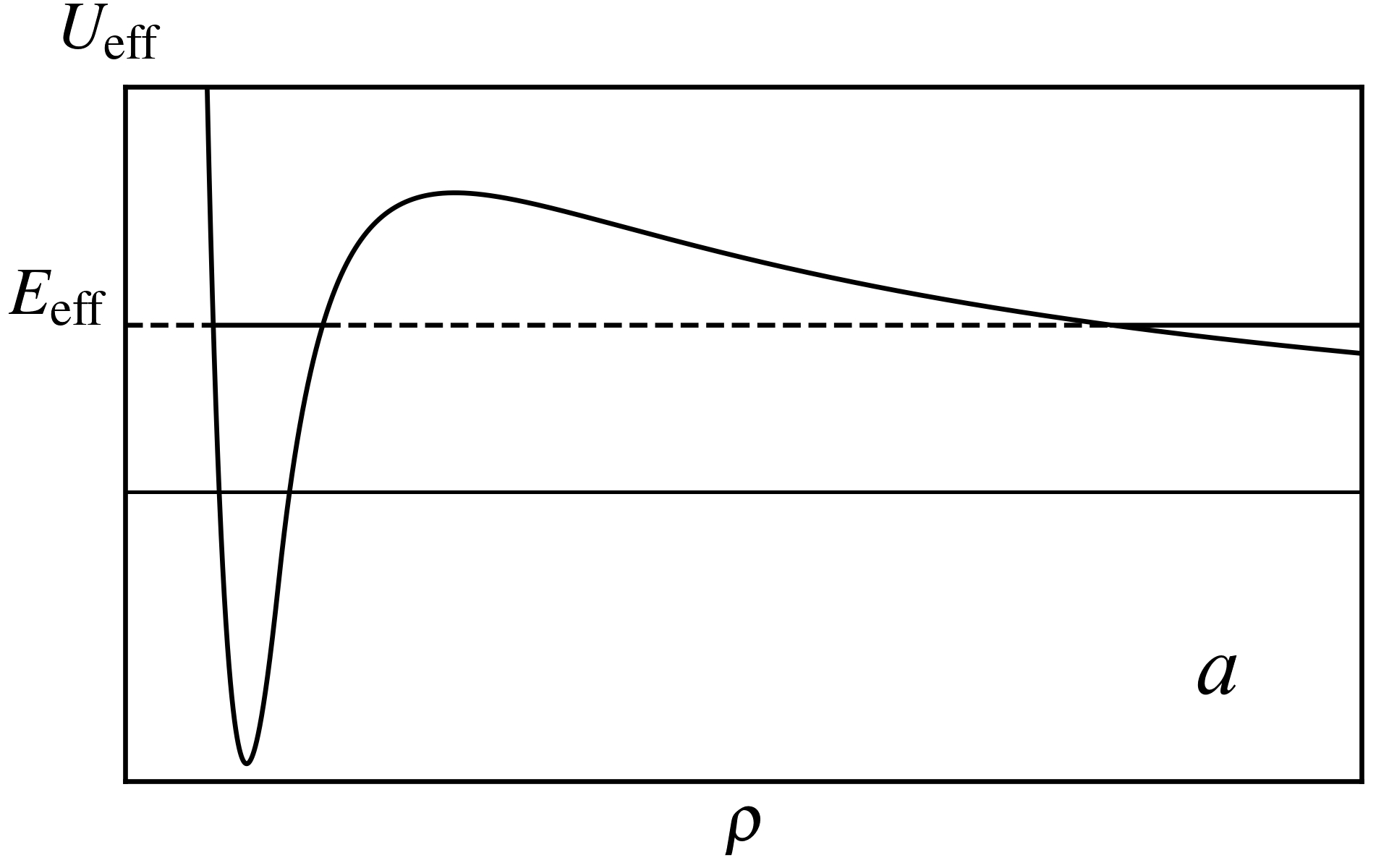}
    \end{subfigure}
    \hfill
    \begin{subfigure}{0.47\textwidth}
        \centering
        \includegraphics[width=1.0\textwidth]{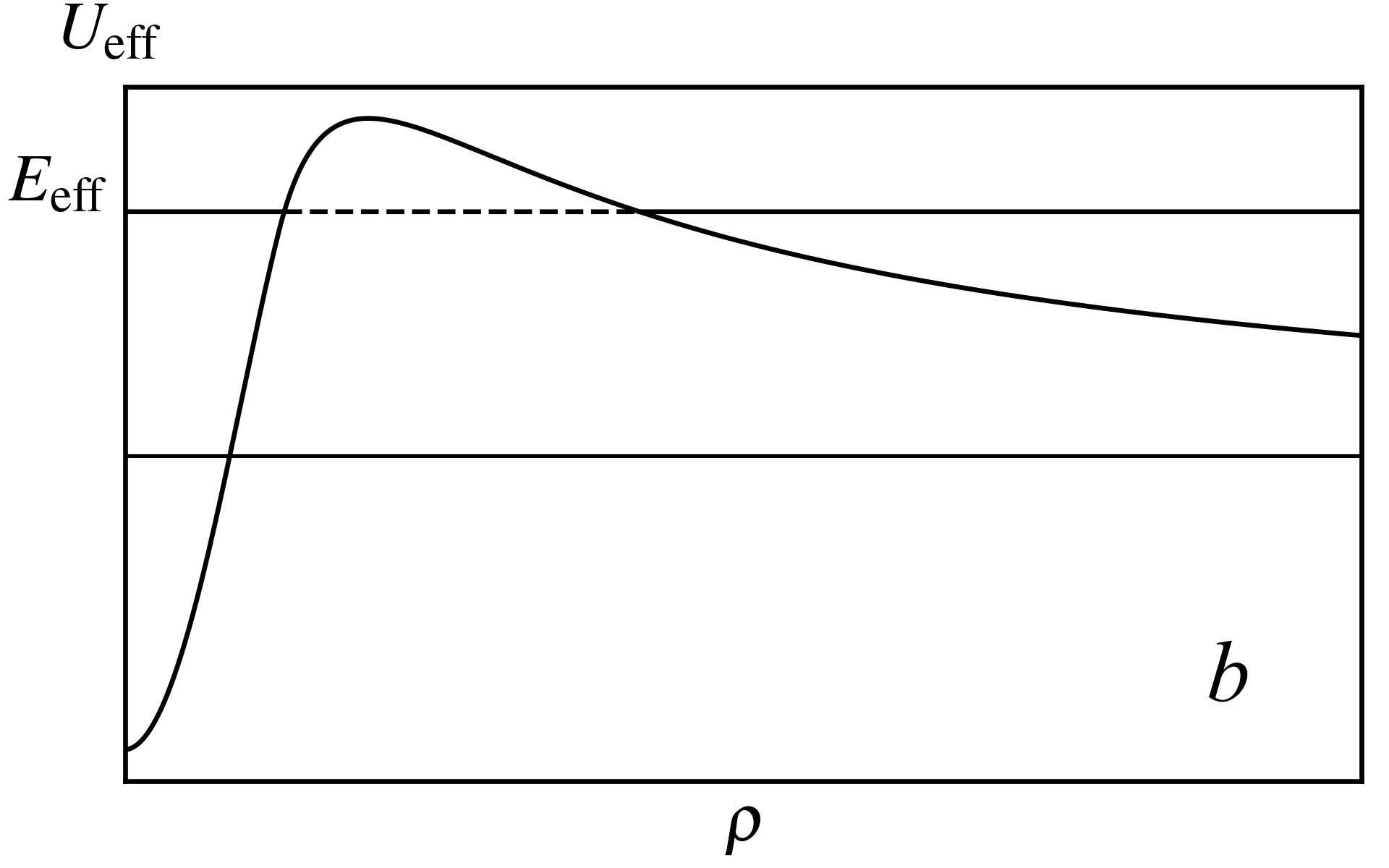}
    \end{subfigure}
    \caption{Effective potential (\ref{eq132}) for states with $\eps < -1$; (\textit{a}) with total momentum $J \ne 0$ and (\textit{b}) with total momentum $J = 0$.}
    \label{fig12}
\end{figure}

Let's illustrate it in the quasiclassical approximation for quasistationary states near the lower continuum of the Dirac equation solutions, see \cite{MurPopovVoskresensky1978JETPLett, MurPopov1978YadFiz}. For the quasiclassical momentum $p(\rho) = \sqrt{2(E_\text{eff} - U_\text{eff})}$, the effective energy is $E_\text{eff} = \frac{1}{2}(\eps^2 - 1)$, and for effective potential we have
\begin{equation}\label{eq132}
    U_\text{eff}(\rho; \eps; J) = -\frac{1}{2} V^2(\rho) + \eps V(\rho) + \frac{J^2}{2\rho^2},
\end{equation}
see Fig.~\ref{fig12} (Fig.~5 in \cite{KuleshovMurEtAl2017JETP}) for attractive potential with the \enquote{Coulomb tail}. Effective potential (\ref{eq132}) corresponds to attraction at short distances for both particles and antiparticles. At the same time, because of the repulsion of antiparticles at large distances ($V(\rho) < 0$,\, $\eps < -1$,\, $\rho \gg R$), in an effective potential (\ref{eq132}) it occurs a low-penetrable Coulomb barrier, so the width of quasistationary level is small, $\gamma \ll \eps_0$, and
\begin{equation}\label{eq133}
    k = k_0' - ik_0'',\quad k_0' = \sqrt{\eps_0^2 - 1} > 0,\quad k_0'' = \frac{\gamma}{2k_0'} > 0.
\end{equation}
So, at large distances with exponential accuracy we have
\begin{equation}\label{eq134}
    \Psi_{\eps, J}(\rho) \sim \exp(ik\rho) = \exp(ik_0'\rho + k_0''\rho),\quad
        \rho \gg \frac{1}{|k|},
\end{equation}
i.e. the Gamov wave function divergent at infinity. The opposite sign of $\gamma$ would mean the presence of a square integrable solution of the equation (\ref{eq64}), i.e. the bound state with complex energy which is not compatible with self-adjointness of the operator $H_{\theta_\tau}$. The unusual sign at width in the expression (\ref{eq131}) ensures the displacement of the discrete level to the unphysical sheet at $q > q_\text{cr}$ and the consistency of the single-particle approximation \cite{MurPopov1976TheorMathPhys_FermionCase}.

\begin{figure}[t]
    \centering
    \includegraphics[width=0.6\textwidth]{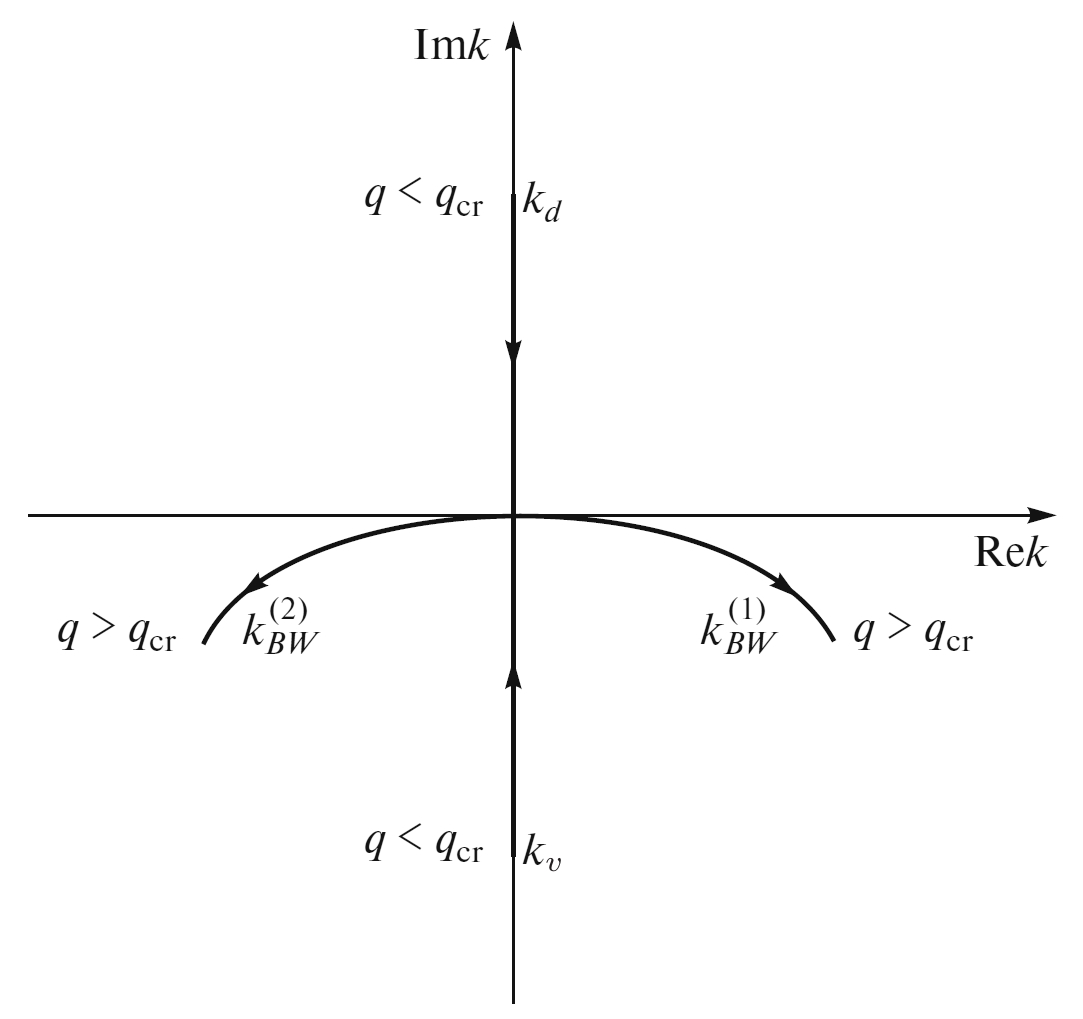}
    \caption{Motion of poles of the $S$-matrix in the complex plane $k = \sqrt{\eps^2 - 1}$ near the boundary of the lower continuum $\eps = -1$. The arrows indicate the direction of motion for increasing charge $q = Ze^2 / \hbar v_F$.}
    \label{fig13}
\end{figure}

The trajectories of the poles of the partial scattering matrix $S_J(k; q)$ in the plane of the complex variable $k = \sqrt{\eps^2 - 1}$ near the boundary of the lower continuum with an increase of the charge $q$ are shown in Fig.~\ref{fig13} (Fig.~6 in \cite{KuleshovMurEtAl2017JETP}). At $q < q_\text{cr}(J)$ the energy of the discrete level falls within the range $-1 < \eps_d < 1$, hence $k_d = i\lambda_d$,\, $\lambda_d > 0$ and the pole of the $S$-matrix corresponding to it is located on the imaginary axis of the upper $k$ half-plane, i.e. on the first (physical) sheet.

With an increase in the charge, this level approaches the boundary of the lower continuum, $\eps = -1$, whereas the virtual level $k_v = -i\lambda_v$,\, $\lambda_v > 0$ located on the second (unphysical) sheet moves toward it also along the imaginary axis. At $q = q_\text{cr}(J)$ they collide and transform to a pair of divergent at $q > q_\text{cr}(J)$ Breit--Wigner poles $k_\text{BW}^\Par{1,2}$ located on the unphysical sheet \footnote{compare with Fig.~\ref{fig5} in \ref{part421}. The difference between scattering on the short-range potential and potential with the Coulomb tail is that in first case the width of the quasistationary state is provided by a centrifugal barrier, and in second~--- by the Coulomb barrier which does not depend on $J$.}. In the overcritical range, the pole $k_\text{BW}^\Par{1} \equiv k_\text{BW}$ with energy (\ref{eq131}) which is the nearest to the positive real semiaxis, corresponds to the quasistationary state of a hole.

Indeed, according to Feynman, we should treat electrons with negative energy as those moving back in time, so that the time-dependent Gamov wavefunction
\[
    \Psi(t) \sim e^{-i\eps t} \equiv \exp[-i(-\eps)(-t)] =
    \exp[-i\eps_0(-t) - \tfrac{1}{2}\gamma (-t)]
\]
decreases with an increase in $(-t)$. Such electrons correspond to holes with positive energy in the Dirac sea. This restores the conventional interpretation of the quasistationary states with negative energy in the lower continuum.

This conclusion can be approached from the other side. Accounting for the modification (\ref{eq117}) of the Coulomb potential, Hamiltonian (\ref{eqC3}) of the set (\ref{eq64}) is self-adjoint operator and its eigenfunctions form the complete set. Since the electronic spectrum, $-1 < \eps < \infty$, does not overlap\footnote{At $Z = Z_\text{cr}$ the electron state with $\eps_d = -1$ is discrete, and states with $\eps < -1$ belong to the continuous spectrum. That's why even in this case electronic spectrum does not overlap with the hole one, $\ol{\eps} = -\eps > 1$.} with the hole spectrum, $1 \le \ol{\eps} < \infty$,\, $\ol{\eps} = -\eps$ at $\eps \le -1$. Therefore, the second quantization according to Furry \cite{Furry1951PhysRev} is an allowable procedure based on the
given complete set of functions, as in the case of short-range potential, see section \ref{part423}. Then, the complex
energy of the hole quasistationary state is given by the expression
\begin{equation}\label{eq135}
    \eps_p = \eps_0 - \tfrac{i}{2}\gamma,\quad \eps_0 > 0,\quad \gamma > 0,
\end{equation}
which is a usual expression for the energy of a quasistationary level. Such quasistationary states may manifest themselves as resonances in the scattering of holes on the supercritical impurity.
%%%%%%%%%%%%%%%%%%%%%%%%%%%%%%%%%%%%%%%%%%%%%%%%%%%%%%%%%%%%%%%%%%%%%%%%%%%%%%%%
\subsubsection{Resonant hole scattering by supercritical impurity}\label{part433}
%%%%%%%%%%%%%%%%%%%%%%%%%%%%%%%%%%%%%%%%%%%%%%%%%%%%%%%%%%%%%%%%%%%%%%%%%%%%%%%%
\begin{figure}[t]
    \centering
    \includegraphics[width=0.6\textwidth]{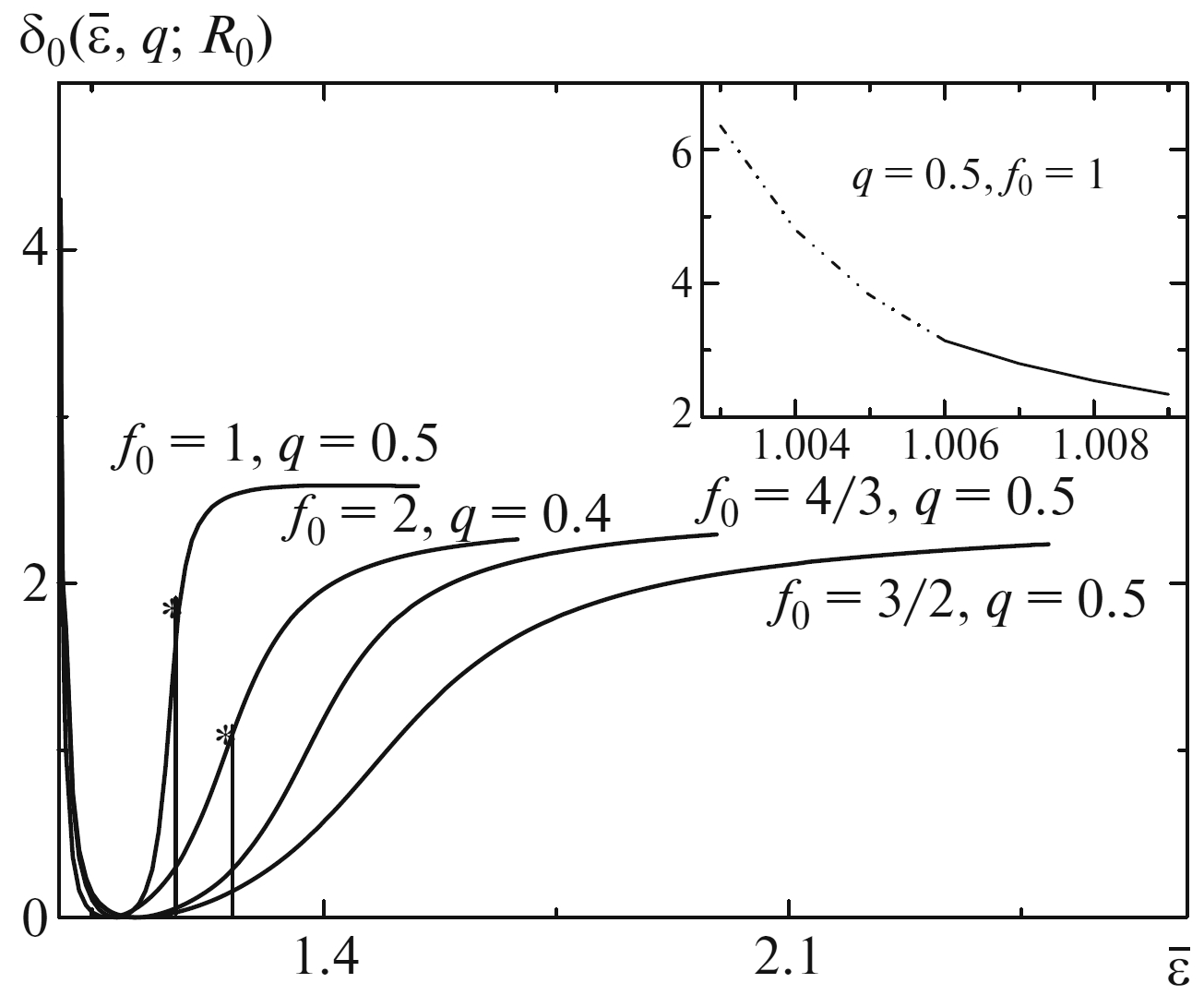}
    \caption{Scattering phases $\delta_0(\ol{\eps}, q ; R_0)$ for the states with total momentum $J = 0$ as a function of the hole energy for several shapes of the cutoff function for $R = 1/20$. The numbers at the curves are the values of charge $q = Z \alpha_F$ and $f_0$. The asterisks indicate the positions of resonances in the scattering of holes by impurity (see Fig.~\ref{fig10}). The dot-and-dash curve in the inset represents the asymptotics (\ref{eq137}).}
    \label{fig14}
\end{figure}

Since the ground level is the first to descend to the boundary of the lower continuum, we begin the discussion of scattering of holes with energy $\ol{\eps} = -\eps > 1$ with states with a total angular momentum $J = 0$ corresponding to a half-integer orbital angular momentum $M = -1/2$. The results of calculating the \cite{KuleshovMurEtAl2017JETP} phase $\delta_0(\ol{\eps}, q; R)$ for various forms of the trimming function are presented in Fig.~\ref{fig14} (Fig.~7 in \cite{KuleshovMurEtAl2017JETP}).

\begin{figure}[t]
    \begin{subfigure}{0.47\textwidth}
        \centering
        \includegraphics[width=1.0\textwidth]{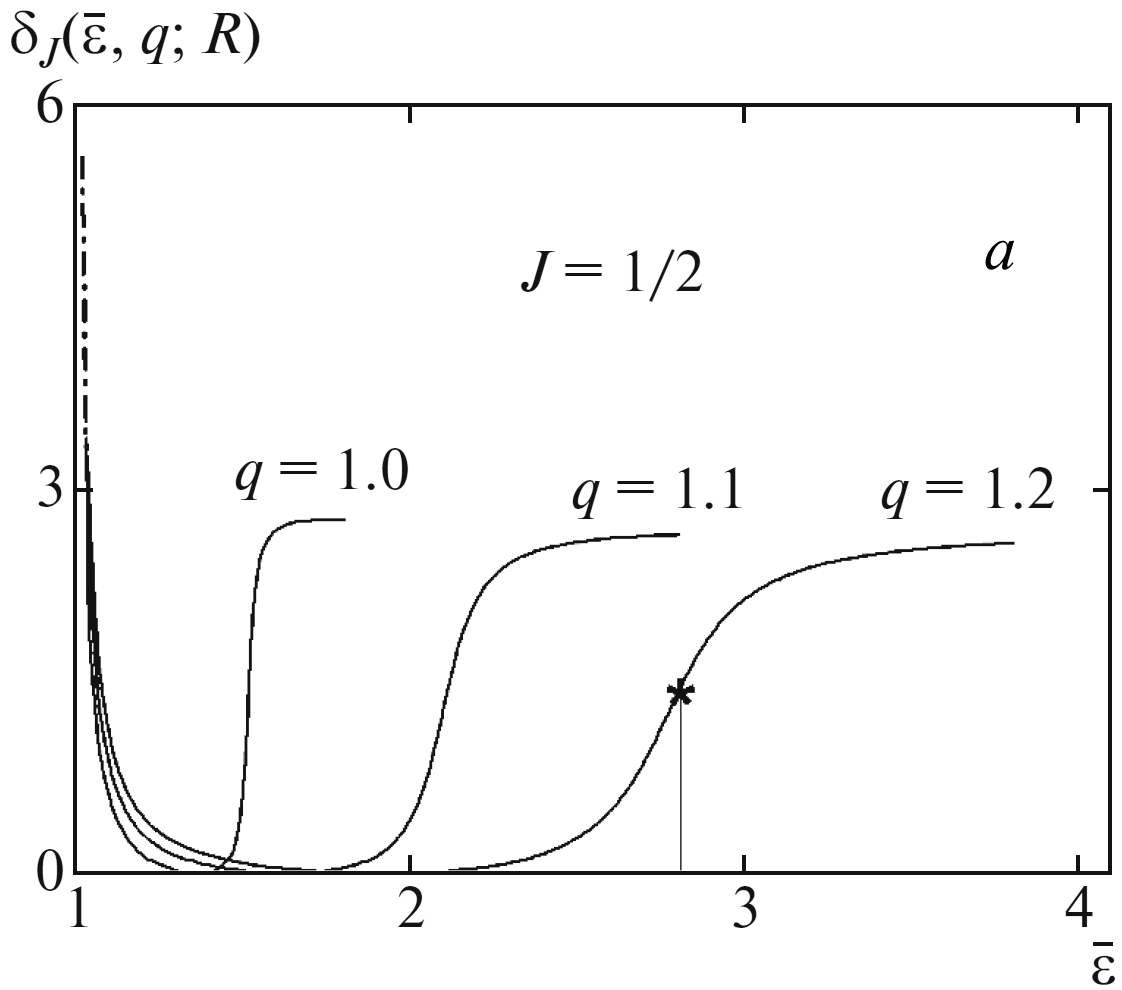}
    \end{subfigure}
    \hfill
    \begin{subfigure}{0.47\textwidth}
        \centering
        \includegraphics[width=1.0\textwidth]{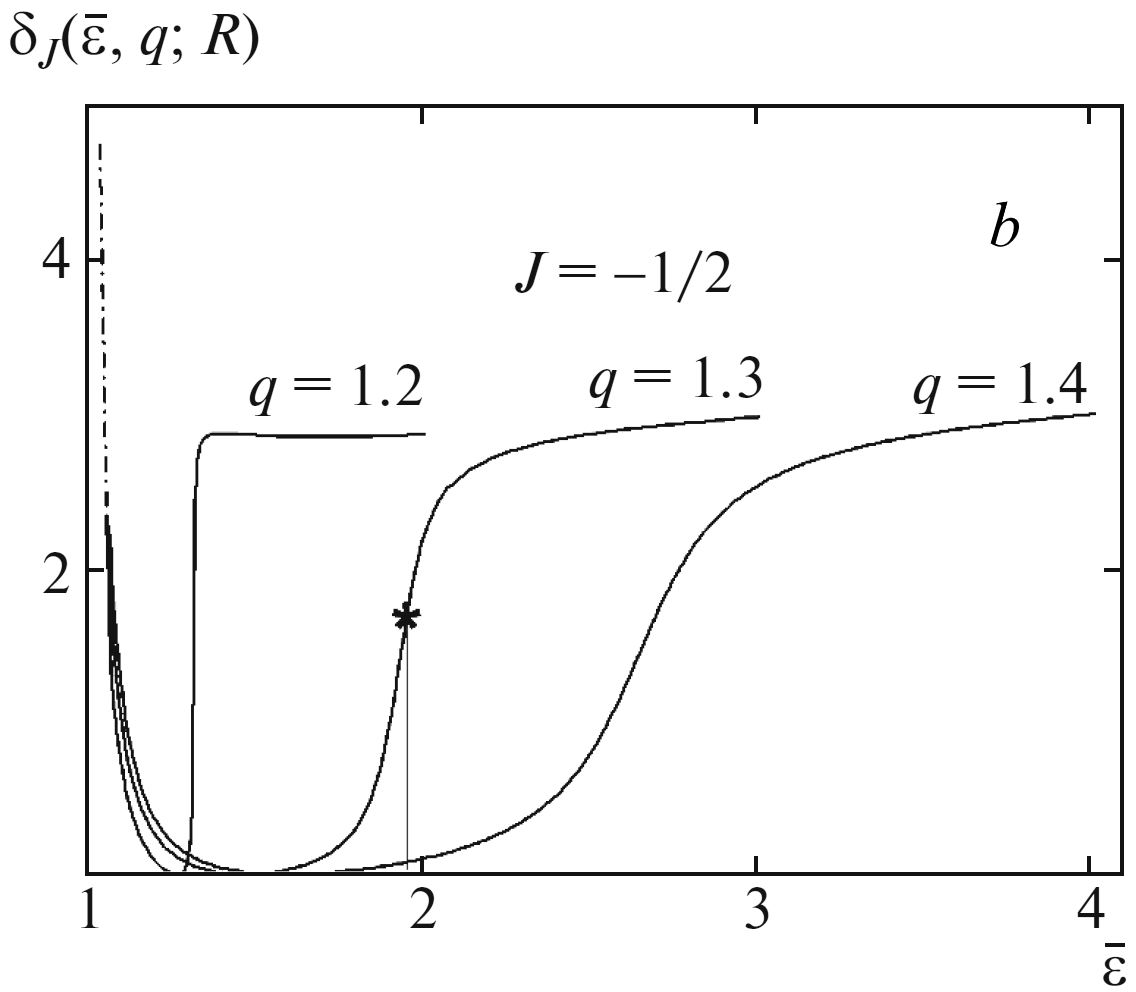}
    \end{subfigure}

    \begin{subfigure}{0.47\textwidth}
        \centering
        \includegraphics[width=1.0\textwidth]{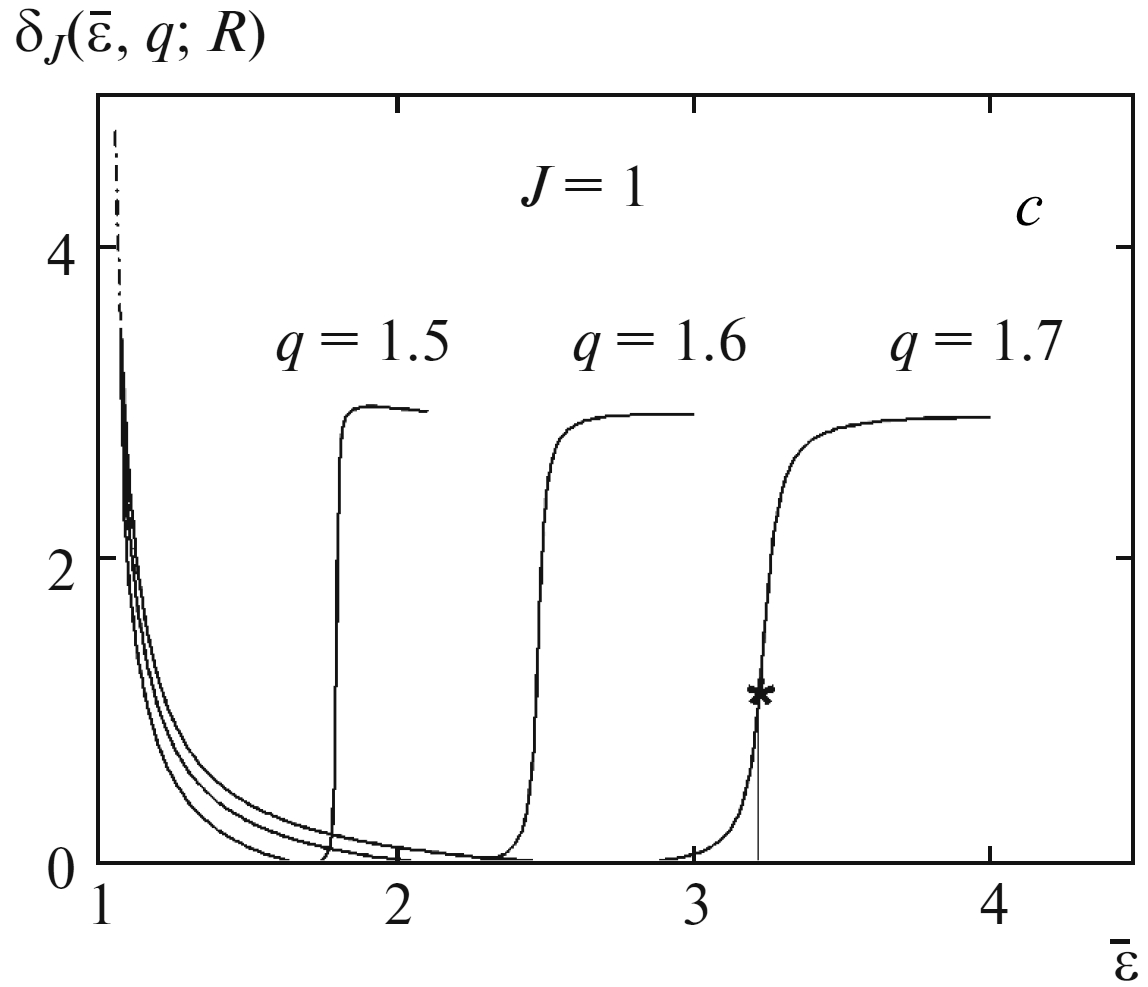}
    \end{subfigure}
    \hfill
    \begin{subfigure}{0.47\textwidth}
        \centering
        \includegraphics[width=1.0\textwidth]{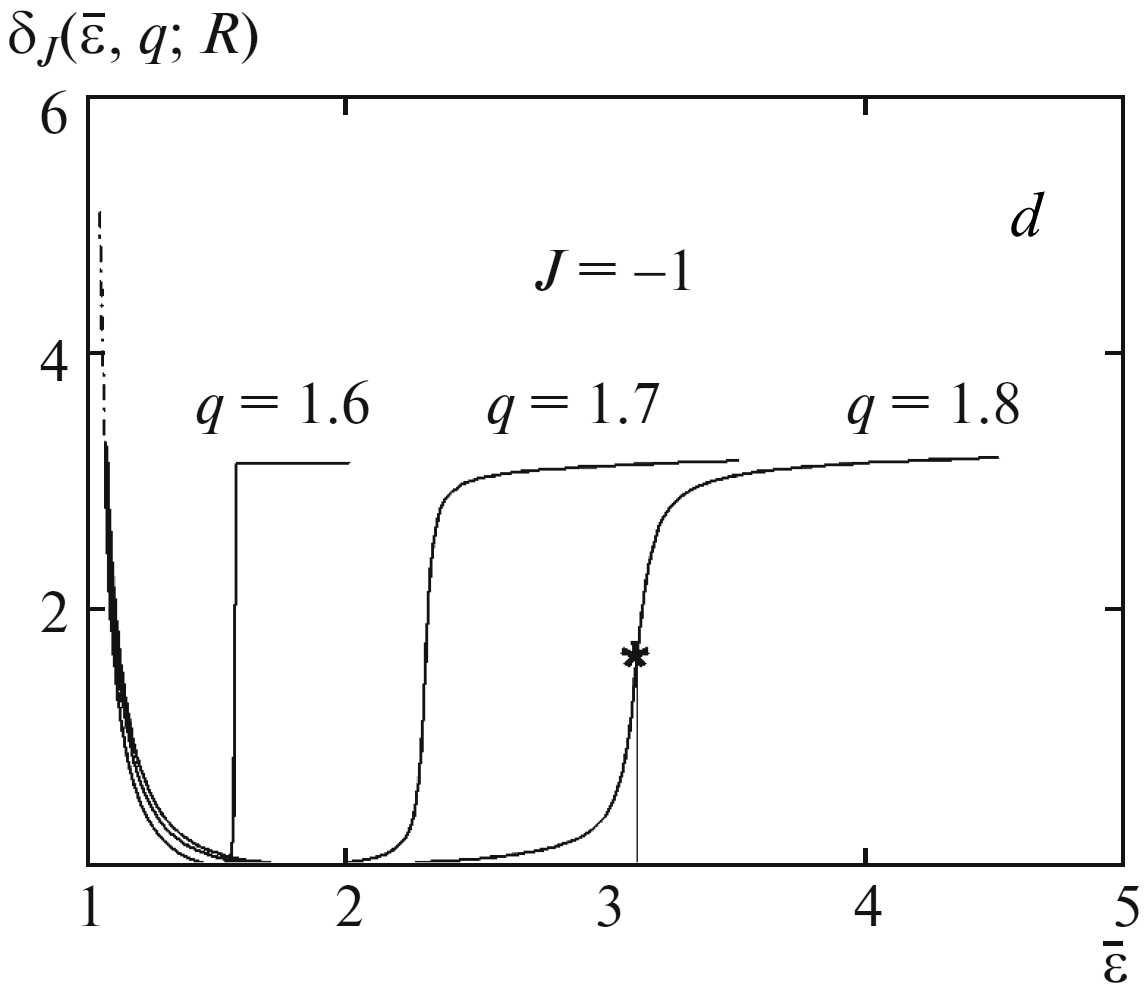}
    \end{subfigure}
    \caption{Scattering phases $\delta_J(\ol{\eps}, q; R)$ as a function of the hole energy for the values of the total momentum (\textit{a}) $J = 1/2$, (\textit{b}) $J = -1/2$, (\textit{c}) $J = 1$, and (\textit{d}) $J = -1$. The asterisks indicate the resonances in the scattering of holes by impurity; the positions and widths of the resonances are shown in Fig.~\ref{fig11}. The dot-and-dash curve corresponds to the asymptotics (\ref{eq137}).}
    \label{fig15}
\end{figure}

The scattering phases $\delta_J(\ol{\eps}, q; R)$ with $J = \pm 1/2$ and $J = \pm 1$ for several impurity charge values $q = Z\alpha_F$ are shown in Fig.~\ref{fig15} (Fig.~8 in \cite{KuleshovMurEtAl2017JETP}). The specific values of $q$ are chosen so that the integer values of the impurity charge $Z$ are as close as possible to the critical values $Z_\text{cr}$ shown in Table~\ref{tab2} for graphene on the SiC substrate, when $\alpha_F = 0.4$ is considered, for definiteness, that effective dielectric constant $\epsilon = 1$.

As in the nonrelativistic theory of scattering, see Chap.~13 in the monograph \cite{Taylor1972}, quasistationary states can manifest themselves as resonances in hole scattering. The positions $\eps_0^*$ of these resonances and their widths $\gamma^*$ are shown above in Fig.~\ref{fig10} and \ref{fig11}. If the hole energy $\ol{\eps} > 0$ falls in the region of a sharp change in the scattering phase, then a resonance arises in its scattering, and the partial cross section corresponds to the Breit-Wigner formula
\begin{equation}\label{eq136}
    \sigma_J(\ol{\eps}) = \sin^2\delta_J =
        \frac{(\gamma^*/2)^2}{(\ol{\eps} - \eps^*)^2 + (\gamma^*/2)^2},
\end{equation}
in complete analogy with scattering by a short-range potential, cf. with (\ref{eq116}) in the \ref{part423} section. Moreover, with an increase in supercriticality, the phase change becomes smoother, see Fig.~\ref{fig14} and \ref{fig15}.

For the asymptotic behavior of the scattering phases for small values of the wave vectors, $k = \sqrt{\eps^2 - 1} \ll 1$, we have \cite{KuleshovMurEtAl2017JETP}
\begin{equation}\label{eq137}
    % Исправлена опечатка в KuleshovMurEtAl2017JETP - неправильный знак перед последним членом
    \delta_J(k; q) \simeq \frac{q}{k} \qty(\ln\frac{q}{k} - 1) + \frac{\pi}{4} -
        \frac{1}{2}kq\ln k + O(k),\quad k \ll 1.
\end{equation}
Since such values of $k$ correspond to large distances, $\rho \gg 1$, in this case the scattering phases are the same for all values of the total angular momentum $J$, since the centrifugal potential \enquote{dies out}. In addition, at such distances the Coulomb tail of the potential $V_R(\rho)$ dominates and the scattering phases are independent of the regularization (\ref{eq117}). Asymptotics (\ref{eq137}) are shown by dashed lines in Fig.~\ref{fig14} and \ref{fig15}.

It should be noted that the effect of supercharged Coulomb impurity on a system of two-dimensional massive Dirac particles in graphene with a gap in the electronic spectrum is considered in \cite{PereiraKotovCastroNeto2008PhysRevB}. It discusses, in particular, the screening of an impurity charge by electrons generated together with holes from the Dirac Sea, in complete analogy with the spontaneous production of electron-positron pairs in $Z > Z_\text{cr}$ in the relativistic Coulomb problem discussed in \cite{GershteinZeldovich1970JETP, Popov1970JETPLett_SovJNuclPhys}, a review of \cite{ZeldovichPopov1972SovPhysUsp} , the monograph \cite{GreinerMullerRafelski1985} and the numerous literature references indicated therein.

However, as shown in \cite{KuleshovMurEtAl2015PhysicsUspekhi}, the one-particle approximation for the Dirac equation is valid not only for $Z \le Z_\text{cr}$, but also for $Z > Z_\text{cr}$, so that spontaneous production of $e^+e^-$ pairs is absent. A similar statement applies to the two-dimensional effective Dirac equation with a gap in the electronic spectrum, so that the mechanism for screening the charge of a supercharged impurity indicated in \cite{PereiraKotovCastroNeto2008PhysRevB} according to the scenario described in the review \cite{ZeldovichPopov1972SovPhysUsp} cannot be realized \cite{KuleshovMurNarozhny2015JETPLett, KuleshovMurEtAl2017JETP}.

The conclusion about the absence of spontaneous production of pairs of antiparticle particles in the Coulomb problem for $Z > Z_\text{cr}$ is based on the unitarity of the partial elastic scattering matrix. Since the amplitudes of the incident and diverging waves with the total angular momentum $J$ are the same, inelastic channels with this $J$, including spontaneous pair production, are absent. This supercritical radial Coulomb problem differs significantly from the situation in the case of the \enquote{Klein paradox} \cite{Klein1926ZPhys, SauterZPhys1931_69_1932_73} or a constant uniform electric field \cite{HeisenbergEuler1936ZPhys, Schwinger1951PhysRev}, when the upper and lower continua overlap. Qualitatively, this difference is due to the fact that a homogeneous field breaks the virtually generated pair, while in the radial case both the particle and antiparticle are attracted to the center at small distances, see the effective semiclassical potential (\ref{eq132}), and annihilate.

As already mentioned in the \ref{part423} section, the authors of \cite{GodunovMachetVysotsky2017EurPhysJ} disagree with these arguments, who believe that it is the resonant scattering by the supercritical nucleus that indicates the spontaneous production of $e^+e^-$ pairs according to the mechanism described in the review of \cite{ZeldovichPopov1972SovPhysUsp} and monographs \cite{GreinerMullerRafelski1985}. However, as shown in the \ref{part42} section using an example of a deep narrow rectangular well, resonance scattering occurs near both the lower and upper continua, when a discrete level is immersed or pushed into the continuous spectrum when the depth of the well changes. The modified Coulomb potential (\ref{eq117}) differs from the short-range potential (\ref{eq90}) by the \enquote{Coulomb tail}, which provides a small resonance width that is not related to the presence of a centrifugal barrier, but does not qualitatively change the picture, see Fig.~\ref{fig12}. Therefore, there is no reason to believe that resonance scattering near the lower continuum boundary is associated with spontaneous pair production, i.e. that the one-particle Dirac equation is not applicable in the supercritical region, $Z > Z_\text{cr}$.

Moreover, in the work of \cite{WangWongShytov2013Science} it was actually shown that the single-particle approximation for the effective two-dimensional Dirac equation describing the electronic properties of gapless graphene in the presence of supercharged $Z > Z_s = |J|\alpha_F^{-1}$ impurity at $J = 1/2$ agrees with the experimental data on the spectra of current-voltage characteristics obtained by scanning tunneling spectroscopy. As emphasized in the \cite{Morgenstern2011PhysStatSolB} review, this method allows one to determine the electronic structure of graphene near a charged impurity, if it is described by a single-particle wave function. Since in the gapless effective Dirac equation, i.e. for massless charge carriers in graphene, there is no discrete spectrum, then the singular impurity charge $q_s = Z_s \alpha_F = |J|$ plays a special role here, see \ref{appendixC}, and not its critical value, as in the massive case. A qualitative explanation of this can be obtained again in the semiclassical approximation.

As for massive fermions, the expression (\ref{eq132}) for the effective potential $U_\text{eff}(\rho; \eps; J)$ is still valid in the considered problem, but the effective energy in the massless case is $E_\text{eff} = \frac{1}{2}\eps^2$. If in (\ref{eq132}) we take the Coulomb potential of attraction, $V_C(\rho) = -q/\rho$ as $V(\rho)$, then when $q > q_s = |J|$ in the effective potential there arises the well-known \cite{LandauLifshitz3_1981Butterworth} in the nonrelativistic theory of “fall on the center”. Related to this is the need to modify the Coulomb potential at small distances, as in the relativistic Coulomb problem \cite{PomeranchukSmorodinsky1945JPhysUSSR}.

As well as there, with $q > q_s$, effective attraction occurs at small distances of both electrons and holes, and with $\eps < -1$ the Coulomb barrier is hardly permeable for the latter near the lower continuum boundary, see Fig.~\ref{fig12}. Resonant scattering of a massless hole with an energy $\ol{\eps} = -\eps$ appears when its energy $\ol{\eps}$ is close to the position of the quasidiscrete level, $\ol{\eps} \simeq \eps_0$, when quantized taking into account the permeability of the barrier \cite{MurPopov1990JETPLetters, PopovMurSergeev1991JETP}.

Unfortunately, in the gapless case, the scattering phase for states with angular momentum $J = 0$ is a smooth function of the hole energy $\ol{\eps} = k$. So, for example, for rectangular cropping, we have
\[
    \delta_0(k; R) = Z\alpha_F \qty[\ln(2kR) - 1],
\]
which does not lead to a peak in the current-voltage characteristics. At the same time, in graphene with a gap in the electronic spectrum with a small value of supercriticality, such a phase is resonant, see Fig.~\ref{fig10}. Therefore, the measurement of the $dI/dV$ spectra near the Dirac point can answer the question of whether the value $M = -1/2$ is realized in graphene with a gap in the electronic spectrum, i.e. half-integer quantization of the orbital angular momentum. The same conclusion applies to other integer values of the total angular momentum $J$ corresponding to half-integer values of the orbital angular momentum in graphene.
%%%%%%%%%%%%%%%%%%%%%%%%%%%%%%%%%%%%%%%%%%%%%%%%%%%%%%%%%%%%%%%%%%%%%%%%%%%%%%%%
\section{Conclusions}\label{conclusions}
%%%%%%%%%%%%%%%%%%%%%%%%%%%%%%%%%%%%%%%%%%%%%%%%%%%%%%%%%%%%%%%%%%%%%%%%%%%%%%%%
\textbf{1}. Fractional values of orbital angular momentum may occur in classical and quantum two-dimensional problems where variables are separable in cylindrical coordinates. Such values are concerned with the topological phase $\theta$ acquired by a wave function $\Psi(\varphi)$ under a full rotation, $0 \le \varphi \le 2\pi$, see (\ref{eq2}) in section \ref{introduction}. This non-trivial topological phase is associated with multi-valued irreducible representations of 2D rotational group SO(2). In T-invariant quantum systems only integer and half-integer values may occur.

\textbf{2}. D. van Vleck paid attention to the existence of half-integer momenta in molecules in his paper \cite{Vleck1929PhysRev}. In \cite{BuschDevEckelEtAl1998PhysRevLett} the half-integer quantization of the orbital angular momentum associated with the Berry geometrical phase \cite{Berry1984ProcRSocLondA, Simon1983PhysRevLett} equal to $\pi$ was experimentally observed. This phase is acquired by a nuclear wave function under a pseudo-rotation around the equilateral configuration of Na\textsubscript{3} molecule. In molecules, the Berry phase is associated with the monopole induced by the electronic subsystem \cite{MoodyShapereWilczek1986PhysRevLett}.

\textbf{3}. Geometric (topological) Berry phase \cite{Berry1984ProcRSocLondA} usually results from the evolution determined by a time-dependent Hamiltonian. In two-dimensional axial systems topological phase itself detaches a generator $L_\theta$ from the family of self-adjoint operators and, hence, determines the rotational dynamics of the system, i.e. the unitary operator of finite rotations, see (\ref{eq3}). Thus, half-integer quantization of the orbital angular momentum in circular quantum dots does not require the existence of monopole.

\textbf{4}. The issue about the relation between quantization of orbital angular momentum in two- and three-dimensional Euclidean spaces was first addressed by Pauli in \cite{Pauli1939HelvPhysActa}, see also section \ref{introduction} in the present paper. Moreover, the specific value of the topological phase $\theta$ ($0$ or $\pi$) in circular quantum dots with different numbers of electrons is fixed by the Pauli exclusion principle. Thus, the phase in the boundary condition (\ref{eq2}) may legitimately be called the Pauli’s topological phase.

\textbf{5}. Hence, the experimental data \cite{SchmidtTewordtEtAl1995PhysRevB} for the ground-state energy of few-electron circular quantum dots in a strong magnetic field show the existence of the non-trivial Pauli’s topological phase.

To confirm its existence in a gapped graphene, it is essential to determine its electronic structure near the impurity with $Z > Z_\text{cr}$, or with $Z > Z_s$ in a gapless case, using the method of scanning tunnel spectroscopy, see \cite{WangWongShytov2013Science, Morgenstern2011PhysStatSolB}. The alternative method is to measure few Coulomb energy levels using photoelectronic spectroscopy with high angular resolution \cite{ZhouGweonEtAl2007NatureMaterials, SeyllerBostwickEtAl2008PhysStatSolB}.

Furthermore, in these experiments it can be determined if the overcritical impurity is screened by electron-hole pair creation (following \cite{PereiraKotovCastroNeto2008PhysRevB, ZeldovichPopov1972SovPhysUsp}) or not.
%%%%%%%%%%%%%%%%%%%%%%%%%%%%%%%%%%%%%%%%%%%%%%%%%%%%%%%%%%%%%%%%%%%%%%%%%%%%%%%%
\section*{Acknowledgements}
%%%%%%%%%%%%%%%%%%%%%%%%%%%%%%%%%%%%%%%%%%%%%%%%%%%%%%%%%%%%%%%%%%%%%%%%%%%%%%%%
We are grateful to A.M. Fedotov, B.M. Karnakov, and V.P. Yakovlev for fruitful discussions, and especially to our friend and teacher N.B. Narozhny, who passed away leaving us at an early stage of this work. The work was partially supported by the Russian Foundation for Basic Research (Grant 19-02-00643a). Yu.E.L. was supported by the Russian Foundation for Basic Research (Grants 20-02-00410 and 20-52-00035).
%%%%%%%%%%%%%%%%%%%%%%%%%%%%%%%%%%%%%%%%%%%%%%%%%%%%%%%%%%%%%%%%%%%%%%%%%%%%%%%%
\appendix
\section{The translation generator on an interval}\label{appendixA}
%%%%%%%%%%%%%%%%%%%%%%%%%%%%%%%%%%%%%%%%%%%%%%%%%%%%%%%%%%%%%%%%%%%%%%%%%%%%%%%%
It’s known for a long time \cite{Neumann1955Princeton}, that the differential operator of infinitesimal translation on an interval $[a,b]$,
\[
    S = -i \dv{q},\quad a \le q \le b,
\]
becomes a self-adjoint operator $S_\theta = S_\theta^+$ on the Hilbert space $\Hilbert = \Ltwo{[a,b]}$ of square-integrable on $[a,b]$ wave functions with hermitian scalar product
\[
    \Scalar{\Phi}{\Psi} = \Int{a}{b} \Phi^*(q) \Psi(q) \dd{q}
\]
under the following boundary condition
\begin{equation}\label{eqA1}
    \Psi(b) = e^{i\theta}\Psi(a),\quad 0 \le \theta < 2\pi.
\end{equation}

Indeed, following \cite{Wightman1967Cargese} let us define hermitian (symmetric) operator
\begin{equation}\label{eqA2}
    S \chi(q) = -i \chi'(q),\quad
    \Domain{S} = \qty{\chi \in \Hilbert,\, \chi' \in \Hilbert;\, \chi \in \Cinf{[a,b]}},
\end{equation}
where $\Cinf{[a,b]}$ is a class of infinitely differentiable finite functions \cite{Richtmyer1978Springer} on $[a,b]$. If the functions $\chi_1(q)$ and $\chi_2(q)$ are in the domain $\Domain{S}$ of the operator $S$, then
\[
    \Scalar{\chi_1}{S\chi_2} - \Scalar{S\chi_1}{\chi_2} = -i\chi_1^*(q)\chi_2(q)\Big|_a^b = 0.
\]
According to the definition of the conjugate $S^+$ of the operator $S$
\begin{equation}\label{eqA3}
    \Scalar{S^+X}{\chi} = \Scalar{X}{S\chi},\quad
    \chi \in \Domain{S},\quad X \in \Domain{S^+} = \Hilbert,
\end{equation}
we have $S \subset S^+$, i.e. the operator $S$ is not self-adjoint but only symmetric, or hermitian.

Because differentiation operator $S$ is an infinitesimal translation operator and $\chi(q)$ are infinitely differentiable functions, then the equation
\begin{equation}\label{eqA4}
    e^{i\alpha S} \chi(q) = \chi(q + \alpha)
\end{equation}
is valid for all sufficiently small $\alpha$. This smallness is defined by the vicinity of support of the function $\chi(q)$ to the boundary of an interval $a \le q \le b$. At large $\alpha$ the translation operator (\ref{eqA4}) shifts the function $\chi(q)$ to the boundary of this interval. One can not say anything about what happens in this case, knowing the operator $S$ only in its domain and additional considerations are needed.

To conserve the norm, i.e. the integral
\begin{equation}\label{eqA5}
    \Scalar{\Psi}{\Psi} = \Int{a}{b} |\Psi(q)|^2 \dd{q},
\end{equation}
it is necessary that everything that moves through one of the boundaries be returned through the other. Since under the integral sign in (\ref{eqA5}) there is a module $|\Psi(q)|$, than the function passing through the boundary can change the phase. Because of the quantum mechanical superposition principle, this phase should be the same for all functions. It is convenient to assume that the ends of the interval are connected forming a circle. Then the problem reduces to circular motion, and at the junction point the phase of wave functions which is essentially topological (geometric, see (\ref{eqA1})) is allowed to occur.

So, there should be a one-parameter family of self-adjoint operators, $S_\theta = S_\theta^+$,
\begin{equation}\label{eqA6}
    S_\theta \Psi(q) = -i\Psi'(q),\quad
    \Domain{S_\theta} = \qty{\Psi \in \Hilbert,\, \Psi' \in \Hilbert;\,
        \Psi(b) = e^{i\theta}\Psi(a)},
\end{equation}
because in this case
\[
    \Scalar{\Phi}{S_\theta \Psi} - \Scalar{S_\theta \Phi}{\Psi} = -i \Phi^*(q) \Psi(q)\Big|_a^b = 0,
\]
if both functions $\Phi$, $\Psi$ are in the domain of operator $S_\theta$. In this case, instead of (\ref{eqA4}), a unitary finite shift operator arises,
\begin{equation}\label{eqA7}
    U_\theta(\alpha) = e^{i\alpha S_\theta},\quad
    U_\theta(\alpha) \Psi(q) = \Psi(q + \alpha),
\end{equation}
the generator of which is a self-adjoint operator $S_\theta$.

To find out if there are any other self-adjoint extensions of the Hermitian operator $S$, one must find its defect indices $(n_+, n_-)$. According to the von Neumann general theory of unbounded operators, they are equal to the number of linearly independent solutions of equations
\begin{equation}\label{eqA8}
    S^+ X_\pm = \pm iX_\pm.
\end{equation}
According to the definition of a conjugate operator (\ref{eqA3}), for any $X \in \Hilbert$ the right-hand side of this equality defines a linear functional on $\chi$, and this functional coincides with $-iX'(q)$ as a distribution in the sense of Schwarz, see, e.g., \cite{Wightman1967Cargese, Richtmyer1978Springer}.

This means that when solving equations (\ref{eqA8}) one should consider the operator $S^+$ as a differentiation operator and find a solution with in the form of distributions with infinitely differentiable on $[a,b]$ test functions $\chi \in \Cinf{[a,b]}$.

In our case
\begin{equation*}
\begin{aligned}
    &X_+ = c_+ e^{-q},\quad c_+ = e^{i\gamma_+} \frac{e^{(b+a)/2}}{\sqrt{\sinh(b-a)}},\quad
        \Scalar{X_+}{X_+} = 1,\\
    &X_- = c_- e^{q},\quad c_- = e^{i\gamma_-} \frac{e^{-(b+a)/2}}{\sqrt{\sinh(b-a)}},\quad
        \Scalar{X_-}{X_-} = 1,
\end{aligned}
\end{equation*}
so $n_+ = n_- = 1$. Since none of the defect indices is zero, then the symmetric (hermitian) extension of operator $S$ may be defined by the following equation \cite{Wightman1967Cargese}
\[
    S_\theta [\chi + c(\chi_+ + \chi_-)] = S\chi + ic(\chi_+ - \chi_-),
\]
where $c$ is an arbitrary complex number and defect indices $(n_+ - 1, n_- - 1) = (0,0)$, are zeros because of (\ref{eqA8}).

Indeed, to calculate, e.g., $\wt{\chi}_+$, we have the equation
\[
    S\chi + i\wt{c} (\wt{\chi}_+ - \wt{\chi}_-) = i\chi + i\wt{c} (\wt{\chi}_+ + \wt{\chi}_-).
\]
This implies
\[
    -i\chi' = i\chi + 2i\wt{c}\, \wt{\chi}_-.
\]
Then at either end of the interval, for example at $q = b$, we have
\[
    \wt{\chi}_-(b) = e^{i\wt{\gamma}_-} e^{(b+a)/2} / \sqrt{\sinh(b-a)} \ne 0,
\]
that leads to $\wt{c} = 0$, due to $\chi'(b) = \chi(b) = 0$ because the function $\chi(q)$ belongs to a class $\Cinf{[a,b]}$. So, there are no \enquote{additional} solutions $\wt{\chi}_+(q)$; the same is valid for $\wt{\chi}_-(q)$.

It is easy to show that the functions
\[
    \Psi(q) = \chi(q) + c [\chi_+(q) + \chi_-(q)]
\]
meet the boundary condition (\ref{eqA1}). Thus, operators $S_\theta$ form the unique one-parametric family of self-adjoint extensions of the operator $S$ defined in (\ref{eqA2}). The eigenfunctions of self-adjoint operator $S_\theta$,
\begin{equation}\label{eqA9}
    \Psi_{\lambda_m}(q) = \frac{1}{\sqrt{b-a}} e^{i\lambda_m q},\quad
    \lambda_m = \frac{2\pi}{(b-a)} (m + \delta),\quad
    m = 0, \pm 1, \pm 2, \ldots,\quad 0 \le \delta = \frac{\theta}{2\pi} < 1,
\end{equation}
form \cite{Neumann1955Princeton} the complete orthonormal set of functions in the space $\Ltwo{[a,b]}$, see also problem 1.29 in \cite{GalitskiKarnakovKogan2013Oxford}.

Let us note that according \cite{Smirnov1964Pergamon} for the range $\Range{S_\theta}$ of $S_\theta$ we have
\begin{equation}\label{eqA10}
   \Range{S_{\theta \ne 0}} = \Ltwo{[a,b]},\quad
   \Range{S_{\theta = 0}} = \Range{\Ol{S}} \subset \Ltwo{[a,b]},
\end{equation}
where $\Ol{S} = S^{++}$ is the closure of the operator $S$. The last equation discriminates a single-valued self-adjoint extension of $S$ from the family of multiple-valued self-adjoint extensions.

Note, if the system is invariant with respect to the reversal of the direction of motion, according to Wigner \cite{Wigner1959AcadPress} the complex conjugate functions must also meet the boundary condition (\ref{eqA1}),
\[
    e^{-i\theta} = \qty(\frac{\Psi(b)}{\Psi(a)})^* = \frac{\Psi^*(b)}{\Psi^*(a)} = e^{i\theta}.
\]
Hence, only two values of the topological phase $\theta$ are possible in T-invariant quantum systems \cite{KowalskiPodlaskiEtAl2002PhysRevA},
\begin{equation}\label{eqA11}
    \text{1) } \theta = 0,\, \delta = 0,\quad
    \text{2) } \theta = \pi,\, \delta = 1/2,
\end{equation}
These values correspond to single- and double-valued representations of two-dimensional rotation group $O(2)$, i.e. integer and half-integer quantization of orbital angular momentum operator.
%%%%%%%%%%%%%%%%%%%%%%%%%%%%%%%%%%%%%%%%%%%%%%%%%%%%%%%%%%%%%%%%%%%%%%%%%%%%%%%%
\section{The generalization of plane wave expansion and half-integer orbital angular momenta}\label{appendixB}
%%%%%%%%%%%%%%%%%%%%%%%%%%%%%%%%%%%%%%%%%%%%%%%%%%%%%%%%%%%%%%%%%%%%%%%%%%%%%%%%
Following the authors of \cite{MorseFeshbach1953McGrawHill}, eq. (11.2.38), we consider the function\footnote{This function differs from $u(\rho, \varphi)$ in section \ref{part22} by the substitution $i \to -i$, $u(\rho, \varphi) = \ol{v}(\rho, \varphi)$, i.e. by the complex conjugation.}
\begin{equation}\label{eqB1}
    v(\rho, \varphi) = \frac{1}{2} \Sum{n=0}{\infty}
        \eps_n i^\frac{n}{2} J_\frac{n}{2} (k\rho) \cos\qty(\frac{n}{2}\varphi),
\end{equation}
where $\eps_n$ is the Neumann coefficient, $\eps_0 = 1$, $\eps_n = 2$ at $n = 1, 2, \ldots$, and $J_\nu(x)$ is the Bessel function.  Given the value of the series from paragraph 5.7.10 of \cite{PrudnikovBrychkovMarichev_2_1986_GordonAndBreach}, and the formulas for Fresnel integrals and its generalizations in section 9.10 of \cite{BatemanErdelyi1_2_1953McGrawHill}, we come to
\begin{equation}\label{eqB2}
    v(\rho, \varphi) = \frac{1}{2} e^{ik\rho \cos\varphi} \qty{1 +
        \erf\qty[e^{i\frac{\pi}{4}} \sqrt{2k\rho} \cos\qty(\frac{\varphi}{2})]}
\end{equation}
Here $\erf(z) = \frac{2}{\sqrt{\pi}} \Erf(z)$ is the probability integral (error function)
\begin{equation}\label{eqB3}
    \erf(z) = -\erf(-z),\quad \ol{{\erf(z)}} = \erf(\ol{z}),
\end{equation}
and if $|z| \to \infty$ we have
\begin{equation}\label{eqB4}
    \erf(z) = 1 - \frac{1}{\sqrt{\pi z^2}} e^{-z^2} \qty[1 + O\qty(\frac{1}{|z|^2})],\quad
        -\frac{3\pi}{4} < \arg z < \frac{3\pi}{4}.
\end{equation}

It can be shown that
\begin{equation}\label{eqB5}
    v(\rho, \varphi) = \sqrt{\frac{i}{\pi}} e^{ik\rho\cos\varphi} F\qty[\sqrt{2k\rho} \cos\qty(\frac{\varphi}{2})],
\end{equation}
where
\[
    F(z) = \Int{-\infty}{z} e^{-it^2} \dd{t} =
        \sqrt{\frac{\pi}{2}} \qty[\frac{1}{\sqrt{2}}e^{-i\frac{\pi}{4}} + C(z^2) - iS(z^2)]
\]
is an integral associated with Fresnel integrals \cite{BatemanErdelyi1_2_1953McGrawHill} $C(z^2)$ and $S(z^2)$. The equation (\ref{eqB5}) is in full agreement with\footnote{With substitutions $i \to -i$, $F(z) \to \Phi(z)$. In the monograph \cite{MorseFeshbach1953McGrawHill} equation (\ref{eq18}), contrary to (\ref{eqB2}), is derived by a method not related to the direct summation of the series.} the formula (11.2.42) from \cite{MorseFeshbach1953McGrawHill}.

By symmetry (\ref{eqB3}) we have
\begin{equation}\label{eqB6}
    v(\rho, \varphi) + v(\rho, \varphi + 2\pi) = e^{ik\rho\cos\varphi},
\end{equation}
what also follows directly from the definition (\ref{eqB1}) and the expansion of a plane wave in cylindrical coordinates \cite{BatemanErdelyi1_2_1953McGrawHill},
\begin{equation}\label{eqB7}
    e^{ikx} = e^{ik\rho\cos\varphi} = \Sum{n=0}{\infty} \eps_n i^n J_n(k\rho) \cos(k\rho) =
        \Sum{m=-\infty}{\infty} i^m J_m(k\rho) e^{im\varphi},\quad
    m = 0, \pm 1, \pm 2, \ldots
\end{equation}
It's worth to emphasize that according to (\ref{eqB2}) and (\ref{eqB4}) the function $v(\rho, \varphi)$ asymptotically (at large distances),
\begin{equation}\label{eqB8}
    v(\rho, \varphi) \simeq e^{ik\rho\cos\varphi} -
        \frac{1}{\sqrt{i8\pi k\rho}} \frac{e^{-ik\rho}}{\cos\qty(\frac{\varphi}{2})},\quad
    k\rho \gg 1,\quad -\pi < \varphi < \pi,
\end{equation}
coincides with the plane wave $\exp(ikx)$ only if the angle $\varphi \ne \pm \pi$, when $v(\rho, \pm \pi) = \frac{1}{2} \exp(ikx)$, because $\erf(0) = 0$.

Note that according to (\ref{eqB7}) the summation over even values of $n = 2l$ (integer orbital angular momenta, $M = m = \pm l$) in (\ref{eqB1}) gives the plane wave, or more precisely, its half. At the same time, after summation over odd values $n = 2l+1$ (half-integer orbital angular momenta, $M = m + 1/2 = \pm l + 1/2$), we get the function
\begin{equation}\label{eqB9}
\begin{gathered}
    w(\rho, \varphi) = 2\Sum{l=0}{\infty} i^{l + 1/2} J_{l + 1/2}(k\rho) \cos[(l+1/2)\varphi] =
        \Sum{m = -\infty}{\infty} i^{|M|} J_{|M|}(k\rho) e^{iM\varphi},\\
    M = m + 1/2,\quad m = 0, \pm 1, \pm 2, \ldots
\end{gathered}
\end{equation}
Acting like when obtaining equalities (\ref{eqB2}), we have
\begin{equation}\label{eqB10}
    w(\rho, \varphi) = e^{ik\rho\cos\varphi}
        \erf\qty[e^{i\frac{\pi}{4}}\sqrt{2k\rho} \cos\qty(\frac{\varphi}{2})]
    \underset{k\rho \gg 1}{\simeq} e^{ik\rho\cos\varphi}.
\end{equation}
According to (\ref{eqB2})
\begin{equation}\label{eqB11}
    v(\rho, \varphi) = \frac{1}{2} \qty[e^{ik\rho\cos\varphi} + w(\rho, \varphi)],
\end{equation}
so both integer and half-integer orbital angular momenta give asymptotically equal contribution to the function $v(\rho, \varphi)$.
%%%%%%%%%%%%%%%%%%%%%%%%%%%%%%%%%%%%%%%%%%%%%%%%%%%%%%%%%%%%%%%%%%%%%%%%%%%%%%%%
\section{Boundary conditions for the radial Dirac equation}\label{appendixC}
%%%%%%%%%%%%%%%%%%%%%%%%%%%%%%%%%%%%%%%%%%%%%%%%%%%%%%%%%%%%%%%%%%%%%%%%%%%%%%%%
The problem is posed correctly if equation (\ref{eq64}) is supplemented by physically acceptable boundary conditions. In the present case, it means that the operator associated with differential operator $H_D$ is the self-adjoint one $\wt{H}$ acting in the Hilbert space $\Hilbert = \Ltwo{\Reals_+}$ of the square integrable functions with Hermitian scalar product
\begin{equation}\label{eqC1}
    \Scalar{\Psi_2}{\Psi_1} = \Int{0}{\infty} \Psi_2^+(\rho) \Psi_1(\rho) \dd{\rho} =
        \Int{0}{\infty} \qty(F_2^* F_1 + G_2^* G_1) \dd{\rho}
\end{equation}
and the norm
\begin{equation}\label{eqC2}
    \Scalar{\Psi}{\Psi} \equiv \norm{\Psi}^2 =
        \Int{0}{\infty} \qty(|F(\rho)|^2 + |G(\rho)|^2) \dd{\rho} < \infty.
\end{equation}

Since any operator $\wt{H}$, associated with $H_D$, is unbounded then according to von Neumann theory one must specify its domain $\Domain{\wt{H}}$. The minimum conditions for linear operator associated with $H_D$ are following
\begin{equation}\label{eqC3}
    H\Psi = H_D\Psi,\quad
    \Domain{H} = \qty{\Psi \in \Ltwo{\Reals_+},\, H_D\Psi \in \Ltwo{\Reals_+}}.
\end{equation}

According to restriction (\ref{eqC2}) and first of the conditions (\ref{eqC3}) the functions $F(\rho)$ and $G(\rho)$ are square integrable,
\begin{equation}\label{eqC4}
    \Int{a}{\infty} |F(\rho)|^2 \dd{\rho} < \infty,\quad
    \Int{a}{\infty} |G(\rho)|^2 \dd{\rho} < \infty,
\end{equation}
at any $a \ge 0$. However, this does not ensure that these functions vanish at $\rho \to \infty$. The examples are
\[
    f_1(\rho) = \exp\qty(-\rho^4\sin^2\rho),\quad
    f_2(\rho) = \rho^2\exp\qty(-\rho^8\sin^2\rho),
\]
and the second of them is not even limited, see section 5.6 in \cite{Richtmyer1978Springer}. Moreover, if their derivatives are also square integrable,
\begin{equation}\label{eqC5}
    \Int{a}{\infty} \abs{\dv{F}{\rho}}^2 \dd{\rho} < \infty,\quad
    \Int{a}{\infty} \abs{\dv{G}{\rho}}^2 \dd{\rho} < \infty,
\end{equation}
the following equations are valid
\begin{equation}\label{eqC6}
    F(\infty) = G(\infty) = 0,
\end{equation}
see section 5.6 of monograph \cite{Richtmyer1978Springer}.

In the case of potentials vanishing at infinity, for the Dirac operator (\ref{eq64}) we have
\[
    H_D^\Par{\infty} = \mqty(1 & \dv{\rho} \\ -\dv{\rho} & -1), \quad \rho \to \infty.
\]
Then for large $a$ and $b$ we get
\begin{equation}\label{eqC7}
\begin{gathered}
    \Int{a}{b} \qty(H_D\Psi)^+ \qty(H_D\Psi) \dd{\rho} =
        \Int{a}{b} \qty(H_D^\Par{\infty}\Psi)^+ \qty(H_D^\Par{\infty}\Psi) \dd{\rho} =\\
    = \Int{a}{b} \qty(\abs{\dv{F}{\rho}}^2 + \abs{\dv{G}{\rho}}^2 + |F|^2 + |G|^2) \dd{\rho} +
        \qty(F^*G + G^*F)\Big|_a^b.
\end{gathered}
\end{equation}
Let $a$ and $b$ go to infinity independently. Due to continuity of $F(\rho)$ and $G(\rho)$ on any finite interval, the integrated term in the RHS of (\ref{eqC7}) vanishes. However, the LHS of this equality vanishes due to the second condition of (\ref{eqC3}). Because of inequalities (\ref{eqC4}) the same is valid for integrals of $|F|^2$ and $|G|^2$. So the remaining integral vanishes too. This means that at random fixed lower limit the integral converges
\begin{equation}\label{eqC8}
    \Int{a}{b} \qty(\abs{\dv{F}{\rho}}^2 + \abs{\dv{G}{\rho}}^2) \dd{\rho} < \infty,
\end{equation}
along with integrals (\ref{eqC5}), that leads to boundary conditions (\ref{eqC6}) at infinity.

Let's now discuss boundary conditions at the origin. The partial integration with regard to equations (\ref{eqC6}) gives:
\begin{equation}\label{eqC9}
    \Int{0}{\infty} \Psi_2^+ (H_D\Psi_1) \dd{\rho} - \Int{0}{\infty} (H_D\Psi_2)^+ \Psi_1 \dd{\rho} =
        \lim_{\rho \to 0} \qty[F_1(\rho) G_2^*(\rho) - F_2^*(\rho) G_1(\rho)].
\end{equation}
For self-adjoint operator the integrated term (the RHS of this equation) should vanish. Thus, the fact whether the operator $\wt{H}$, associated with the differential operator $H_D$, is self-adjoint or not is defined by the wave functions behaviour at $\rho \to 0$, and, consequently, by the potential form at short distances.

Now consider the Coulomb attractive potential which is important for applications,
\begin{equation}\label{eqC10}
    V_C(\rho) = -\frac{q}{\rho},\quad q > 0.
\end{equation}
Radial functions of the Dirac equation (\ref{eq64}) satisfy the set of equations
\begin{equation}\label{eqC11}
\begin{aligned}
    &\dv{F}{\rho} - \frac{J}{\rho}F + \qty(1 + \eps + \frac{q}{\rho})G = 0,\\
    &\dv{G}{\rho} + \frac{J}{\rho}G + \qty(1 - \eps - \frac{q}{\rho})F = 0,
\end{aligned}
\end{equation}
where angular momentum $J$ may be either half-integer or integer, including zero, $J = 0, \pm 1/2, \pm 1, \pm 3/2, \pm 2, \ldots$

The solutions $\Psi_{\eps, J}(\rho)$ of this set are eigenfunctions of the Dirac Hamiltonian $H_D$, and correspond to eigenvalues $\eps$. They form a complete set if only the Hamiltonian $\wt{H}$ associated with $H_D$ is self-adjoint. According to (\ref{eqC9}) boundary conditions at the origin are required. To obtain them one should consider the asymptotic of the solution (\ref{eqC11}) at short distances which for the relativistic Coulomb problem was discussed in the work of Case \cite{Case1950PhysRev}, and in section 10.17 of \cite{Richtmyer1978Springer}.

At $\rho \to 0$ such an asymptotic is defined by the single parameter $\sigma = \sqrt{J^2 - q^2}$. Given the value of angular momentum $J$ this asymptotic changes with the growth of charge.

\textbf{1}. If\footnote{In \cite{KuleshovMurEtAl2017JETP} a misprint is made: in an inequality one should set $\sigma = 0$, and then it coincides with the one given in p.1.} $0 < q < |J|$, then
\begin{equation}\label{eqC12}
    \Psi_\sigma(\rho \to 0) = C_\sigma \qty{u_\sigma \mqty(1 \\ g_\sigma)\rho^\sigma +
        u_{-\sigma}\mqty(1 \\ g_{-\sigma})\rho^{-\sigma}},\quad
    qg_{\pm \sigma} = J \mp \sigma,\quad \sigma = \sqrt{J^2 - q^2} > 0.
\end{equation}
Inserting this into RHS of (\ref{eqC9}) and considering that for self-adjoint operator $\Domain{\wt{H}^+} = \Domain{\wt{H}}$, we get
\begin{equation}\label{eqC13}
    \frac{u_\sigma}{u_{-\sigma}} = \qty(\frac{u_\sigma}{u_{-\sigma}})^* = \tan\theta_\sigma,\quad
    -\frac{\pi}{2} \le \theta_\sigma \le \frac{\pi}{2},\quad \sigma < |J|.
\end{equation}
This boundary condition determines the one-parametric family of self-adjoint operators,
\begin{equation}\label{eqC14}
    H_{\theta_\sigma} \Psi = H_D\Psi,\quad
    \Domain{H_{\theta_\sigma}} = \qty{\Psi \in \Ltwo{\Reals_+},\, H_D\Psi \in \Ltwo{\Reals_+};\,
        \theta_\sigma},
\end{equation}
where $\theta_\sigma$ is given in (\ref{eqC13}).

If $0 < q \le \sqrt{J^2 - 1/4}$, i.e. $1/2 \le \sigma < |J|$, then to satisfy the first condition in (\ref{eqC3}), one should set $u_{-\sigma} = 0$, that corresponds to $\theta_\sigma = \pm \pi/2$. The RHS of (\ref{eqC9}) vanishes automatically and the operator $H$ is self-adjoint without any additional conditions. In this range of $q$ only (\ref{eqC3}) realize so-called \enquote{built-in} boundary condition \cite{Richtmyer1978Springer}.

Note that the constraint $\sigma \ge 1/2$ for the ground state of hydrogen-like atom, $J = -\kap = 1$, means, that the nucleus charge is $Z < \sqrt{3}/2\alpha \approx 118.5$, see section 10.17 in \cite{Richtmyer1978Springer}. At the same time the condition $\sigma > 0$ in the case of point Coulomb potential leads to well known constraint, $Z < \alpha^{-1} \approx 137$, see, e.g., section 13.1 in \cite{AkhiezerBerestetskii1965QED}.

\textbf{2}. In a particular case $q = q_s \equiv |J|$ we have the expansion with logarithms,
\begin{equation}\label{eqC15}
    \Psi_0(\rho \to 0) = C_0 \qty{u_0\mqty(1 \\ g_0) + \wt{u}_0\mqty(1 \\ \wt{g}_0)\ln\rho},\quad
    qg_0 = J - \frac{\wt{u}_0}{u_0},\quad q\wt{g}_0 = J.
\end{equation}
Acting like in the previous range of $q$, we have
\begin{equation}\label{eqC16}
    \frac{u_0}{\wt{u}_0} = \qty(\frac{u_0}{\wt{u}_0})^* = \tan\theta_0,\quad
    -\frac{\pi}{2} \le \theta_0 \le \frac{\pi}{2},\quad \sigma = 0.
\end{equation}
This boundary condition determines the family of self-adjoint operators $H_{\theta_0}^+ = H_{\theta_0}$, see (\ref{eqC14}) with $\sigma = 0$.

\textbf{3}. If $q > |J|$, then we again arrive to the expansion (\ref{eqC12}), where one should set $\sigma = i\tau$. This leads to the boundary condition
\begin{equation}\label{eqC17}
    \qty(\frac{u_\tau}{u_{-\tau}})^* = \qty(\frac{u_\tau}{u_{-\tau}})^{-1},\quad
    \frac{u_\tau}{u_{-\tau}} = e^{2i\theta_\tau},\quad \Im\theta_\tau = 0,\quad
    \tau = \sqrt{q^2 - J^2},
\end{equation}
and to the family of self-adjoint operators
\begin{equation}\label{eqC18}
    H_{\theta_\tau} \Psi = H_D\Psi,\quad
    \Domain{H_{\theta_\tau}} = \qty{\Psi \in \Ltwo{\Reals_+},\, H_D\Psi \in \Ltwo{\Reals_+};\,
        \theta_\tau},
\end{equation}

One-parameter families of self-adjoint radial Dirac Hamiltonians $H_{\theta_\sigma}$ and $H_{\theta_\tau}$ were derived in a different way in \cite{VoronovGitmanTyutin2007TheorMathPhys, VoronovGitmanEtAl2016TheorMathPhys} for integer $J = -\kap = \pm 1, \pm 2, \ldots$ in a three-dimensional Coulomb problem and in \cite{KhalilovLee2011TheorMathPhys} for half-integer values of total angular momentum $J = \pm 1/2, \pm 3/2, \ldots$ in 2D case.

The boundary conditions (p.1--3) define the complete set of wave functions and the energy spectrum of radial Coulomb problem (\ref{eqC11}) for any value of angular momentum, including $J = 0$, and for any charge $q$.
%%%%%%%%%%%%%%%%%%%%%%%%%%%%%%%%%%%%%%%%%%%%%%%%%%%%%%%%%%%%%%%%%%%%%%%%%%%%%%%%
\biboptions{compress,merge,elide}
%\bibliographystyle{model1a-num-names} 
%\bibliography{../my_bibliography.bib}

%%%%%%%%%%%%%%%%%%%%%%%%%%%%%%%%%%%%%%%%%%%%%%%%%%%%%%%%%%%%%%%%%%%%%%%%%%%%%%%%
\end{document}